\documentclass{aa}

\usepackage{natbib}
\usepackage{txfonts}
\usepackage{longtable}
\usepackage{color}

\bibpunct{(}{)}{;}{a}{}{,}

\begin{document}

\def\RG{$R_{\rm G}$}
\def\kms{km\,s$^{-1}$}
\def\teff{$T_{\rm eff}$}
\def\logg{$\log{g}$}
\def\loggf{$\log{gf}$}
\def\vt{$\upsilon_t$}
\def\logP{$\log{P}$}
\def\potex{$\chi_{\rm ex}$}
\def\dexkpc{dex\,kpc$^{-1}$}

\titlerunning{On the $\alpha$-element gradients of the Galactic thin disk using Cepheids}
\authorrunning{Genovali et al.}

\title{On the $\alpha$-element gradients of the Galactic thin disk using Cepheids
\thanks{Based on spectra collected with the UVES spectrograph available at the ESO Very Large Telescope (VLT), Cerro Paranal, Chile (ESO Proposals: 081.D-0928(A), PI: S. Pedicelli; 082.D-0901(A), PI: S. Pedicelli; 089.D-0767(C), PI: K. Genovali).}$^,$\thanks{Tables 1 and 3 are only fully available in electronic form at the CDS via anonymous ftp to cdsarc.u-strasbg.fr (130.79.128.5) or via http://cdsweb.u-strasbg.fr/cgi-bin/qcat?J/A+A/}}

\author{
K. Genovali\inst{1} \and
B. Lemasle\inst{2} \and
R. da Silva\inst{1} \and
G. Bono\inst{1,3} \and
M. Fabrizio\inst{4} \and
M. Bergemann\inst{5,6} \and
R. Buonanno\inst{1,4} \and
I. Ferraro\inst{3} \and \\
P. Fran\c cois\inst{7,8} \and
G. Iannicola\inst{3} \and
L. Inno\inst{1,9} \and
C.D. Laney\inst{10,11} \and
R.-P. Kudritzki\inst{12,13,14} \and
N. Matsunaga\inst{15} \and
M. Nonino\inst{16} \and \\
F. Primas\inst{9} \and
M. Romaniello\inst{9} \and
M.A. Urbaneja\inst{17} \and
F. Th\'evenin\inst{18}
}

\institute{
Dipartimento di Fisica, Universit\`a di Roma Tor Vergata, via della Ricerca Scientifica 1, 00133 Rome, Italy
\and Anton Pannekoek Institute for Astronomy, University of Amsterdam, Science Park 904, PO Box 94249, 1090 GE, Amsterdam, The Netherlands
\and INAF -- Osservatorio Astronomico di Roma, via Frascati 33, Monte Porzio Catone, Rome, Italy
\and INAF -- Osservatorio Astronomico di Teramo, via Mentore Maggini s.n.c., 64100 Teramo, Italy
\and Max-Planck-Institut f\"ur Astronomy, D-69117, Heidelberg, Germany 
\and Institute of Astronomy, University of Cambridge, Madingley Road, CB3 0HA, Cambridge, UK
\and GEPI, Observatoire de Paris, CNRS, Universit\'e Paris Diderot, Place Jules Janssen, 92190 Meudon, France
\and UPJV, Universit\'e de Picardie Jules Verne, 33 rue St. Leu, 80080 Amiens, France
\and European Southern Observatory, Karl-Schwarzschild-Str. 2, 85748 Garching bei M\"unchen, Germany
\and Department of Physics and Astronomy, N283 ESC, Brigham Young University, Provo, UT 84601, USA
\and South African Astronomical Observatory, PO Box 9, Observatory 7935, South Africa
\and Institute for Astronomy, University of Hawaii, 2680 Woodlawn Drive, Honolulu, HI 96822, USA
\and Max-Planck-Institute for Astrophysics, Karl-Schwarzschild-Str.1, D-85741 Garching, Germany
\and University Observatory Munich, Scheinerstr. 1, D-81679 Munich, Germany
\and Department of Astronomy, School of Science, The University of Tokyo, 7-3-1 Hongo, Bunkyo-ku, Tokyo 113-0033, Japan
\and INAF -- Osservatorio Astronomico di Trieste, via G. B. Tiepolo 11, 34143, Trieste, Italy
\and Institute for Astro- and Particle Physics, University of Innsbruck, Technikerstr. 25/8, A-6020 Innsbruck, Austria
\and Laboratoire Lagrange, CNRS/UMR 7293, Observatoire de la C\^ote d'Azur, Bd de l'Observatoire, CS 34229, 06304 Nice, France
}

\date{Received / accepted}

%
%

\abstract{We present new homogeneous measurements of Na, Al and three $\alpha$-elements (Mg, Si, Ca) for 75 Galactic Cepheids. The abundances are based on high spectral resolution ($R$ $\sim$ 38\,000) and high signal-to-noise ratio (S/N $\sim$ 50-300) spectra collected with UVES at ESO VLT. The current measurements were complemented with Cepheid abundances either provided by our group (75) or available in the literature, for a total of 439 Galactic Cepheids. Special attention was given in providing a homogeneous abundance scale for these five elements plus iron (Genovali et al. 2013, 2014). In addition, accurate Galactocentric distances (\RG) based on near-infrared photometry are also available for all the Cepheids in the sample (Genovali et al. 2014). They cover a large fraction of the Galactic thin disk (4.1 $\leq$ \RG\ $\leq$ 18.4~kpc). We found that the above five elements display well defined linear radial gradients and modest standard deviations over the entire range of \RG. Moreover, the [element/Fe] abundance ratios are constant across the entire thin disk; only the Ca radial distribution shows marginal evidence of a positive slope. These results indicate that the chemical enrichment history of iron and of the quoted five elements has been quite similar across the four quadrants of the Galactic thin disk. The [element/Fe] ratios are also constant over the entire period range. This empirical evidence indicates that the chemical enrichment of Galactic Cepheids has also been very homogenous during the range in age that they cover ($\sim$10-300~Myr). Once again, [Ca/Fe] vs. \logP\ shows a (negative) gradient, being underabundant among youngest Cepheids. Finally, we also found that Cepheid abundances agree quite well with similar abundances for thin and thick disk dwarf stars and they follow the typical Mg--Al and Na--O correlations.}


\maketitle

%
%
\section{Introduction}
\label{intro}

Massive and intermediate-mass stars play a crucial role in different fields of modern astrophysics. They are the main contributors in the UV and in the NIR emission of unresolved stellar systems \citep{Crowther2007}. Moreover, they also play a crucial role in the chemical enrichment of the interstellar medium \citep{Venturaetal2002,Decressinetal2007,Maeder2009,Karakasetal2012}. Blue and red supergiants are also fundamental stellar tracers to constrain the metallicity gradients in external early type galaxies \citep{Urbanejaetal2005,Kudritzkietal2008,Bresolin2011,Evansetal2011}.

The distribution of the elements along the disk of external spiral galaxies generally follows a radial gradient with metallicity increasing towards the center. This can be seen for instance in
M31 \citep[e.g.,][]{Sandersetal2012,ZuritaBresolin2012},
M33 \citep{Magrinietal2010a},
M81 \citep[e.g.,][]{Kudritzkietal2012,Stanghellinietal2014},
NGC~300 \citep[e.g.,][]{Stasinkaetal2013},
NGC~3621 \citep[e.g.,][]{Kudritzkietal2014}, 
NGC~5668 \citep[e.g.,][]{Marinoetal2012},
or in large samples such as \citet{Pilyuginetal2014a} and \citet{Sanchezetal2014}.
The Milky Way shows an iron gradient comparable to other galaxies, in particular a well investigated negative iron gradient \citep[see, e.g.,][and references therein]{LuckLambert2011,Genovalietal2013}. In general, shallower [O/H] gradients \citep{VilaCostasEdmunds1992,EdmundsRoy1993,Zaritskyetal1994,MartinRoy1994} have been found in barred galaxies in comparison with the ones in non-barred galaxies \citep[but see also][]{Pilyuginetal2014b,Sanchezetal2014}. 

The gradient in the Milky Way has been well investigated using different stellar tracers of different age and nature: H\,II regions, O/B-type stars, Cepheids, open clusters (OC), and planetary nebulae (PNe). Several authors
investigated, together with the iron gradient, also the $\alpha$-element abundances along the disk \citep[e.g.][]{Yongetal2006,Lemasleetal2007,Lucketal2011,LuckLambert2011,Lemasleetal2013} and provided gradients based on Cepheids spanning from $-$0.039 to $-$0.095~\dexkpc, depending on the element. In particular, \citet{Korotinetal2014} using a non-LTE abundance analysis of the oxygen triplet in the near-infrared give an [O/H] gradient of $-$0.058~\dexkpc.

[O/H] gradients spanning from $-$0.030 to $-$0.08~\dexkpc\ had been inferred from H\,II regions \citep[e.g.,][]{Quirezaetal2006,Rudolphetal2006,Balseretal2011}, very similar to that obtained from O-B1 stars, $-$0.03 and $-$0.07~\dexkpc\ \citep[see, e.g.,][]{SmarttRolleston1997,DaflonCunha2004,Rollestonetal2000}. In particular, \citet{DaflonCunha2004} provide abundance gradients based on $\alpha$ elements in OB stars spanning from $-$0.032 $\pm$ 0.012~\dexkpc\ for oxygen to $-$0.052 $\pm$ 0.014~\dexkpc\ for magnesium.

Planetary nebulae observed up to Galactocentric distances (\RG) = 10~kpc show either shallow [O/H] gradients, from $-$0.02 to $-$0.085~\dexkpc\ \citep{MacielQuireza1999,Costaetal2004,Henryetal2004,PerinottoMorbidelli2006,PottaschBernard2006,Henryetal2010}, or no evidence of radial gradient \citep{Stanghellinietal2006}.

\citet{Twarogetal1997} first suggested that instead of a break in the slope the radial metallicity gradient could experience an abrupt drop of the order of 0.2-0.3~dex around 10~kpc. This feature was also proposed in other studies \citep[e.g.,][]{DaflonCunha2004,Andrievskyetal2004}. The exact location of the possible jump in metallicity has not been properly defined. \citet{Lepineetal2011} claimed the presence of such a metallicity discontinuity in the gradients of both iron and $\alpha$ elements (an average of O, Si, S, Mg, and Ca) based on Cepheids and associated it with the depression in velocity displayed around 10~kpc in the rotation curve of the Galaxy.

The theoretical scenario that better seems to describe the gradient of iron and $\alpha$ elements in the Milky Way is the so-called inside-out scenario \citep[e.g.,][]{MatteucciFrancois1989,Chiappinietal1997,PrantzosBoissier2000,Chiappinietal2001,MollaDiaz2005,Fuetal2009}. According to this model, the Milky Way primarily forms during two episodes of gas infall (the first giving birth to the halo and the bulge, the second producing the disk) with an almost independent evolution between the halo and the thin disk.

Although the inside-out model is not unique, it reproduces the majority of the observed features of the Milky Way. \citet{Cescuttietal2007} computed new radial gradients for numerous elements and closely reproduced Cepheid-based observations based on the nucleosynthesis prescription provided by \citet{Francoisetal2004}, who adopted the star formation and infall laws based on observed features of the Milky Way. The most important factor in reproducing the [element/Fe] vs. [Fe/H] relations, as well as the solar absolute abundances in the solar neighborhood, is the combination of the yields from supernovae (SNe) type~Ia and II whose progenitors are, respectively, low/intermediate mass and massive stars. In this context, it is worth mentioning that $\alpha$-element abundances play also a crucial
role in constraining the plausibility of the physical assumptions (instantaneous mixing, instantaneous recycling) adopted to compute chemical evolution models \citep{Spitonietal2009}.

The $\alpha$-enrichment of Cepheids is also often investigated by means of abundance gradients scaled to iron. Iron is mainly produced in SNe~Ia, with contributions from SNe~II, whereas $\alpha$-elements are the principal yields of core-collapse SNe, but with contributions of SNe~Ia for Ca and Si. Therefore, the [$\alpha$-element/Fe] ratio is an interesting diagnostic to constrain the chemical enrichment history. A comprehensive study of open
clusters abundance gradients was provided by \citet{Yongetal2012} collecting results from different groups \citep{Bragagliaetal2008,Carraroetal2007a,Carraroetal2007b,Chenetal2003,Frieletal2002,Frieletal2010,Jacobsonetal2008,Jacobsonetal2009,Jacobsonetal2011a,Jacobsonetal2011b,Magrinietal2009,Magrinietal2010b,Pancinoetal2010,Sestitoetal2006,Sestitoetal2007,Yongetal2005}. They found different trends for the [$\alpha$/Fe] ratios based on open clusters. Ca abundances attain solar values over the entire range of Galactocentric distances, while Mg and Si display a marginal evidence of a slope ($\sim$0.01-0.02~\dexkpc). On the other hand, the average [$\alpha$/Fe] ratio attains (see their Fig.~21) enhanced values in the outer disk due to the increase in O and Ti abundances. However Ti abundances, as noted by the anonymous referee, display large variations in different investigations and may be dominated by systematic effects. In general, the large majority of investigations on radial dependence based on OCs found a linear gradient of approximately $-$0.06~\dexkpc in the range of 5-10~kpc. Concerning the outer disk, however, several authors found evidence of a flattening of the gradients ([Fe/H] $\sim$ $-$0.3) for distances larger than 12-14~kpc \citep[][and references therein]{Magrinietal2009,Pancinoetal2010,Yongetal2012}.

The abundance gradients of the elements along the disk of the Milky Way provide strong constraints to chemical evolution models as they are connected to the evolution of the Galaxy disk. In this context, the $\alpha$ elements play a key role since they are good tracers of the chemical enrichment history of stellar populations. In particular, an overabundance of $\alpha$ elements is typically associated with a fast chemical enrichment in which iron played a minor role \citep{Tinsley1979,Matteucci2003}. However, the theoretical and empirical scenario is far from being settled, since the Initial Mass Function \citep[IMF,][]{Caluraetal2010} and the radial gas flows \citep{Spitonietal2013} can also affect the abundance gradients. In particular, gas flows can have similar effects on the gas density distribution, and in turn they can have an impact on the star formation rate \citep[][and references therein]{Colavittietal2008}. Recent chemical evolution models including both radial gas flows and radial stellar migrations \citep{,Kubryketal2014a,Kubryketal2014b,Minchevetal2013,Minchevetal2014} indicate a steady decrease in the slope as a function of time.

In this paper we investigate the gradient associated to three $\alpha$ elements (Mg, Si, Ca), together with Na and Al, focusing on the inner and outer disk regions. The paper is organized as follows: in Sect.~2 we briefly recall the observations and data analysis; in Sect.~3 we present and discuss the radial gradients, and in Sect.~4 the metallicity distributions. In Sect.~5 we discuss some correlations involving the derived abundances, and finally, in Sect.~6 we give a summary of our findings.

%
%
\section{Observations, data reduction and analysis}
\label{obs}

\subsection{Spectroscopic data}
\label{spec_data}

In this work we used the same high-resolution ($R$ $\sim$ 38\,000) and high signal-to-noise (S/N) ratio spectra analyzed by \citet[][hereafter G14]{Genovalietal2014}. The spectra of 75 Galactic Cepheids were collected with the UVES spectrograph at the ESO VLT (Cerro Paranal, Chile). In particular, we collected 122 spectra using two different instrument settings:
$i)$ the former one makes use of the UVES DIC\,2 configuration which allow the blue and red arms to operate in parallel. With this setting, we collected (between October 2008 and April 2009\footnote{081.D-0928(A) and 082.D-0901(A), PI: S. Pedicelli}) 80 spectra of 74 stars, with wavelength ranges of $\sim$3760--4985~\AA, $\sim$5684--7520~\AA, and $\sim$7663--9458~\AA;
$ii)$ the latter one uses the UVES red arm configuration and the cross disperser CD\,3. By adopting this setting we collected (between May and September 2012\footnote{089.D-0767(C), PI: K. Genovali}) 42 spectra of a control sample of 11 Cepheids, with wavelength ranges of $\sim$4786--5750~\AA\ and $\sim$5833--6806~\AA.

\begin{figure*}
\centering
\resizebox{0.9\hsize}{!}{\includegraphics[angle=-90]{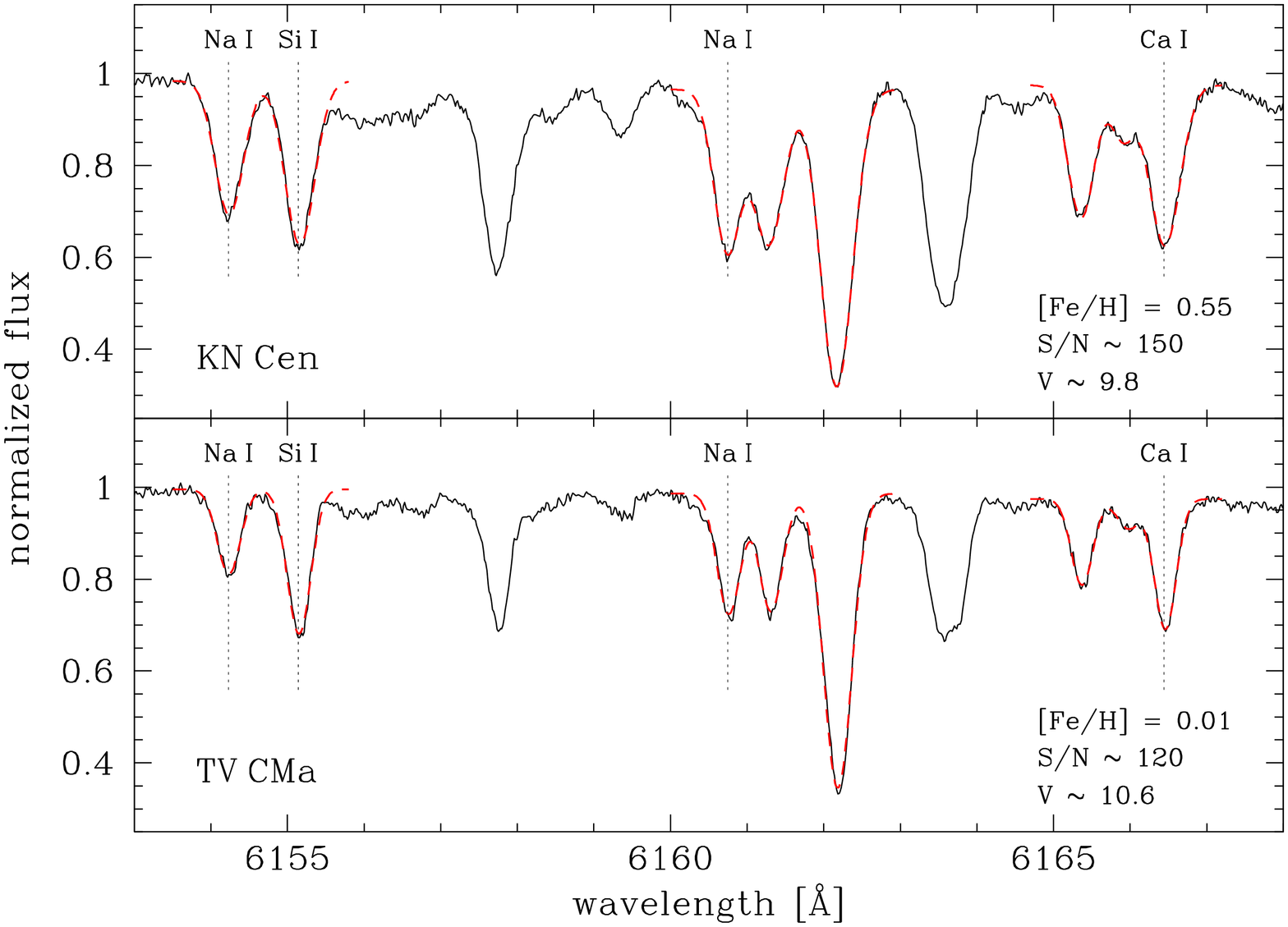}}
\caption{High-resolution ($R \sim 38\,000$) UVES spectrum of KN\,Cen and TV\,CMa. The apparent visual magnitude and the S/N in the spectral range $\lambda \sim 5650-7500$~\AA\ are also labeled. The vertical dashed lines display some of the spectral lines (\ion{Na}{i} 6154.23, \ion{Si}{i} 6155.14, \ion{Na}{i} 6160.75, \ion{Ca}{i} 6166.44~\AA) adopted to estimate the abundances.}
\label{spectra}
\end{figure*}

The 11 Cepheids (V340\,Ara, AV\,Sgr, VY\,Sgr, UZ\,Sct, Z\,Sct, V367\,Sct, WZ\,Sgr, XX\,Sgr, KQ\,Sco, RY\,Sco, V500\,Sco) were used as a control sample since we have collected from four to six spectra each and with both the instrumental configurations (with the exception of V500\,Sco, which has 4 spectra collected only with the second instrument setting). The S/N ratios are typically better than $\sim$100 for all the \'echelle orders in the case of the first instrumental configuration (see examples in Fig.~\ref{spectra}), and ranges from $\sim$50 to roughly 300 for the second one. All the spectra were reduced using the ESO UVES pipeline Reflex v2.1 \citep{Ballesteretal2011}.

The stars BB\,Gem and GQ\,Ori were initially present in the sample of G14. BB\,Gem still has an uncertain status as it is classified either as a classical or as a type~II Cepheid in different catalogs \citep[see e.g.][]{Harris1985,Loomisetal1988,Bersieretal1997}. It has been excluded from the current investigation, since its uncertain classification could lead to a wrong distance determination or to a wrong interpretation of the abundance pattern. The spectrum of GQ\,Ori has low S/N, it is not good enough to provide a reliable abundance determination, and it was also excluded.

\subsection{Atmospheric parameters and abundances}
\label{meth}

We adopted the same iron abundances and atmospheric parameters derived by G14. The iron abundances are based on the equivalent widths (EW) of about 100-200 \ion{Fe}{i} and about 20-40 \ion{Fe}{ii} lines, the number of lines
depending on the spectral range used. The number of lines also varies according to the metallicity and to the spectral type of the star. To determine the atmospheric parameters, we set a limit of EW $<$ 120~m\AA\ in order to remain in the linear part of the curve of growth. For the objects where the number of weak lines was too small, we increased the limit to 180~m\AA. This slightly increases the uncertainties affecting the correlated atmospheric parameters, namely the effective temperature and the microturbulent velocity. For more details on the impact that typical uncertainties on effective temperature, surface gravity, and microturbulent velocity have on the iron abundances, see Table~2 of G14.

The effective temperature (\teff) of individual spectra was estimated using the line depth ratio (LDR) method \citep{KovtyukhGorlova2000}. The estimated values of \teff\ were validated to make sure that the \ion{Fe}{i}
abundances do not depend on the excitation potential (\potex), i.e., the slope of [\ion{Fe}{i}/H] vs. \potex\ should be as close to zero. The surface gravity (\logg) was derived through the ionization equilibrium between \ion{Fe}{i} and \ion{Fe}{ii} lines, and the microturbulent velocity (\vt) by minimizing the slope in the [\ion{Fe}{i}/H] vs. EW plot.

Concerning the Na, Al, and $\alpha$ (Mg, Si, Ca) elements, we used the linelist provided by \citet{Lemasleetal2013}, with the same atomic parameters (\potex\ and \loggf) listed in their Table~A.1, but with small differences in the number of lines. We used the six \ion{Al}{I} lines, two lines for \ion{Na}{I} instead of three (the line at 5688.21~\AA\ is often too strong), the nine \ion{Ca}{I} lines (the \ion{Ca}{II} line was not included), and 14 \ion{Si}{I} lines (5665.56~\AA\ was included but used only for the P89 stars). For magnesium, we preferred to adopt only the abundances provided by the line \ion{Mg}{I} at 5711.09~\AA\ because those at 8712.69 and 8736.02~\AA\ are not available in the wavelength range of the 11 control sample stars. Note that the line at 5711.09~\AA\ is affected by non-LTE effects. However, \citet{Merleetal2011} found that the non-LTE correction to the EWs is smaller than 10\% in the quoted stellar parameter regime.

To minimize any systematic bias in the continuum estimate due to the subjectivity of the operator, three of us have independently performed EW measurements on a sample of selected lines. The EWs for several elements were also measured using the {\it Automatic Routine for line Equivalent widths in stellar Spectra} \citep[ARES,][]{Sousaetal2007}, and double-checked using the {\it splot} task of IRAF\,\footnote{{\it Image Reduction and Analysis Facility}, distributed by the National Optical Astronomy Observatories (NOAO), USA.}. The internal dispersion is smaller than 6~m\AA\ and there is no evidence of systematics.

The abundances were derived with {\it calrai}, a spectrum synthesis code originally developed by \citet{Spite1967} and regularly improved since then. The code computes synthetic spectra over a large grid of stellar atmospheres in which we can interpolate, using 1-D, hydrostatic, plane-parallel, and spherical LTE model atmospheres \citep[MARCS,][]{Gustafssonetal2008}. For all the elements studied here, we assumed the standard solar abundances provided by \citet{Grevesseetal1996}, namely A(Fe)$_\odot$ = 7.50, A(Na)$_\odot$ = 6.33, A(Al)$_\odot$ = 6.47, A(Mg)$_\odot$ = 7.58, A(Si)$_\odot$ = 7.55, and A(Ca)$_\odot$ = 6.36.

Table~\ref{table_ab_alpha_spec} lists the abundances obtained from individual spectra. Columns 3 and 4 show the iron abundances and the number of \ion{Fe}{i} and \ion{Fe}{ii} lines used by G14. The other columns show our results for the abundances of Na, Al, and $\alpha$ elements, together with the number of lines used. In Table~\ref{table_ab_alpha_mean} we list the mean abundances computed for the stars with multiple spectra.

\subsection{Data available in the literature}
\label{liter}

We compared our abundance estimates with the results provided by similar studies available in the literature:
\citet[][hereafter LEM]{Lemasleetal2013}, \citet[][LII]{Lucketal2011}, \citet[][LIII]{LuckLambert2011}, and \citet[][YON]{Yongetal2006}. By comparing the stars in common among the different data sets, we evaluated the systematic difference among them. The mean difference between our measurements and those of LEM, LII, and LIII are quite small, about 0.13~dex in modulus or smaller for all the elements investigated here. The difference for Mg is on average 0.23~dex more abundant in our study than in LII's. The differences are larger in the comparison with YON's measurements, which have abundances systematically smaller than ours and range from 0.18~dex for Mg up to 0.31~dex for Fe. The details on these comparisons are listed in Table~\ref{table_diff_ab}, where we show the zero-point differences obtained by G14 for the iron abundance ratios together with our determinations for the other elements. Each pair of data sets was chosen aiming to maximize the number of stars in common between them. In order to provide a homogeneous abundance scale for Galactic Cepheids, we applied these zero-point differences to the quoted data sets, putting them in the same scale of our current sample. The element abundances available in the literature and the rescaled values are listed in columns from 2 to 15 of Table~\ref{table_liter}.

The priority in using the abundances from the literature follows the same approach adopted by G14: firstly, we adopt the abundances provided by our group, i.e., this study and then results from LEM, and finally those provided by the other studies, namely LIII, LII, and YON in this order. We ended up with a sample of 439 Cepheids, with a homogeneous abundance scale for Fe, Na, Al, Mg, Si, and Ca.

%
%
\section{$\alpha$-element gradients}
\label{alpha_dist}

In this section we investigate the radial gradients of Na, Al, and three $\alpha$ elements (Mg, Ca, Si) across the Galactic disk using our sample of 75 classical Cepheids plus a sample of 364 Cepheids available in literature. 
The iron gradient for our sample was provided by G14 together with a homogenous iron scale for the remaining objects. The current approach when compared with similar investigations available in the literature has two indisputable advantages:
{\it i)} We use accurate elemental abundances based on high-resolution, high signal-to-noise spectra. Moreover, the approach to constrain the intrinsic parameters (surface gravity, effective temperature, microturbulent velocity) is identical and we also use similar line lists;
{\it ii)} The individual Cepheid distances were estimated using the same near-infrared Period-Wesenheit relations. The key advantages of distances based on this diagnostic is that they are independent of uncertainties affecting reddening corrections and minimally affected by the metallicity dependence \citep{Innoetal2013}.

\begin{figure}
\centering
\resizebox{\hsize}{!}{\includegraphics{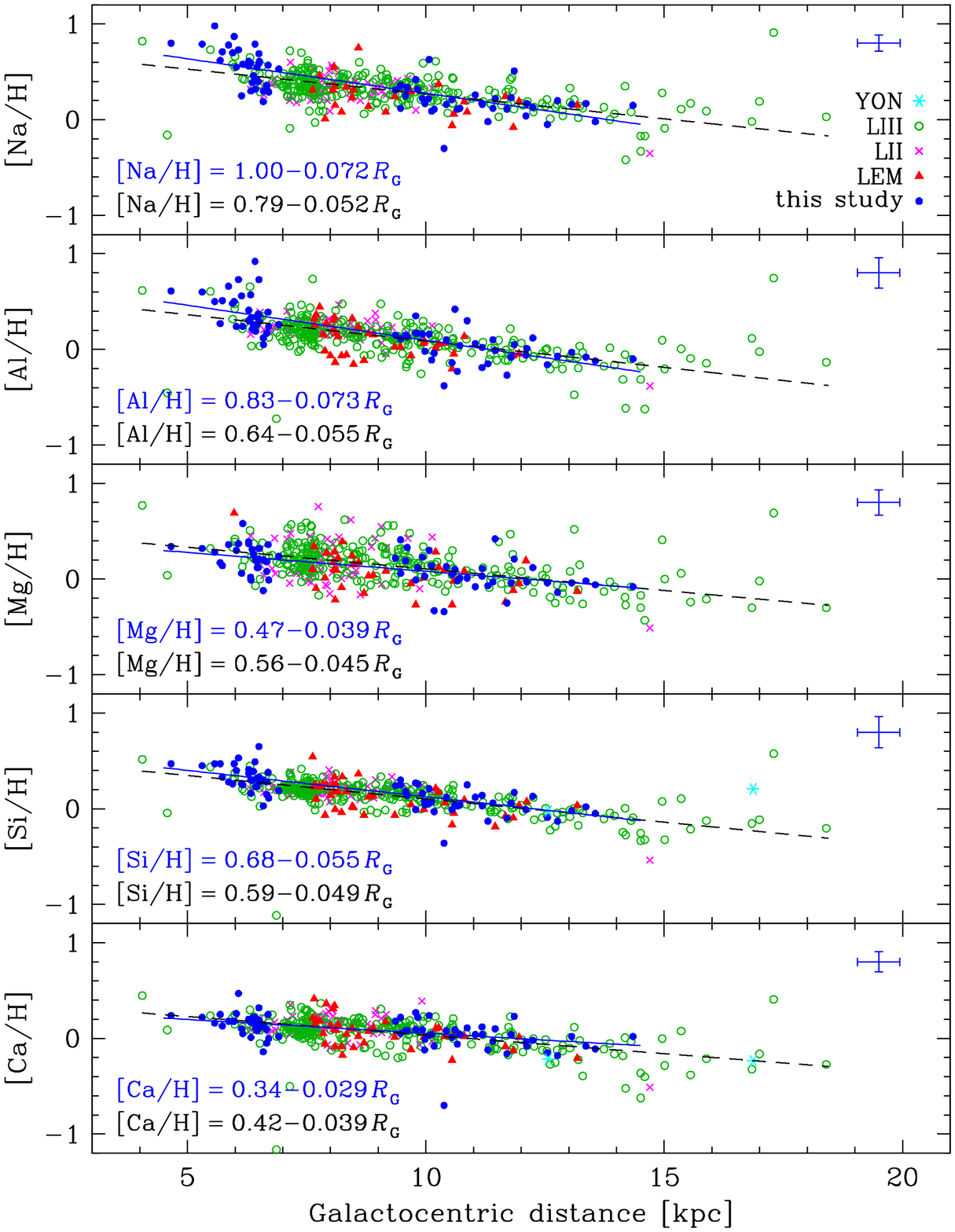}}
\caption{Abundances of Na, Al, and $\alpha$ elements as a function of \RG. Our results (filled blue circles) are compared with those of \citet[][YON, cyan asterisks]{Yongetal2006}, \citet[][LII, magenta crosses]{Lucketal2011}, \citet[][LIII, open green circles]{LuckLambert2011}, and \citet[][LEM, red triangles]{Lemasleetal2013}. The blue solid line shows the linear regression of our Cepheid sample, while the black dashed line the linear regression of the entire Cepheid sample. The blue error bars display the mean spectroscopic error of the current sample. The abundances available in the literature have similar errors.}
\label{xh_Gdist}
\end{figure}

Figure~\ref{xh_Gdist} shows the abundances scaled to hydrogen of Na, Al, and of the three $\alpha$ elements as a function of \RG\ for the final sample. Objects plotted in this figure include the current 75 Cepheids (filled blue circles) plus 38 Cepheids from LEM (red triangles), 263 from LIII (open green circles), 61 from LII (magenta crosses), and two from YON (cyan asterisks). The individual Galactocentric distances of the Cepheids in the total sample have been estimated by G14. They adopt the Near-Infrared Period-Wesenheit relations derived by \citet{Innoetal2013} and assume a solar Galactocentric distance of 7.94 $\pm$ 0.37 $\pm$ 0.26~kpc \citep[see][and references therein]{Groenewegenetal2008,Matsunagaetal2013}. The \RG\ values are listed in our Table~\ref{table_ab_alpha_mean} and in their Table~4. The typical final uncertainty on the distances is $\sim$5\% and is mainly due to the Period-Wesenheit zero-point calibration \citep[for more details see][]{Innoetal2013}.

The abundances from the literature plotted in Fig.~\ref{xh_Gdist} were corrected by adopting the zero-point differences listed in Table~3. This figure also shows the linear Least Squares fits to the current sample of 75 Cepheids (blue solid line) and to the entire sample (439, black dashed line). A {\it biweight} procedure \citep{Beersetal1990} was adopted to remove outliers from the data fitted. The slopes and the zero-points of these two radial gradients are labeled in the panels. The slopes and the zero-points of the fits based on the entire sample together with their uncertainties and standard deviations are listed in columns 2 to 4 of Table~\ref{table_slopes}.

A glance at the data plotted in this figure shows that the occurrence of a radial gradient is solid for the five investigated elements. The main difference is that the standard deviations scale with the number of lines adopted to estimate the abundances. Indeed, the dispersion of the Mg gradient, based on a single line, is almost a factor of two larger than the dispersion of the Si gradient, based on 14 lines. The radial gradients of the five investigated elements attain, within the errors, similar slopes. This finding supports similar results by \citet{Lemasleetal2007,
Lemasleetal2013} and by LII+LIII. The above result becomes even more compelling if we take account of the similarity with the iron radial gradient ($-$0.060$\pm$0.002~\dexkpc) showed by the same Cepheids (G14). The abundances of Na, Al and of the three $\alpha$-elements seem to show a flattening for distances larger than $\sim$13~kpc. Thus supporting a similar trend found using open clusters \citep{Yongetal2012}. However, firm conclusions are hampered by the increased spread in abundance and by the paucity of the sample in the outermost disk regions.

The current slopes also agree, within 1$\sigma$, with similar estimates available in the literature. In our results the slopes range from $-$0.055 $\pm$ 0.003 for [Al/H] to $-$0.039 $\pm$ 0.002~\dexkpc\ for [Ca/H]. The slopes estimated by LII+LIII for the same elements range from $-$0.053 $\pm$ 0.004 to $-$0.040 $\pm$ 0.004~\dexkpc, while those estimated by LEM range from $-$0.046 $\pm$ 0.013 to $-$0.044 $\pm$ 0.012~\dexkpc. The reader interested in a more detailed analysis of the difference among the different data sets is referred to columns 6 to 9 of Table~\ref{table_slopes}.

\subsection{Comparisons with independent radial gradients}
\label{comp}

\begin{figure}
\centering
\resizebox{\hsize}{!}{\includegraphics{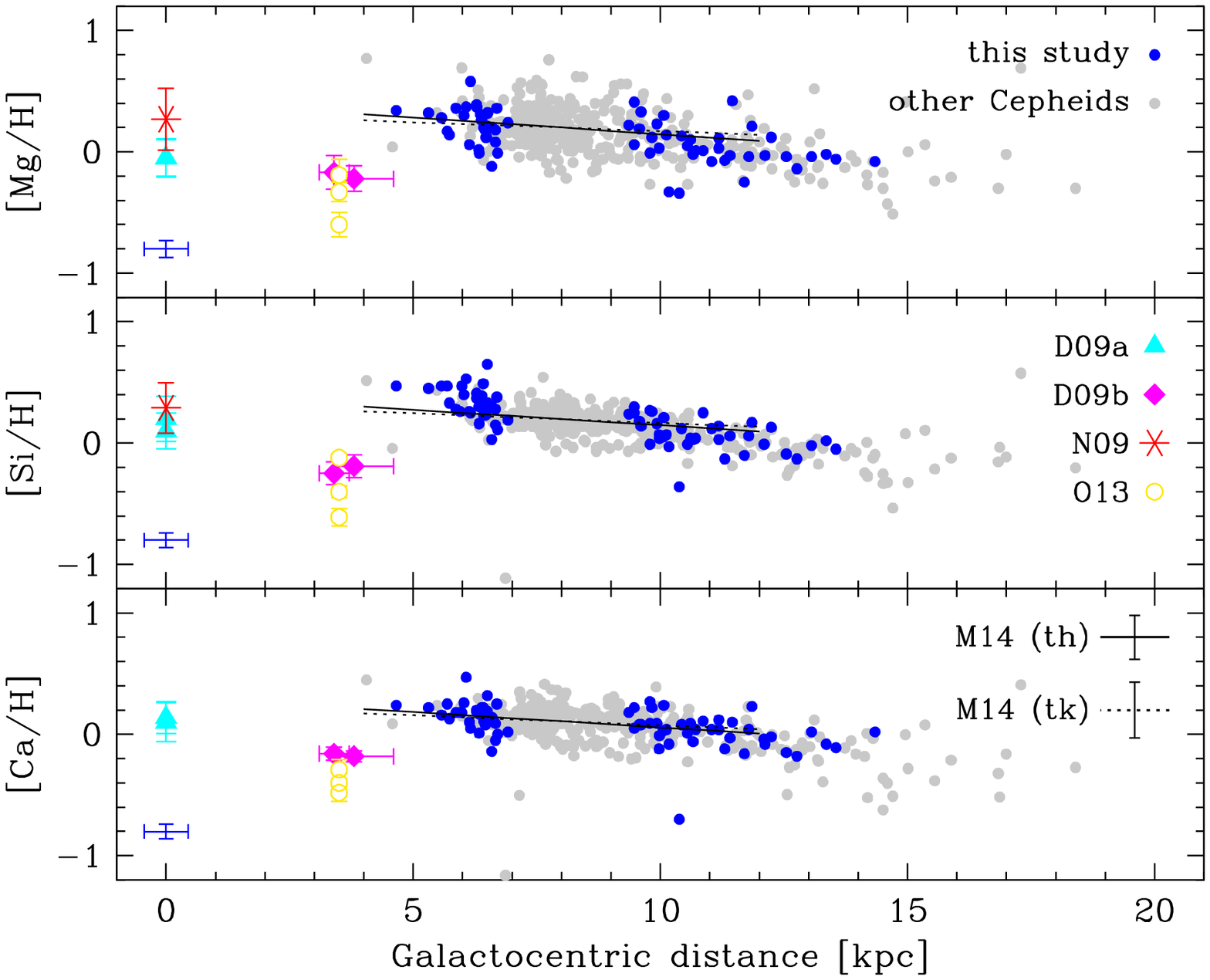}}
\caption{Abundances of $\alpha$ elements as a function \RG. Cepheid stars (filled circles) are compared with RSGs in the Galactic center analyzed by \citet[][D09a, cyan triangles]{Daviesetal2009a}, RSGs in the Scutum cluster analyzed by \citet[][D09b, magenta diamonds]{Daviesetal2009b} and by \citet[][O13, yellow open circles]{Origliaetal2013}, and with the mean abundance of two Luminous Blue Variables (LBVs) in the Quintuplet cluster measured by \citet[][N09, red asterisk]{Najarroetal2009}. The radial gradients derived by \citet[][M14]{Mikolaitisetal2014} for thin (solid line) and thick (dotted line) disk stars in the Gaia-ESO survey are also shown.}
\label{xh_Gdist_SG05}
\end{figure}

Figure~\ref{xh_Gdist_SG05} shows the comparison between the slopes of the $\alpha$-element gradients based on Cepheids and similar abundances for Galactic field stars. We focussed our attention on the $\alpha$-element abundances provided by \citet{Daviesetal2009a,Daviesetal2009b} for two red supergiants (RSGs) located in the Galactic center and two RSGs in the Scutum cluster, plus two Luminous Blue Variables belonging to the Quintuplet cluster measured by \citet{Najarroetal2009}. Note that their $\alpha$-element and iron abundances were rescaled to the abundances of the solar mixture adopted in the current investigation \citep{Grevesseetal1996}. Data plotted in this figure indicate that young stellar tracers located in the innermost Galactic regions attain $\alpha$-element abundances that are, at fixed Galactocentric distance, lower than the radial gradients of Galactic Cepheids. To further constrain this difference we extrapolated the Cepheids gradient to the Galactocentric distance of the above targets. We found that the difference $\Delta$[$\frac{1}{3}$([Mg/H]+[Si/H]+[Ca/H])] between the expected Cepheid gradient and the Scutum stars is 0.3--0.6~dex and becomes of the order of 0.1--0.4~dex for the targets located in the Galactic center.

Note that $\alpha$-element abundances provided by these authors are based on a different approach (spectrum-synthesis vs. EWs). To constrain the possible occurrence of systematic differences in the abundances we also plotted the $\alpha$-element abundances recently provided by \citet{Origliaetal2013} using high spectral resolution ($R \sim 50\,000$) NIR (Y,J,H,K) spectra collected with GIANO at the Telescopio Nazionale Galileo (TNG). Note that these authors observed the same targets observed by \citet{Daviesetal2009b}. The comparison shows, once again, that RSGs located in the near end of the Galactic bar have $\alpha$ abundances lower than classical Cepheids located in the inner edge of the Galactic thin disk.

Figure~\ref{xh_Gdist_SG05} also shows the comparison with the radial gradients provided by \citet[][hereafter M14]{Mikolaitisetal2014} using thin (solid black line) and thick (dotted black line) disk field stars observed within the Gaia-ESO survey. We adopted their {\it clean} sample, which only includes dwarfs, with \logg\ $>$ 3.5. To overcome possible differences between Cepheid and spectroscopic distances of field dwarfs, in plotting their radial gradients we adopted the zero-points of our gradients at the solar Galactocentric distance. The comparison shows that the slopes agree quite well over the entire range of Galactocentric distances covered by the Gaia-ESO survey (4-12~kpc).

\subsection{Age dependence of the [$\alpha$/H] ratios}
\label{dep_xh_logP}

The above findings rely on the Galactocentric distance as independent variable. This is a crucial parameter to identify possible radial trends. However, one of the key issues in dealing with the chemical enrichment history of the thin disk is the age dependence. Detailed chemo-dynamical models suggest a strong dependence of the metallicity gradients for ages older than 1-3 Gyr \citep[][]{Kubryketal2014a,Kubryketal2014b,Minchevetal2013,Minchevetal2014}. This means a steady decrease in the metallicity with increasing age. Moreover, they also predict a strong dependence on stellar migrations.

Classical Cepheids can play a crucial role in this context, since there are solid theoretical and empirical arguments suggesting that the pulsation period is strongly anti-correlated with age \citep{Bonoetal2005}. An increase in stellar mass causes, for intermediate-mass stars during central helium burning phases, an increase in the mean luminosity of the blue loop. Plain physical arguments rooted on the Stefan-Boltzman relation, on the fundamental pulsation relation and on the mass-luminosity relation indicate that the increase in luminosity causes a decrease in surface gravity, and in turn, an increase in the pulsation period.  

\begin{figure}
\centering
\resizebox{\hsize}{!}{\includegraphics{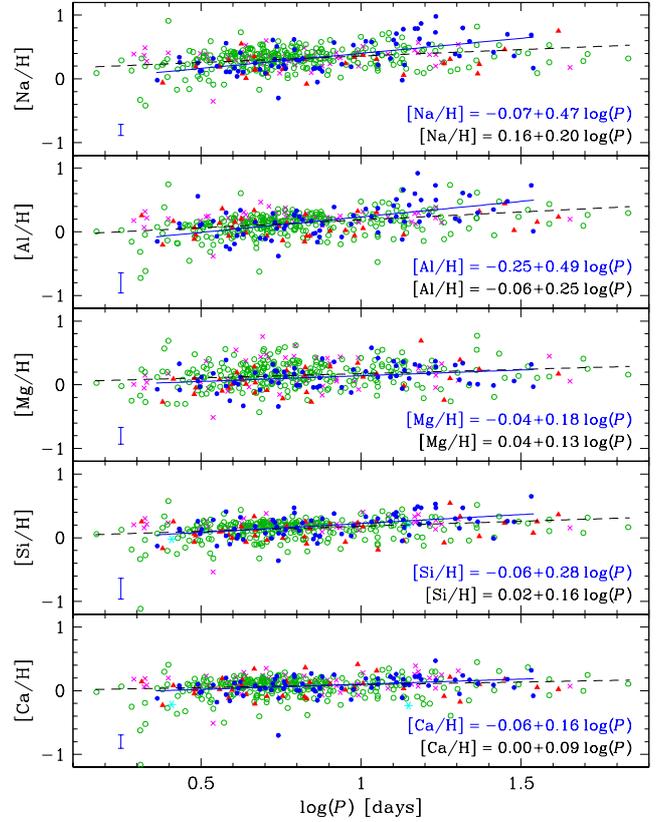}}
\caption{Abundances of Na, Al, and $\alpha$ elements as a function of the pulsation period. Symbols and colors are the same as in Fig.~\ref{xh_Gdist}.}
\label{xh_logP}
\end{figure}

To constrain the age dependence of the metallicity gradients, Fig.~\ref{xh_logP} shows the same elemental abundances plotted in Fig.~\ref{xh_Gdist}, but as a function of the logarithmic period. Data plotted in this figure show a well defined positive gradient as a function of the pulsation period. The slopes of the $\alpha$-elements attain similar
values, while for Na and Al they are systematically larger (see labelled values). This evidence further supports the hydrostatic nature of both Na and Al due to their steady increase with pulsation period (stellar mass). This finding agrees quite well with yields predicted by nucleosynthesis models \citep{Arnett1971,LimongiChieffi2012}.

\subsection{Radial gradient of [element/Fe]}
\label{alpha_Fe}

\begin{figure}
\centering
\resizebox{\hsize}{!}{\includegraphics{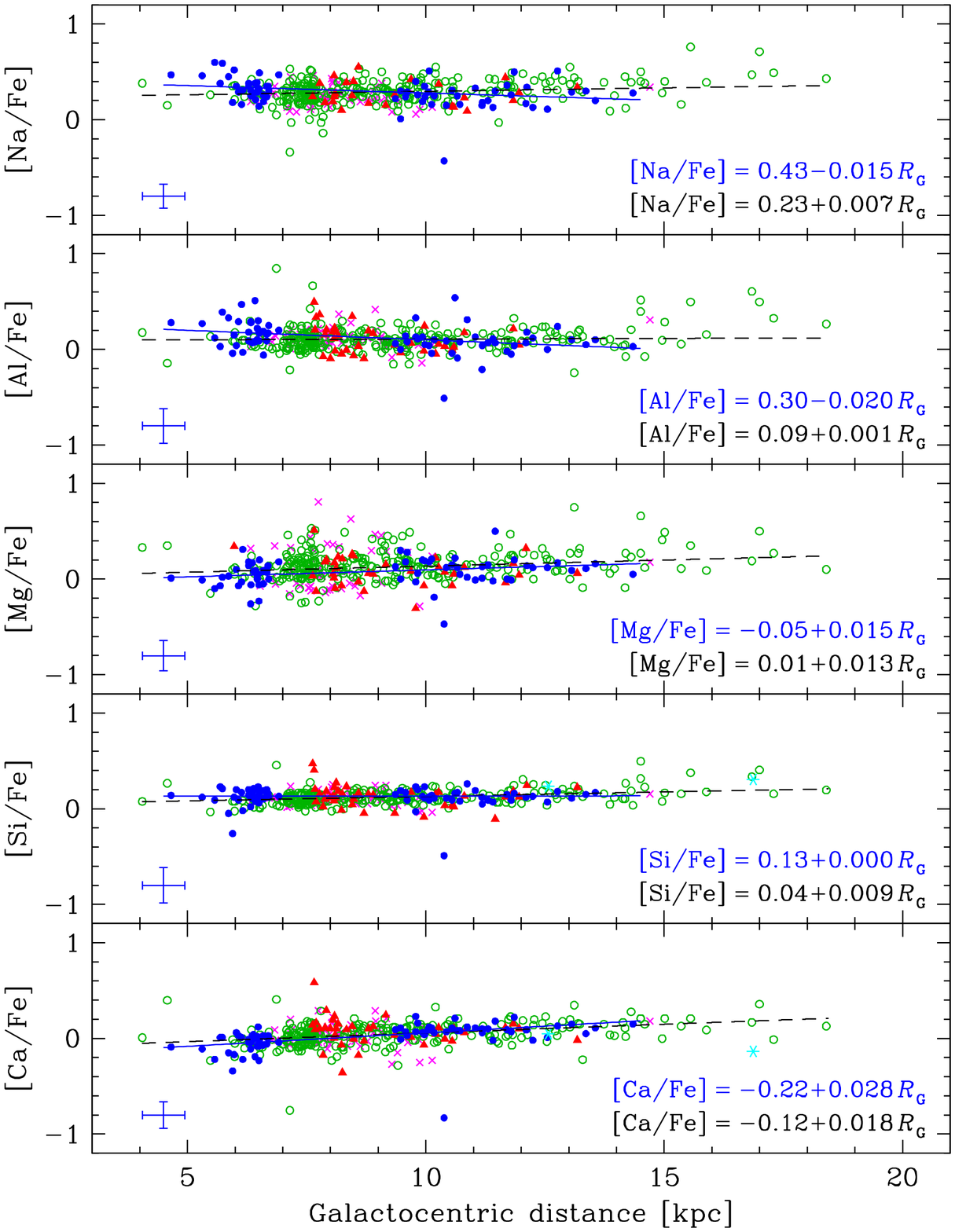}}
\caption{The same as Fig.~\ref{xh_Gdist}, but the abundances are scaled to iron.}
\label{xfe_Gdist}
\end{figure}

The similarity of the slopes between iron and the current elements suggested to investigate the [element/Fe] radial gradients as a function of the Galactocentric distance. Data plotted in Fig.~\ref{xfe_Gdist} show that the ratio is on average quite flat across the entire thin disk. We performed a {\it biweight} linear Least Squares fit over the entire sample and we found that the slopes are vanishing for Na, Al, and Si. There is a mild evidence of positive slope for Mg and Ca. However, the standard deviation of the former element is a factor of two larger than for Si (0.15 vs. 0.06~dex). Thus suggesting that the slope has to be cautiously treated. The positive slope for Ca appears more solid (see values in Table~\ref{table_diff_ab}), and indeed, the slope attains the largest value among the investigated elements.

The above findings bring forward a few interesting consequences:

{\it i)} The radial gradients of the [element/Fe] ratios are slightly positive if not zero. This evidence indicates that the chemical enrichment history of iron and of the other five elements has been quite similar across the Galactic thin disk. This finding is also supported by the small standard deviations of the quoted ratios. The standard deviation of Si, the element with the most accurate measurements, can be explained as the consequence of
the intrinsic error. These results become even more compelling if we take account of the range in iron abundance covered by Galactic Cepheids ($\sim$1~dex).

{\it ii)} The [$\alpha$/Fe] ratios are typically considered as tracers of the star formation activity. This means that the [$\alpha$/Fe] ratios are also tracers of stellar mass, and in particular, of gas and dust mass. The lack of a clear negative gradient, between the high (inner) and the low (outer) density regions of the thin disk, is suggesting that the [$\alpha$/Fe] radial gradients are affected by other parameters such as radial migration of stars or radial gas flows.

{\it iii)} There is evidence of a mild enhancement in the investigated elements and of an increase in the intrinsic scatter when moving toward the outer disk. The synthesis of Na, Al and Mg takes mainly place in massive stars during hydrostatic central He, C or Ne burning phases and current prediction suggest that their production is metallicity dependent \citep{WoosleyWeaver1995,LimongiChieffi2012}. However, these findings further support the evidence that the ratios [Na/Fe] and [Al/Fe] are constant across the Galactic thin disk \citep{McWilliametal2013}.

However, the number of Cepheids with Galactocentric distance larger than 13~kpc is limited. New identifications of classical Cepheids in the outer disk from the ongoing long-term photometric surveys are clearly required to
further constrain the trend in the outer disk. The same applies for new homogenous spectroscopic measurements.

\subsection{Comparisons with independent radial gradients}
\label{comp_Fe}

\begin{figure}
\centering
\resizebox{\hsize}{!}{\includegraphics{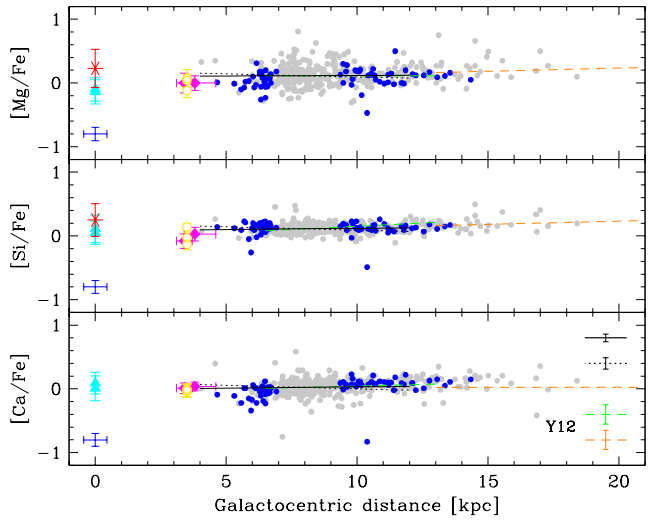}}
\caption{The same as Fig.~\ref{xh_Gdist_SG05}, but the abundances are scaled to iron. The abundance trends derived by \citet[][Y12]{Yongetal2012} for open clusters (green and orange dashed lines) are also shown.}
\label{xfe_Gdist_SG05}
\end{figure}

To validate the above findings concerning the flat trend of [$\alpha$/Fe] radial gradients, Fig.~\ref{xfe_Gdist_SG05} shows the comparison with similar data available in the literature. The magenta, cyan, red and yellow symbols mark the same young stars plotted in Fig.~\ref{xh_Gdist_SG05}. Interestingly enough, [$\alpha$/Fe] is solar and quite similar to the mean ratio of the entire Cepheid sample. This evidence further supports the working hypothesis that the Galactic thin disk experienced a homogeneous chemical enrichment history during the last 300~Myr. The lack of a clear enhancement in the innermost Galactic regions is also supporting the mild correlation of the [$\alpha$/Fe] ratio with the star formation rate. Indeed, recent NIR \citep{Matsunagaetal2011} and MIR/FIR \citep{Ramirezetal2008,
Gerinetal2015} investigations indicate that the Galactic bar and the Galactic center are very efficient star forming regions.

A comparison with radial gradients provided by M14, for thin (solid black line) and thick (dotted black line) disk field stars in the Gaia-ESO survey further supports the above conclusions. Indeed, there is a very good agreement with the [$\alpha$/Fe] ratios of both thin and thick disk stars. Once again, in plotting their radial gradients we adopted the zero-points of our gradients at the solar Galactocentric distance.

The above figure also shows the comparison with the $\alpha$-element abundances for open clusters provided by \citet{Yongetal2012}. The green and the orange dashed lines show the gradients for inner (\RG\ < 13~kpc) and outer (\RG\ > 13~kpc) disk objects, respectively. In plotting these gradients we adopted the zero-points of our gradients at the solar Galactocentric distance. The agreement is quite good over the entire range of distances. This finding becomes even more compelling if we account for the fact that the open clusters adopted by Yong et al. cover the entire range of ages typical of thin disk stellar populations, namely from $\sim$0.5 to $\sim$10~Gyr (see their figures 25 and 26).

\subsection{Age dependence of the [$\alpha$/Fe] ratios}
\label{dep_xfe_logP}

\begin{figure}
\centering
\resizebox{\hsize}{!}{\includegraphics{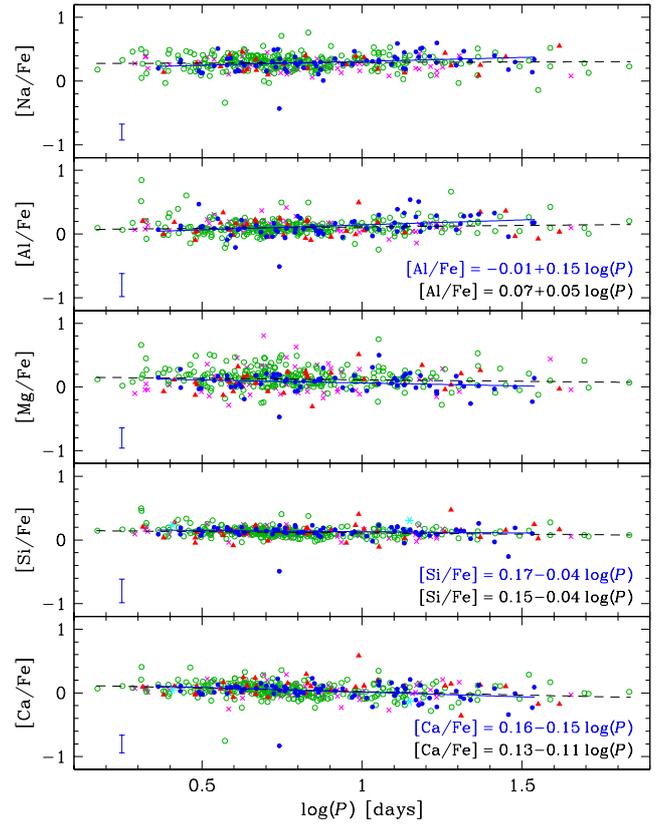}}
\caption{Abundances of Na, Al, and $\alpha$ elements scaled to iron as a function of the logarithmic pulsation period. For Na and Mg the equations are not shown because the slopes have no statistical significance. Symbols and colors are the same as in Fig.~\ref{xh_Gdist}.}
\label{xfe_logP}
\end{figure}

To constrain the age dependence of the [$\alpha$/Fe] abundance ratios, Fig.~\ref{xfe_logP} shows the same elemental abundances plotted in Fig.~\ref{xfe_Gdist}, but as a function of the logarithmic period. A glance at the data plotted in this figure shows that the ratios are approximately constant over the entire period range. Note that the difference in age between short (\logP $\sim$ 0.3) and long-period (\logP $\sim $1.8) is of the order of 300~Myr \citep{Bonoetal2000,Andersonetal2014}. These findings strongly suggest that the chemical enrichment of Galactic Cepheids has been homogenous both in space and in time.

To constrain the period/age dependence on a more quantitative basis, we performed {\it biweight} linear Least Squares fits to the above data. The zero-points, the slopes, their uncertainties, and the standard deviations
are listed in the bottom lines of Table~\ref{table_slopes}. We found that Na, Al, Mg, and Si either do not show evidence of a slope or the slope is marginal (see Fig.~\ref{xfe_logP}). On the other hand, Ca shows once again a
(negative) gradient, suggesting that Ca is underabundant among youngest Cepheids. In passing we note that the nucleosynthesis of Si, Ca, and Fe mostly takes place during the SNe~II explosive events (these elements, in
particular Fe, are also produced in SNe~Ia). This might explain the flat distribution of Si, but the negative trend showed by Ca when moving from older to younger Cepheids would remain unclear. Note that the occurrence of a negative gradient in Ca as a function of period and of a positive gradient of Ca as a function of Galactocentric distance are not correlated, since the Cepheids located in the outer disk have a canonical period distribution (mostly between 2 and 20~days).

%
%
\section{Metallicity distribution}
\label{alpha_feh}

\begin{figure}
\centering
\resizebox{\hsize}{!}{\includegraphics{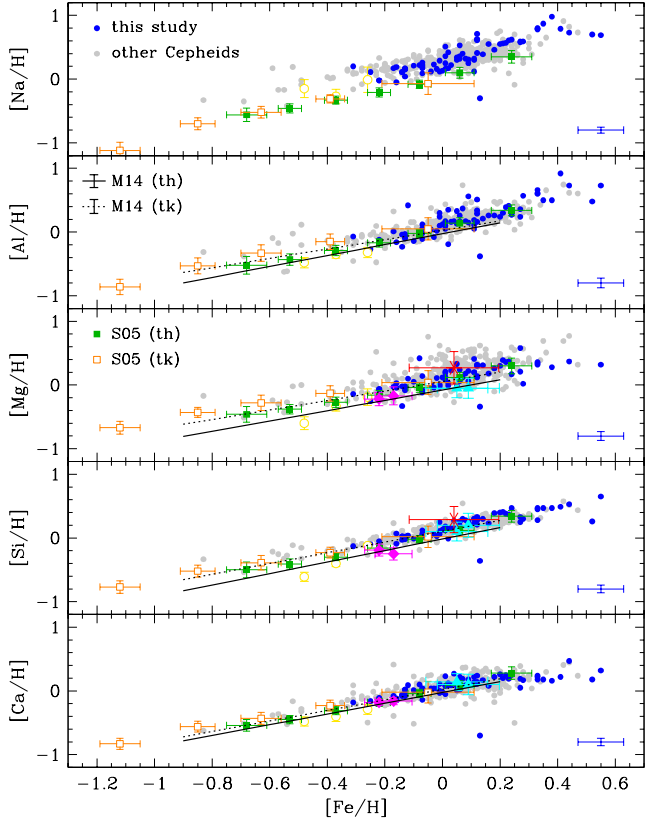}}
\caption{Abundances of Na, Al, and $\alpha$ elements as a function of the metallicity. Cepheid stars (filled circles) are compared with thin disk (filled green squares) and thick disk (open orange squares) field dwarfs from \citet[][S05]{SoubiranGirard2005}. The abundances of red supergiants and LBVs are also plotted when available (symbols and colors are the same as in Fig.~\ref{xh_Gdist_SG05}). The gradients derived by \citet[][M14]{Mikolaitisetal2014} for thin (solid line) and thick (dotted line) disk stars in the Gaia-ESO survey are also shown.}
\label{xh_feh_SG05}
\end{figure}

\begin{figure}
\centering
\resizebox{\hsize}{!}{\includegraphics{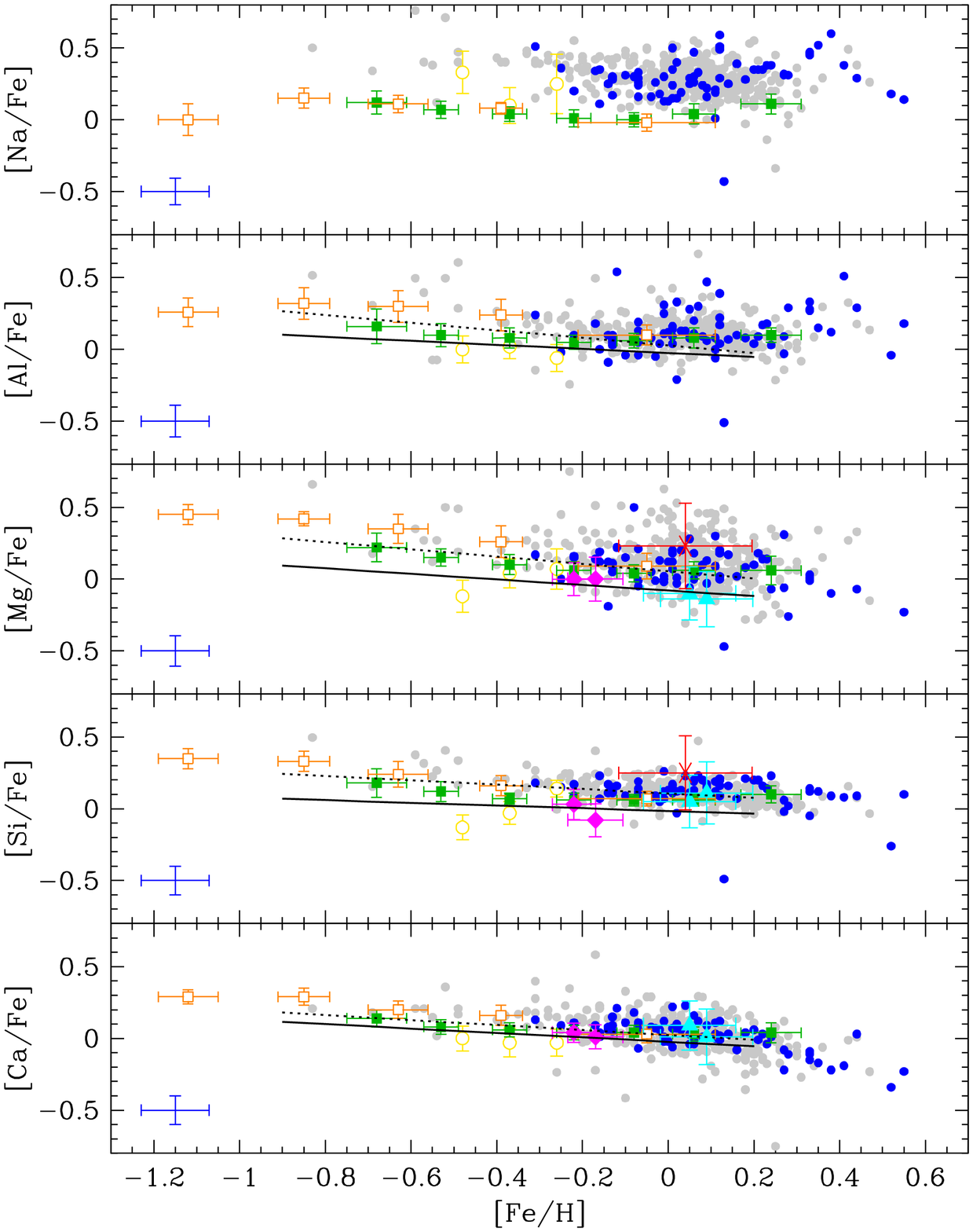}}
\caption{The same as Fig.~\ref{xh_feh_SG05}, but the abundances are
         scaled to iron.}
\label{xfe_feh_SG05}
\end{figure}

The range in age covered by Galactic Cepheids is too short to constrain a possible age dependence on the timescale of the order of a few Gyrs. To further investigate this effect we compared the current Cepheid metallicity distribution with the metallicity distribution of field dwarf stars (743) collected by \citet[][hereafter SG05]{SoubiranGirard2005}. In their sample, 72\% are thin disk stars and 28\% are from the thick disk. Note that the sample collected by SG05 is based on spectroscopic data available in the literature. Their $\alpha$-element and iron abundances were not rescaled to the solar mixture adopted in the current investigation \citep{Grevesseetal1996}. We also perform the same comparison with the results provided by the GAIA-ESO survey \citep{Mikolaitisetal2014,RecioBlancoetal2014}. Data plotted in Fig.~\ref{xh_feh_SG05} shows that the agreement between Cepheids and thin disk dwarfs is quite good over the entire metallicity range covered by the current sample. This outcome applies to the $\alpha$-elements and to Al. On the other hand, the Na in field stars attain values that are the lower envelope of the typical Na abundances of classical Cepheids. The same agreement for $\alpha$-elements and for Al is also shared in the [element/Fe] plane (Fig.~\ref{xfe_feh_SG05}), in which the difference concerning Na is even more clear. This difference appears to be significant and may be caused by non-LTE effects. Indeed, the difference increases when moving from the metal-rich to the metal-poor regime, as expected for non-LTE effects \citep{Fabrizioetal2012,TakedaTakadaHidai2000}. Recent empirical investigations \citep{JohnsonPilachowski2012} are also suggesting an increase in Na enhancement when moving from the base to the tip of the RGB. Moreover, we also found evidence of a positive gradient (slope = 0.20 $\pm$ 0.03~\dexkpc) in [Na/H] versus the logarithmic period. This finding further supports an increase in the difference in Na abundance when moving from long-periods (low surface gravity) to short-periods (high surface gravity).

In this context it is worth mentioning that \citet{Takedaetal2013} performed a detailed abundance analysis of ten Galactic Cepheids using high-resolution spectra. They investigated CNO plus Na elements, since they are solid
tracers of induced mixing during advanced evolutionary phases of intermediate-mass stars. They found a well defined Na overabundance of the order of 0.2~dex over the entire period range covered by their targets (2--16 days). They suggested that the Na enhancement could be due to mixing events that dredge up Na-rich material, produced by the NeNa cycle, into the surface of classical Cepheids. A similar explanation was originally suggested by \citet{Sasselov1986} and by \citet{Denissenkov1994}. We plan to address this specific issue, and in particular, the period (i.e. stellar mass) dependence in a forthcoming paper. 

%
%
\subsection{Al-Mg and Na-O correlations}
\label{correl}

\begin{figure*}[p!]
\centering
\begin{minipage}[t]{0.33\textwidth}
\centering
\resizebox{\hsize}{!}{\includegraphics{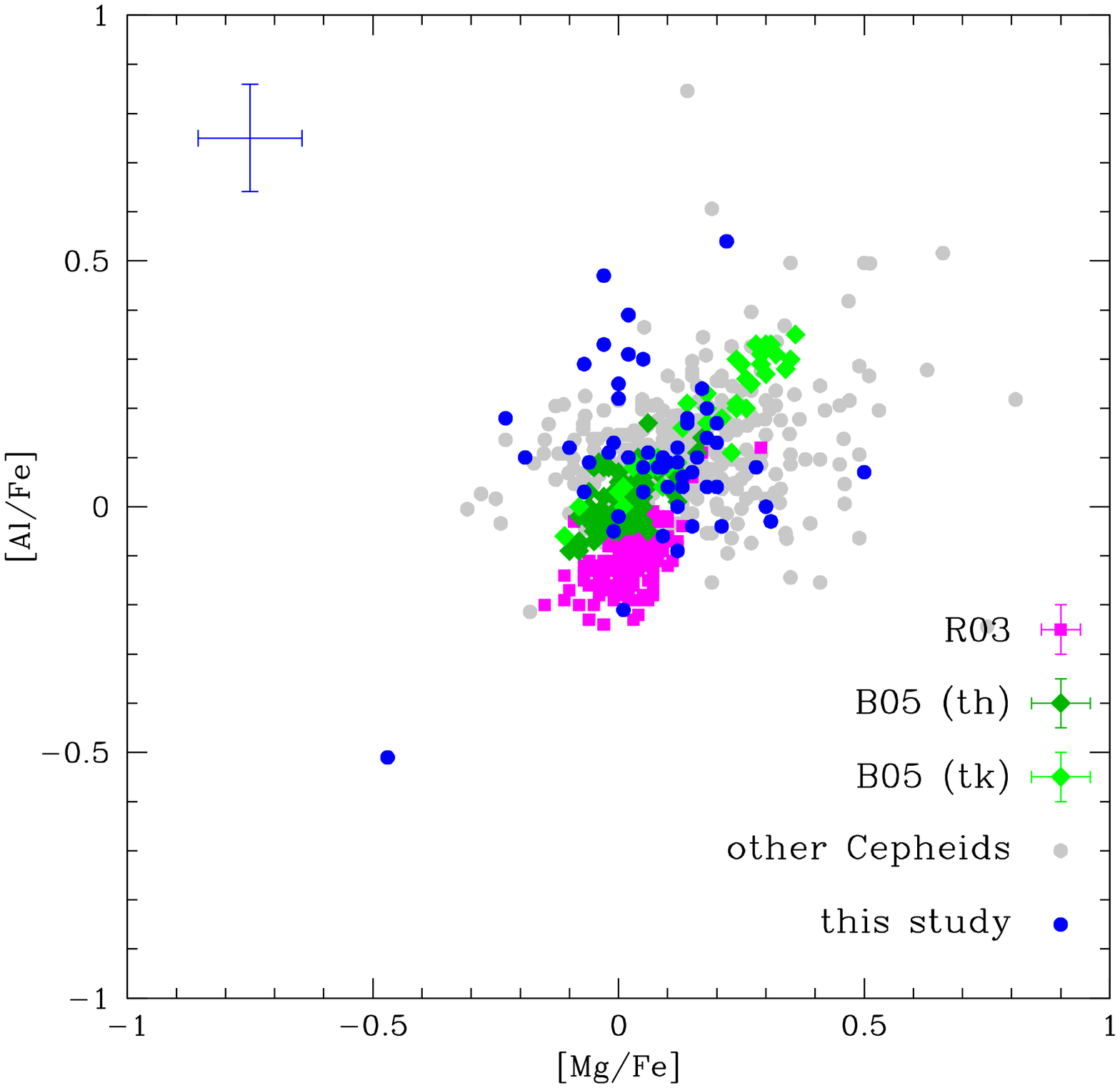}}
\end{minipage}
\begin{minipage}[t]{0.33\textwidth}
\centering
\resizebox{\hsize}{!}{\includegraphics{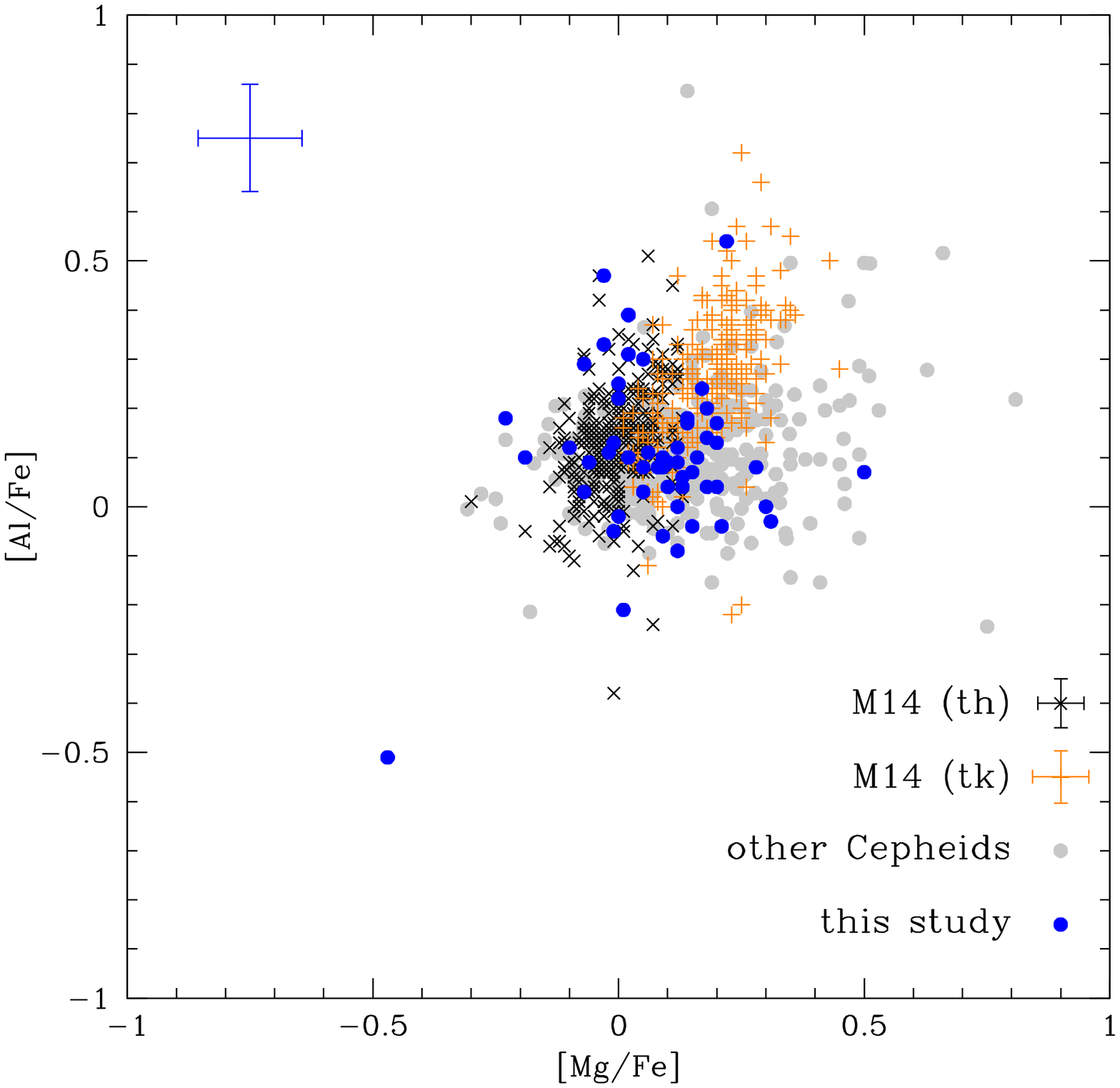}}
\end{minipage}
\begin{minipage}[t]{0.33\textwidth}
\centering
\resizebox{\hsize}{!}{\includegraphics{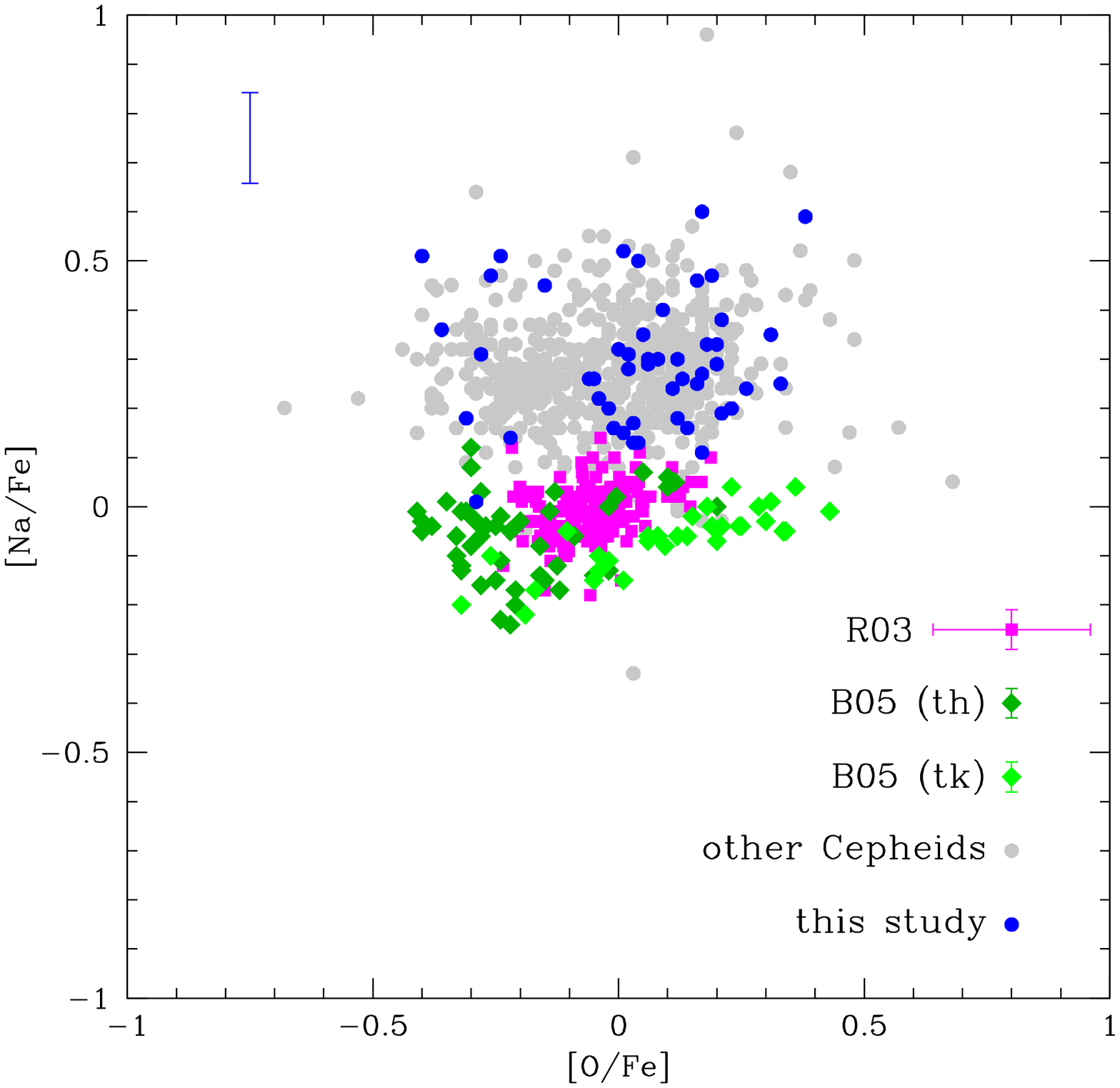}}
\end{minipage}
\caption{Correlations between Al--Mg (left and middle panels) and between Na--O (right panel). Cepheid stars (filled circles) are compared with field dwarfs from the thin disk analyzed by \citet[][R03, magenta squares]{Reddyetal2003}, from the thin (dark green diamonds) and thick (light green diamonds) disks analyzed by \citet[][B05]{Bensbyetal2005}, and from the thin (black crosses) and thick (orange pluses) disks analyzed by \citet[][M14]{Mikolaitisetal2014}.}
\label{corr}
\end{figure*}

To further constrain the recent chemical enrichment of the Galactic thin disk we also investigated the (anti-)correlation between Mg--Al and Na--O. Detailed spectroscopic investigations indicate that evolved and unevolved stars in the Local Group globular clusters display well defined anti-correlation between Na--O, and Mg--Al \citep[e.g.,][]{Carrettaetal2009,Carrettaetal2014}. The left panel of Figure~\ref{corr} show the comparison between the current abundances of Mg and Al for the Cepheid sample (blue dots plus grey dots for Cepheids in the literature) with similar abundances for F- and G-type field dwarf stars collected by \citet[][hereafter B05, 102 objects]{Bensbyetal2005}. The latter sample includes both thin (dark green) and thick (light green) disk stars. Moreover, we also included Mg--Al abundances for field dwarf stars collected by \citet[hereafter R03]{Reddyetal2003}. This sample includes 189 objects and a significant fraction of them are thin disk stars (magenta dots). Note that the $\alpha$-element and iron abundances provided by these authors were rescaled to the abundances of the solar mixture adopted in the current investigation \citep{Grevesseetal1996}. Data plotted in this figure disclose a very good agreement between field dwarfs and Cepheids. However, the sample by R03 shows, at fixed [Mg/Fe] abundance, a well defined underabundance in [Al/Fe]. There are some plausible reasons that could explain the difference. The Mg abundances provided by R03 are based on three lines (4730.04, 6318.71, 7657.61~\AA) while we only use the line at 5711.09~\AA. The agreement with the Mg abundances provided by B05 is due to the fact that they adopted eight different lines including the current one, but not the three Mg lines adopted by R03. The difference may also be caused by different values of \loggf\ adopted by these studies. Nevertheless, Galactic Cepheids display, at fixed [Al/Fe] abundance, a large dispersion in [Mg/Fe] abundances when compared with field dwarfs. However, they clearly follow the Mg--Al correlation typical of field stars.

To confirm the above similarity, the middle panel of Figure~\ref{corr} shows the same comparison, but with the thin (black crosses) and thick (orange pluses) disk field dwarf stars measured by M14. The agreement is once again very good over the entire abundance range covered by the two samples. Data plotted in this panel display that Cepheids, thin and thick disk dwarf stars have similar [Al/Fe] and [Mg/Fe] distributions. This is an interesting finding, since according to R03, B05 and M14 their samples include both old ($\sim$12~Gyr) and intermediate-age (0.5--7~Gyr) field dwarf stars.

The right panel of Figure~\ref{corr} shows the comparison between the abundances of Na and O for the current sample of Cepheids with the same sample of field stars we adopted in the left panel of the same figure. Note that in plotting Cepheid abundances we are using the current Na abundances, but the O abundances for the same objects provided by LII and LIII. To overcome problems introduced by non-LTE effects, we only used oxygen abundances based on the \ion{O}{i} triplet at 6156~\AA\ and on the [\ion{O}{i}] forbidden line at 6300~\AA, the mean values of both provided by LIII, and the triplet-based ones provided by LII. A glance at the data plotted in this panel shows the well defined offset in Na abundances between field dwarfs and Cepheids. The sample by R03 appears to be located between the B05 and the Cepheids. This mild difference might be due to their larger uncertainties in O abundances (see their Sect.~7.2), or to different assumptions on the adopted \loggf\ values.  

%
%
\subsection{Hydrostatic and explosive elements}
\label{hydex}

\begin{figure*}[p!]
\centering
\begin{minipage}[t]{0.33\textwidth}
\centering
\resizebox{\hsize}{!}{\includegraphics{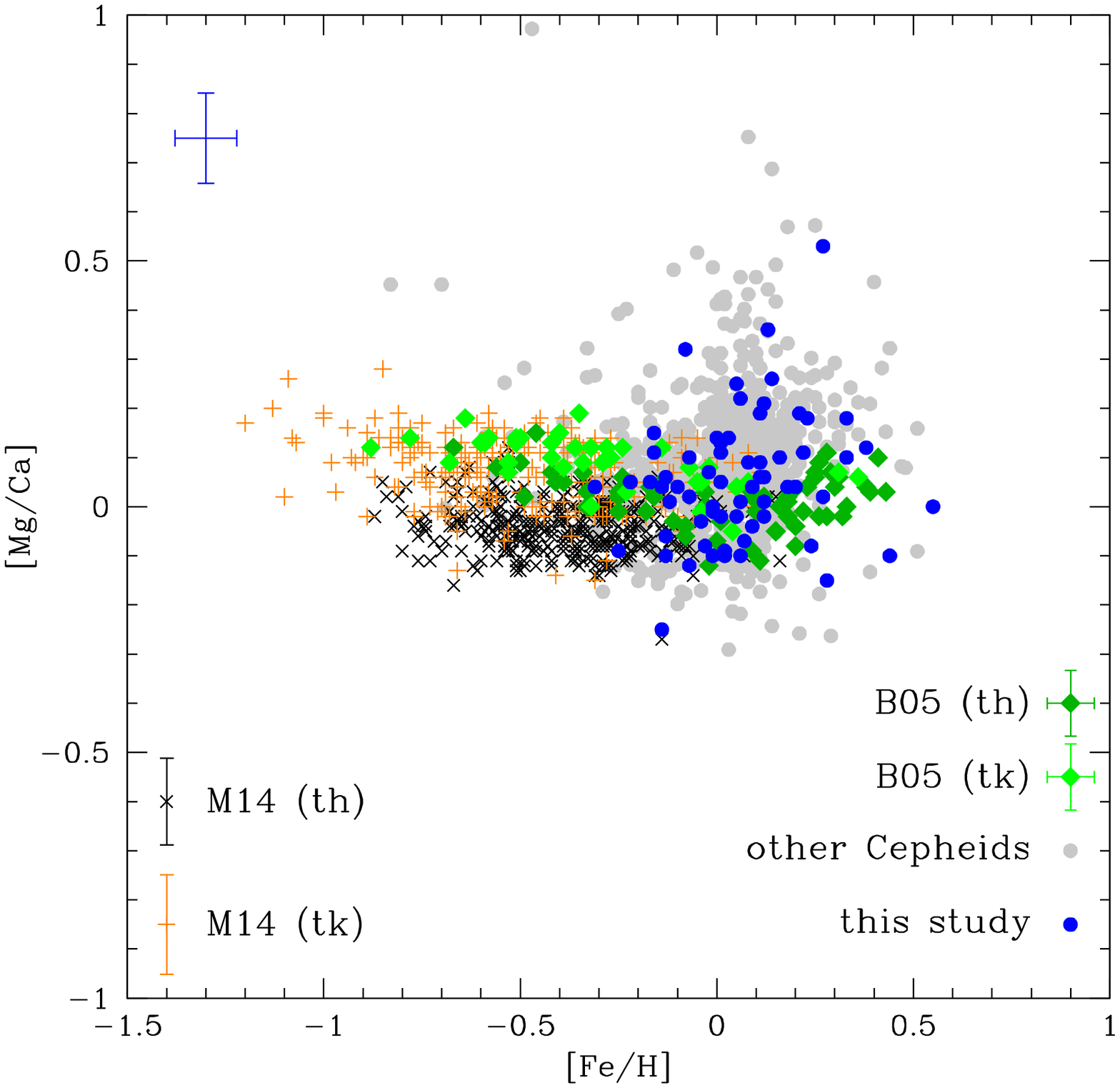}}
\end{minipage}
\begin{minipage}[t]{0.33\textwidth}
\centering
\resizebox{\hsize}{!}{\includegraphics{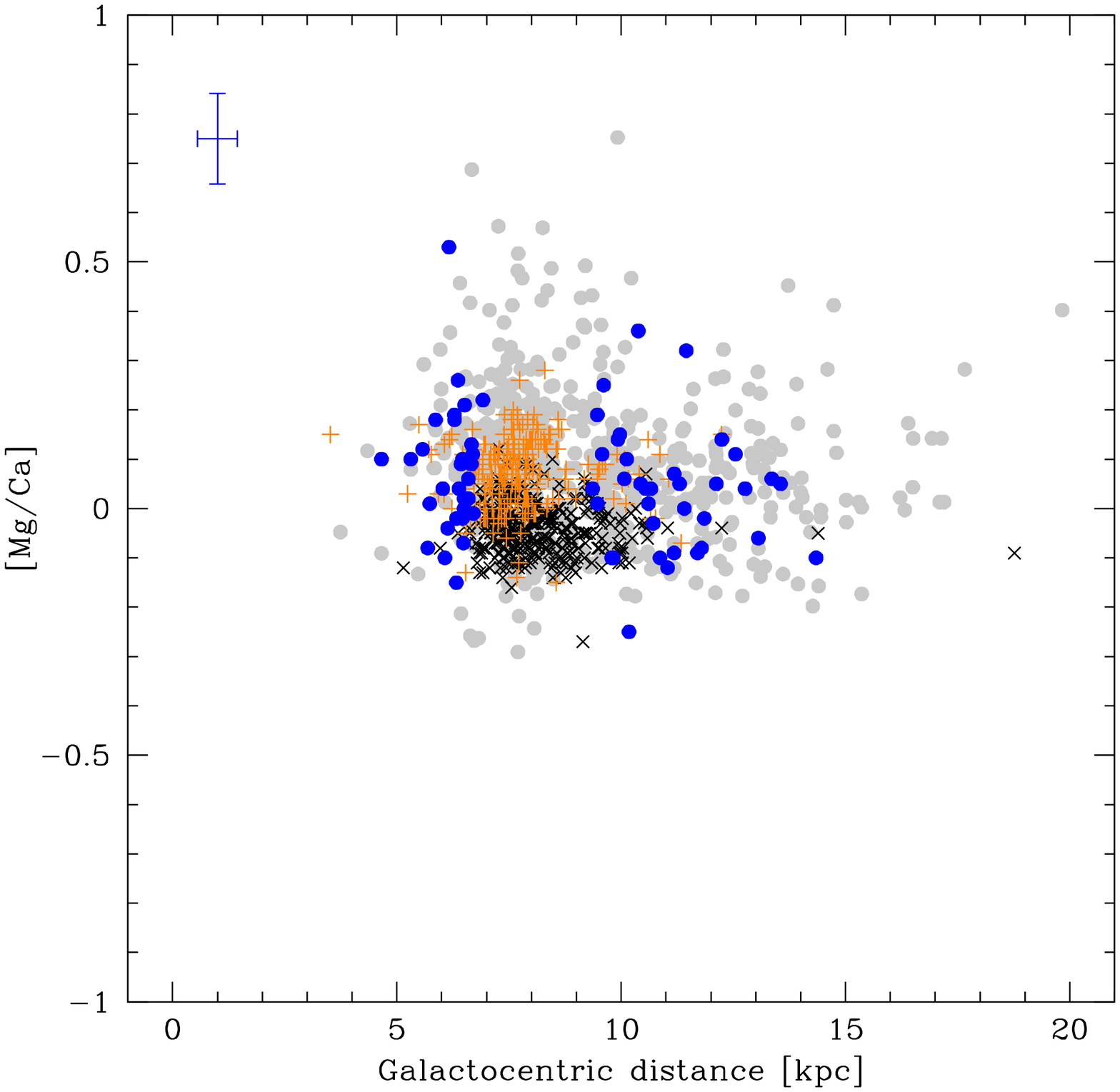}}
\end{minipage}
\begin{minipage}[t]{0.33\textwidth}
\centering
\resizebox{\hsize}{!}{\includegraphics{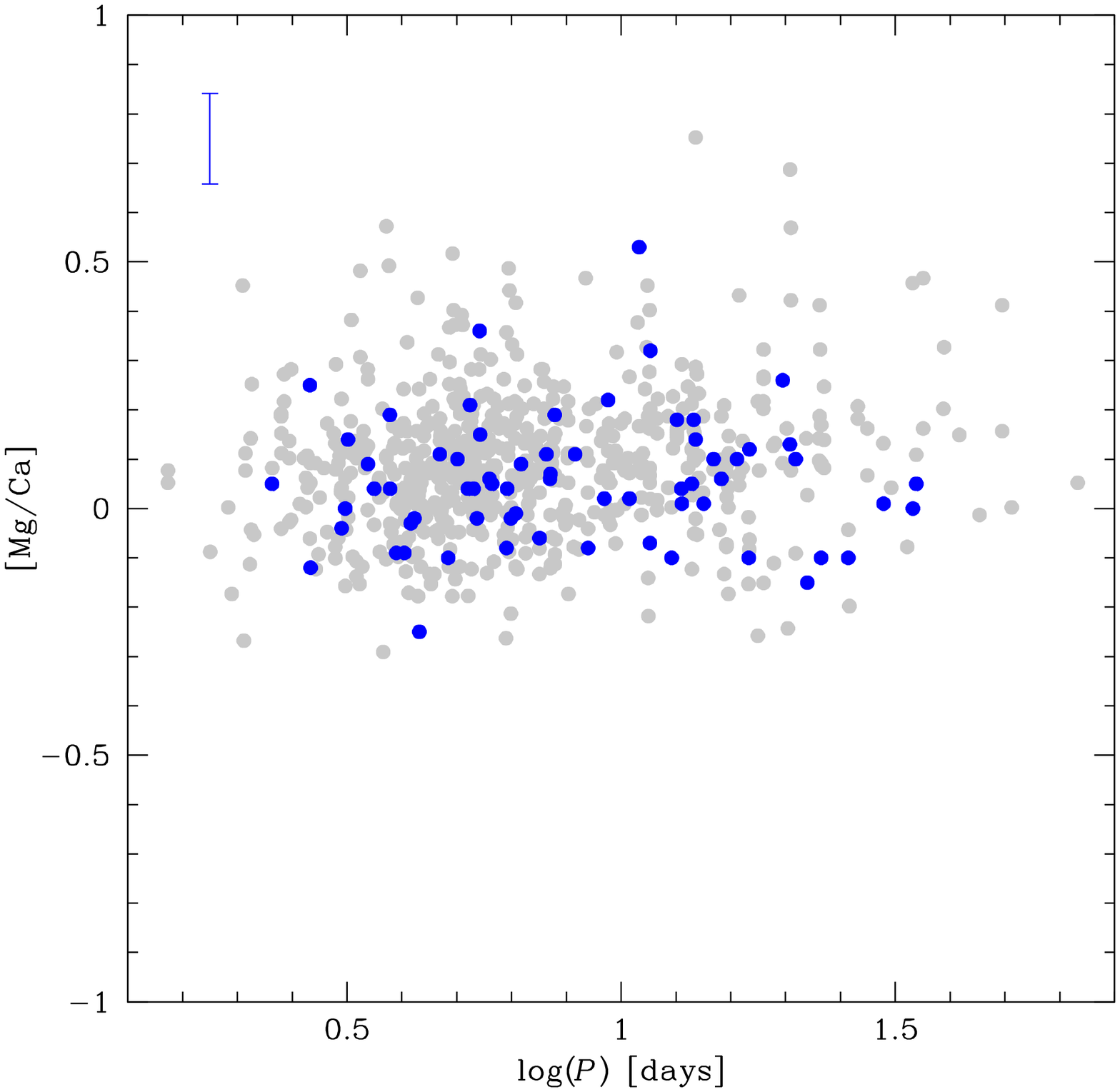}}
\end{minipage}
\caption{Abundance ratios between Mg and Ca as a function of metallicity (left panel), Galactocentric distance (middle panel), and logarithmic pulsation period (right panel). Symbols and colors are the same as in Fig.~\ref{corr}.}
\label{mgca}
\end{figure*}

The chemical enrichment history of the Galactic thin disk can also be constrained by investigating the ratio between hydrostatic and explosive elements. The left panel of Fig.~\ref{mgca} shows [Mg/Ca] as a function of the iron abundance. The Cepheids display typical solar abundances of [Mg/Ca] for iron abundances more metal-poor than the Sun. For iron abundances more metal-rich than the Sun, the spread in [Mg/Ca] abundances increases and becomes of the order of 0.5--0.7~dex. On the other hand, the abundances of thin and thick disk dwarf stars collected either by B05 or by M14 display the typical decreasing trend when moving from metal-poor to more metal-rich objects\footnote{The thin disk stars by M14 appear to have a flat distribution, but we performed a test and we found that the negative slope is significant.}. The same outcome applies to the metal-rich regime, and indeed the metal-rich objects (green diamonds) collected by B05 display a mild increase in the [Mg/Ca] abundance ratio for [Fe/H] $\gtrsim$ 0.15.

The reasons for the above difference are not clear, therefore we investigated a possible radial dependence. The middle panel of Fig.~\ref{mgca} shows both the Cepheid and the M14 [Mg/Ca] abundances as a function of Galactocentric distance. Data plotted in this figure display that a significant fraction of the spread in [Mg/Ca] abundance ($\sim$0.7~dex) showed by metal-rich Cepheids is evident among objects that are located either across or inside the solar circle ($\sim$8~kpc). Evidence for a mild increase in the spread in [Mg/Ca] seems to be also present in the outer disk.

The thin disk stars by M14 display a flat distribution over the entire range of Galactocentric distances they cover. On the other hand, the thick disk stars display a larger spread ($\sim$0.3~dex) in a limited range of Galactocentric distances (6 $\lesssim$ \RG\ $\lesssim$ 8~kpc). Note that in this panel we did not plot the sample by B05, since individual distances are not available. This evidence appears slightly counter intuitive, since the increase in iron typical of the innermost disk regions should also be followed by a steady increase in Ca abundance, since they are explosive elements. This means a steady decrease in the [Mg/Ca] abundance ratio. Oddly enough, we found that the [Mg/Ca] ratio increases when moving towards the inner disk.  

To further constrain this effect we also investigated the age dependence. The right panel of Fig.~\ref{mgca} shows the [Mg/Ca] abundances of the Cepheid sample as a function of the logarithmic period. Data plotted in this figure show no clear trend. The spread in the [Mg/Ca] abundance ratio is almost constant over the entire period range, and therefore, no solid evidence of a dependence on stellar mass.       

%
%
\section{Summary and final remarks}
\label{concl}

We present accurate and homogeneous measurements of Na, Al and three $\alpha$-elements (Mg, Si, Ca) for 75 classical Galactic Cepheids. The current abundances are based on high spectral resolution ($R$$\sim$38\,000) and high signal-to-noise ratio (S/N $\sim$ 50-300) spectra collected with UVES at ESO VLT. The iron abundance of the same Cepheids was already discussed in \citet{Genovalietal2013,Genovalietal2014}. The current sample covers a broad range in pulsation periods (0.36 $\leq$ \logP\ $\leq$ $\sim$ 1.54) and in Galactocentric distances (4.6 $\leq$ \RG\ $\leq$ 14.3~kpc).

The current spectroscopic measurements were complemented with Cepheid abundances available in the literature and based on similar spectra and on a similar approach in performing the abundance analysis. We ended up with a sample of 439 Galactic Cepheids. Among them 140 have measurements estimated by our group (current plus LEM), while the others come from LII, LIII, and YON. The samples have from 16 to 55 Cepheids in common, which allow us to calibrate a homogeneous abundance scale for the quoted five elements plus iron (G14).

The use of homogeneous abundance scales, and accurate and homogenous individual distances based on NIR photometry allowed us to investigate the radial gradients across a significant portion of the Galactic inner and outer disk (4.1 $\leq$ \RG\ $\leq$ 18.4~kpc). The main findings of the current analysis are the following:

{\it i) Radial gradients:}
The five investigated elements display a well defined radial gradient. The slopes range from $-$0.058 for [Al/H] to $-$0.038~\dexkpc\ for [Ca/H]. The negative slopes are linear over the entire range of Galactocentric distances covered by the current sample. Moreover they agree, within the errors, with similar estimates available in the literature. The main difference is that the current gradients display standard deviations on average smaller than
0.16~dex.

{\it ii) Environmental effects:}
The comparison of current abundances with similar measurements for young stars located in the near end of the bar and in the nuclear bulge indicates that the latter attain, within the errors, solar abundances. This means that inner disk Cepheids attain higher $\alpha$ abundances when compared with young stars of the quoted regions. Thus suggesting that the bar plus the nuclear bulge and the inner disk underwent different chemical enrichment histories. However, the above regions attain similar solar [$\alpha$/Fe] ratios.

{\it iii) Spatial homogeneity:}
The [element/Fe] ratios are constant across the entire thin disk with no evidence of radial gradient. This applies to Na, Al, and Si. The radial distribution of Mg and Ca display marginal positive slopes, but they should be treated with caution. The Mg abundances display, at fixed Galactocentric distance, a large spread mainly driven by 
measurement errors.

The above results indicate that the chemical enrichment history of iron and of the other five elements has been quite similar across the Galactic thin disk. Moreover, since the [$\alpha$/Fe] ratios are typically considered tracers of star formation, the lack of a clear negative gradient suggests that the [$\alpha$/Fe] radial gradients are also affected by other parameters. There is also evidence for a mild abundance enhancement for the five elements and of an increase in the intrinsic scatter in the outer disk. However, the number of Cepheids with \RG\ $>$ 13~kpc is limited, and new identifications of classical Cepheids in this region are clearly required.

{\it iv) Temporal homogeneity:}
The ratios [element/Fe] vs. \logP\ are constant over the entire period range for Na, Al, Mg. This empirical evidence together with the well established anti-correlation between age and pulsation period indicates that the chemical enrichment of Galactic Cepheids has also been very homogenous during the last 300~Myr. On the other hand, Ca shows a significant negative gradient, being underabundant among youngest Cepheids.

{\it v) Comparison between giants and dwarfs:}
The comparison between the metallicity distribution of Galactic Cepheids and field dwarf stars shows that the current Al, Mg, Si, and Ca abundances agree quite well with those of the thin disk dwarfs over the entire metallicity range. On the other hand, the Na abundances in field stars are the lower envelope of the typical Na abundances of classical Cepheids. This effect appears real and caused by non-LTE effects, since the difference increases when moving from the metal-rich to metal-poor Cepheids. More recent investigations suggest that an increase in Na abundance in evolved 
intermediate-mass stars might also be caused by mixing events that dredge up Na-rich material, produced by the NeNa cycle, into the surface of classical Cepheids.

{\it vi) Mg--Al and Na--O correlations:}
Evolved and unevolved stars in the Local Group globular clusters show a well defined anti-correlation between Mg--Al and Na--O. On the other hand, classical Cepheid follow the same correlation typical of field stars. However, Cepheid Na abundances show a well defined enhancement when compared with Na abundances for field dwarfs. The difference is mainly caused by evolutionary effects (dredge-up into the surface of Na-rich material) and by non-LTE effects. Moreover, Cepheids, thin and thick disk dwarf stars display very similar [Al/Fe] and [Mg/Fe] abundance distributions. This is an interesting finding, since the different samples of field dwarfs and Cepheids cover a large range in age (0.5--12~Gyr).

{\it vii) Hydrostatic and explosive elements:}
Cepheids in the [Mg/Ca] vs. [Fe/H] plane display a flat solar abundance for iron abundances more metal-poor than the Sun. For iron abundances more metal-rich than the Sun, the spread in [Mg/Ca] abundances increases and becomes of the order of 0.5--0.7~dex. Thin and thick disk stars display in the quoted iron regimes similar trends. The [Mg/Ca] abundance ratio decreases approaching solar iron abundances and increases in the more metal rich regime. Simple chemical enrichment histories would imply a steady decrease in [Mg/Ca] for higher Fe abundances, since Ca and Fe are explosive elements. This trend is mainly caused by objects located either across or inside the solar circle and does not show a clear dependence on the pulsation period (i.e. stellar mass).

The above results bring forward two interesting consequences concerning the chemical enrichment of the Galactic thin disk. The flat and homogeneous trend of the five investigated elements as a function of the Galactocentric distances and of the pulsation period. This suggests similar trends when moving from the inner to the outer disk. Moreover, there is mounting evidence of a marginal difference between young (Cepheids), intermediate- and old age field dwarfs. This means minor spatial and temporal variations across the Galactic thin, and possibly thick, disk. More solid empirical constraints are required to further support the above trends. No doubts that {\it s}- and {\it r}-elements are good diagnostics to constrain possible differences in chemical enrichment between different stellar populations \citep{Cescuttietal2008,Matteuccietal2014}.

The wealth of new results concerning iron and $\alpha$-elements is also opening the path to new empirical constraints on the initial mass function. In particular, the steady increase of the hydrostatic-to-explosive [Mg/Ca] abundance ratio in the innermost, more metal-rich regions of the inner disk is suggesting a more complex relation between the IMF and the nucleosynthesis of SNe~Ia and SNe~II \citep{McWilliametal2013}. More recently, \citet{Kudritzkietal2015} called attention to the chemical enrichment of star forming galaxies using analytical chemical evolution models. They found that their sample can be split into three groups according to the rate either of the galactic wind mass loss or of the accretion mass gain to the star formation rate. They applied their model to the Galaxy and they found that the actual iron gradient would imply a modest accretion rate and a moderate mass-loading factor. The homogeneity of the
above $\alpha$-element gradients can provide new firm constraints on the role that the IMF and the infall rate played in the star formation history of the Galactic thin disk.

\begin{acknowledgements}
This work was partially supported by PRIN-MIUR (2010 LY5N2T) ``Chemical and dynamical evolution of the Milky Way and Local Group galaxies" (P.I.: F. Matteucci).
M.F. acknowledges the financial support from the PO FSE Abruzzo 2007-2013 through the grant "Spectro-photometric characterization of stellar populations in Local Group dwarf galaxies" prot. 89/2014/OACTe/D (P.I.: S. Cassisi).
We also acknowledge an anonymous referee for his/her positive opinions concerning this experiment and for the very pertinent suggestions that improved the content and the readability of the paper.
\end{acknowledgements}

\bibliographystyle{aa}
\bibliography{Genovalietal2015_aph}

\clearpage
\clearpage
\clearpage
\clearpage

\clearpage
\begin{table*}[p]
\centering
\caption[]{Abundances of Fe, Na, Al, and $\alpha$ elements for our sample of
           classical Cepheids derived based on individual spectra.}
\label{table_ab_alpha_spec}
{\scriptsize 
\begin{tabular}{rc r@{ }l c r@{ }l c r@{ }l c r@{ }l c r@{ }l c r@{ }l c}
\noalign{\smallskip}\hline\hline\noalign{\smallskip}
Name & MJD &
\multicolumn{2}{c}{[Fe/H]} & \parbox[c]{1.0cm}{\centering $N_{\rm L}$ (\ion{Fe}{i},\ion{Fe}{ii})} &
\multicolumn{2}{c}{[Na/H]} & $N_{\rm L}$ &
\multicolumn{2}{c}{[Al/H]} & $N_{\rm L}$ &
\multicolumn{2}{c}{[Mg/H]} & $N_{\rm L}$ &
\multicolumn{2}{c}{[Si/H]} & $N_{\rm L}$ &
\multicolumn{2}{c}{[Ca/H]} & $N_{\rm L}$ \\
\noalign{\smallskip}\hline\noalign{\smallskip}
 V340\,Ara & 56137.137 &    0.27 & $\pm$ 0.10 &  (23, 2) &    0.72 & $\pm$ 0.05 &  1	 &    0.71 & $\pm$ 0.08 &  1	 &	   & \ldots	& \ldots &    0.28 & $\pm$ 0.08 &  2	 &    0.17 & $\pm$ 0.04 &  2	 \\
 V340\,Ara & 54708.065 &    0.53 & $\pm$ 0.09 &  (53, 4) &    0.98 & $\pm$ 0.01 &  2	 &    0.59 & $\pm$ 0.14 &  4	 &    0.55 & $\pm$ 0.01 &  1	 &    0.67 & $\pm$ 0.14 & 10	 &    0.41 & $\pm$ 0.12 &  2	 \\
 V340\,Ara & 54709.079 &    0.53 & $\pm$ 0.16 &  (26, 3) &    0.95 & $\pm$ 0.01 &  2	 &    0.71 & $\pm$ 0.06 &  3	 &    0.47 & $\pm$ 0.07 &  1	 &    0.61 & $\pm$ 0.15 & 10	 &    0.36 & $\pm$ 0.12 &  2	 \\
 V340\,Ara & 56138.094 &    0.32 & $\pm$ 0.09 &  (41, 2) &    0.77 & $\pm$ 0.13 &  2	 &    0.55 & $\pm$ 0.18 &  2	 &    0.24 & $\pm$ 0.07 &  1	 &    0.33 & $\pm$ 0.01 &  2	 &    0.17 & $\pm$ 0.05 &  2	 \\
 V340\,Ara & 56139.185 &    0.22 & $\pm$ 0.01 &  (51, 2) &    0.59 & $\pm$ 0.05 &  1	 &    0.50 & $\pm$ 0.12 &  2	 &    0.10 & $\pm$ 0.07 &  1	 &    0.30 & $\pm$ 0.11 &  3	 &    0.14 & $\pm$ 0.06 &  5	 \\
 V340\,Ara & 56152.054 &    0.24 & $\pm$ 0.18 &  (15, 2) &	   & \ldots	& \ldots &	   & \ldots	& \ldots &	   & \ldots	& \ldots &    0.62 & $\pm$ 0.06 &  1	 &    0.17 & $\pm$ 0.06 &  1	 \\
   AS\,Aur & 54845.136 &    0.00 & $\pm$ 0.08 &  (74, 8) &    0.13 & $\pm$ 0.05 &  1	 &    0.12 & $\pm$ 0.40 &  2	 &    0.12 & $\pm$ 0.18 &  1	 &    0.13 & $\pm$ 0.25 & 10	 & $-$0.02 & $\pm$ 0.17 &  3	 \\
   KN\,Cen & 54862.355 &    0.55 & $\pm$ 0.12 &  (14, 3) &    0.69 & $\pm$ 0.04 &  2	 &    0.73 & $\pm$ 0.17 &  4	 &    0.32 & $\pm$ 0.17 &  1	 &    0.65 & $\pm$ 0.24 & 10	 &    0.32 & $\pm$ 0.26 &  2	 \\
   MZ\,Cen & 54584.280 &    0.27 & $\pm$ 0.10 &  (45, 4) &    0.59 & $\pm$ 0.05 &  1	 &    0.36 & $\pm$ 0.18 &  4	 &    0.21 & $\pm$ 0.36 &  1	 &    0.33 & $\pm$ 0.19 &  7	 &    0.19 & $\pm$ 0.12 &  2	 \\
   OO\,Cen & 54585.060 &    0.20 & $\pm$ 0.06 &  (30, 4) &    0.55 & $\pm$ 0.05 &  1	 &    0.24 & $\pm$ 0.19 &  5	 &    0.30 & $\pm$ 0.10 &  1	 &    0.40 & $\pm$ 0.23 &  6	 &    0.26 & $\pm$ 0.12 &  2	 \\
   TX\,Cen & 54862.363 &    0.44 & $\pm$ 0.12 &  (78, 7) &    0.73 & $\pm$ 0.05 &  1	 &    0.73 & $\pm$ 0.16 &  2	 &    0.37 & $\pm$ 0.16 &  1	 &    0.53 & $\pm$ 0.27 &  9	 &    0.47 & $\pm$ 0.11 &  2	 \\
 V339\,Cen & 54584.304 &    0.06 & $\pm$ 0.03 &  (39, 3) &    0.53 & $\pm$ 0.23 &  2	 &    0.26 & $\pm$ 0.11 &  5	 &    0.24 & $\pm$ 0.19 &  1	 &    0.19 & $\pm$ 0.18 & 10	 &    0.02 & $\pm$ 0.07 &  2	 \\
   VW\,Cen & 54862.359 &    0.41 & $\pm$ 0.08 &  (43, 2) &    0.79 & $\pm$ 0.05 &  1	 &    0.92 & $\pm$ 0.08 &  1	 &	   & \ldots	& \ldots &    0.49 & $\pm$ 0.20 &  3	 &    0.22 & $\pm$ 0.01 &  2	 \\
   AO\,CMa & 54839.053 &    0.01 & $\pm$ 0.06 &  (75, 5) &    0.16 & $\pm$ 0.03 &  2	 &    0.10 & $\pm$ 0.15 &  6	 &    0.13 & $\pm$ 0.02 &  1	 &    0.12 & $\pm$ 0.14 & 14	 &    0.08 & $\pm$ 0.10 &  3	 \\
   RW\,CMa & 54839.138 & $-$0.07 & $\pm$ 0.08 &  (83, 5) &    0.22 & $\pm$ 0.08 &  2	 &	   & \ldots	& \ldots &	   & \ldots	& \ldots &    0.05 & $\pm$ 0.04 & 12	 &    0.03 & $\pm$ 0.08 &  2	 \\
   SS\,CMa & 54839.066 &    0.06 & $\pm$ 0.04 &  (57, 5) &    0.32 & $\pm$ 0.04 &  2	 &    0.17 & $\pm$ 0.15 &  6	 &    0.12 & $\pm$ 0.17 &  1	 &    0.26 & $\pm$ 0.18 & 14	 &    0.22 & $\pm$ 0.14 &  3	 \\
   TV\,CMa & 54847.246 &    0.01 & $\pm$ 0.07 &  (89, 6) &    0.25 & $\pm$ 0.02 &  2	 &    0.15 & $\pm$ 0.12 &  6	 &    0.19 & $\pm$ 0.07 &  1	 &    0.18 & $\pm$ 0.14 & 13	 &    0.08 & $\pm$ 0.09 &  3	 \\
   TW\,CMa & 54839.077 &    0.04 & $\pm$ 0.09 &  (38, 4) &	   & \ldots	& \ldots &    0.17 & $\pm$ 0.23 &  2	 &	   & \ldots	& \ldots &    0.27 & $\pm$ 0.16 & 11	 &    0.27 & $\pm$ 0.31 &  2	 \\
   AA\,Gem & 54846.149 & $-$0.08 & $\pm$ 0.05 &  (74, 5) &    0.22 & $\pm$ 0.03 &  2	 & $-$0.01 & $\pm$ 0.15 &  6	 &    0.42 & $\pm$ 0.20 &  1	 &	   & \ldots	& \ldots &    0.10 & $\pm$ 0.22 &  5	 \\
   AD\,Gem & 54846.221 & $-$0.14 & $\pm$ 0.06 &  (70, 7) &    0.11 & $\pm$ 0.13 &  2	 & $-$0.23 & $\pm$ 0.23 &  4	 & $-$0.02 & $\pm$ 0.12 &  1	 &    0.03 & $\pm$ 0.19 & 12	 & $-$0.06 & $\pm$ 0.13 &  2	 \\
   BB\,Gem & 54846.187 & $-$0.09 & $\pm$ 0.04 &  (70, 9) &	   & \ldots	& \ldots &	   & \ldots	& \ldots &	   & \ldots	& \ldots &	   & \ldots	& \ldots &	   & \ldots	& \ldots \\
   BW\,Gem & 54845.122 & $-$0.22 & $\pm$ 0.09 &  (99, 6) & $-$0.02 & $\pm$ 0.00 &  2	 & $-$0.15 & $\pm$ 0.19 &  6	 & $-$0.07 & $\pm$ 0.04 &  1	 & $-$0.13 & $\pm$ 0.14 & 12	 & $-$0.12 & $\pm$ 0.12 &  2	 \\
   DX\,Gem & 54846.196 & $-$0.01 & $\pm$ 0.09 &  (72, 6) &    0.12 & $\pm$ 0.00 &  2	 &    0.10 & $\pm$ 0.16 &  3	 & $-$0.03 & $\pm$ 0.15 &  1	 &    0.06 & $\pm$ 0.11 & 11	 & $-$0.03 & $\pm$ 0.06 &  2	 \\
   RZ\,Gem & 54845.094 & $-$0.16 & $\pm$ 0.03 &  (44, 5) &    0.19 & $\pm$ 0.16 &  2	 &	   & \ldots	& \ldots &    0.03 & $\pm$ 0.07 &  1	 &    0.07 & $\pm$ 0.15 & 12	 & $-$0.12 & $\pm$ 0.06 &  2	 \\
   BE\,Mon & 54846.201 &    0.05 & $\pm$ 0.09 &  (78, 5) &    0.34 & $\pm$ 0.13 &  2	 &    0.13 & $\pm$ 0.24 &  6	 &    0.33 & $\pm$ 0.22 &  1	 &    0.14 & $\pm$ 0.16 & 12	 &    0.08 & $\pm$ 0.12 &  3	 \\
   CV\,Mon & 54846.182 &    0.09 & $\pm$ 0.09 &  (52, 2) &    0.31 & $\pm$ 0.20 &  2	 &    0.15 & $\pm$ 0.16 &  4	 &    0.22 & $\pm$ 0.04 &  1	 &    0.24 & $\pm$ 0.27 & 11	 &    0.18 & $\pm$ 0.24 &  3	 \\
   FT\,Mon & 54845.104 & $-$0.13 & $\pm$ 0.08 &  (61, 8) &    0.15 & $\pm$ 0.19 &  2	 & $-$0.10 & $\pm$ 0.17 &  5	 & $-$0.08 & $\pm$ 0.01 &  1	 &	   & \ldots	& \ldots &    0.02 & $\pm$ 0.01 &  2	 \\
   SV\,Mon & 54845.119 &    0.12 & $\pm$ 0.08 &  (54, 8) &    0.63 & $\pm$ 0.04 &  2	 &    0.16 & $\pm$ 0.06 &  6	 &    0.30 & $\pm$ 0.17 &  1	 &    0.21 & $\pm$ 0.16 & 14	 &    0.24 & $\pm$ 0.20 &  3	 \\
   TW\,Mon & 54796.347 & $-$0.13 & $\pm$ 0.07 &  (75, 6) &    0.16 & $\pm$ 0.06 &  2	 & $-$0.03 & $\pm$ 0.31 &  6	 & $-$0.04 & $\pm$ 0.07 &  1	 & $-$0.02 & $\pm$ 0.14 & 13	 &    0.02 & $\pm$ 0.12 &  3	 \\
   TX\,Mon & 54798.345 & $-$0.03 & $\pm$ 0.05 &  (76, 4) &    0.23 & $\pm$ 0.06 &  2	 & $-$0.08 & $\pm$ 0.11 &  4	 & $-$0.04 & $\pm$ 0.08 &  1	 &    0.06 & $\pm$ 0.13 & 11	 &    0.04 & $\pm$ 0.10 &  2	 \\
   TY\,Mon & 54846.139 &    0.02 & $\pm$ 0.08 &  (85, 6) &    0.17 & $\pm$ 0.09 &  2	 & $-$0.19 & $\pm$ 0.07 &  3	 &    0.03 & $\pm$ 0.11 &  1	 &    0.14 & $\pm$ 0.23 & 11	 &    0.12 & $\pm$ 0.15 &  2	 \\
   TZ\,Mon & 54847.237 & $-$0.02 & $\pm$ 0.07 &  (94, 6) &    0.16 & $\pm$ 0.01 &  2	 &    0.02 & $\pm$ 0.08 &  6	 &    0.11 & $\pm$ 0.05 &  1	 &    0.03 & $\pm$ 0.12 & 12	 &    0.04 & $\pm$ 0.18 &  3	 \\
 V465\,Mon & 54847.241 & $-$0.07 & $\pm$ 0.07 & (107, 6) &    0.26 & $\pm$ 0.08 &  2	 &    0.06 & $\pm$ 0.17 &  6	 & $-$0.08 & $\pm$ 0.17 &  1	 &    0.12 & $\pm$ 0.18 & 12	 &    0.04 & $\pm$ 0.08 &  3	 \\
 V495\,Mon & 54846.167 & $-$0.13 & $\pm$ 0.07 &  (73, 4) &    0.04 & $\pm$ 0.08 &  2	 & $-$0.06 & $\pm$ 0.30 &  6	 &	   & \ldots	& \ldots & $-$0.01 & $\pm$ 0.18 & 12	 & $-$0.04 & $\pm$ 0.12 &  3	 \\
 V508\,Mon & 54847.232 & $-$0.04 & $\pm$ 0.10 & (118, 7) &    0.12 & $\pm$ 0.01 &  2	 &    0.04 & $\pm$ 0.28 &  5	 &    0.01 & $\pm$ 0.05 &  1	 &    0.04 & $\pm$ 0.20 & 13	 &    0.04 & $\pm$ 0.21 &  2	 \\
 V510\,Mon & 54846.153 & $-$0.16 & $\pm$ 0.06 &  (80, 3) & $-$0.05 & $\pm$ 0.03 &  2	 & $-$0.16 & $\pm$ 0.15 &  5	 & $-$0.04 & $\pm$ 0.06 &  1	 & $-$0.09 & $\pm$ 0.15 & 12	 & $-$0.15 & $\pm$ 0.03 &  3	 \\
   XX\,Mon & 54798.335 &    0.01 & $\pm$ 0.08 &  (55, 2) &    0.51 & $\pm$ 0.37 &  2	 &    0.05 & $\pm$ 0.22 &  4	 &    0.21 & $\pm$ 0.07 &  1	 &    0.17 & $\pm$ 0.23 & 12	 &    0.23 & $\pm$ 0.06 &  2	 \\
   GU\,Nor & 54667.205 &    0.08 & $\pm$ 0.06 &  (80, 7) &    0.32 & $\pm$ 0.04 &  2	 &    0.17 & $\pm$ 0.10 &  6	 &    0.18 & $\pm$ 0.19 &  1	 &    0.28 & $\pm$ 0.15 & 13	 &    0.09 & $\pm$ 0.04 &  2	 \\
   IQ\,Nor & 54584.299 &    0.22 & $\pm$ 0.07 &  (63, 7) &    0.57 & $\pm$ 0.07 &  2	 &    0.39 & $\pm$ 0.13 &  5	 &    0.36 & $\pm$ 0.13 &  1	 &    0.38 & $\pm$ 0.16 & 11	 &    0.25 & $\pm$ 0.06 &  3	 \\
   QZ\,Nor & 54863.366 &    0.18 & $\pm$ 0.08 &  (81, 3) &    0.56 & $\pm$ 0.01 &  2	 &    0.30 & $\pm$ 0.18 &  6	 &    0.38 & $\pm$ 0.11 &  1	 &    0.41 & $\pm$ 0.16 & 14	 &    0.17 & $\pm$ 0.15 &  3	 \\
   QZ\,Nor & 54923.345 &    0.23 & $\pm$ 0.07 &  (86, 2) &    0.55 & $\pm$ 0.05 &  2	 &    0.30 & $\pm$ 0.18 &  6	 &    0.40 & $\pm$ 0.16 &  1	 &    0.40 & $\pm$ 0.19 & 14	 &    0.23 & $\pm$ 0.16 &  3	 \\
   RS\,Nor & 54863.361 &    0.18 & $\pm$ 0.08 &  (82, 5) &    0.57 & $\pm$ 0.07 &  2	 &    0.39 & $\pm$ 0.13 &  5	 &    0.36 & $\pm$ 0.13 &  1	 &    0.38 & $\pm$ 0.16 & 11	 &    0.25 & $\pm$ 0.06 &  3	 \\
   SY\,Nor & 54708.061 &    0.27 & $\pm$ 0.10 &  (46, 5) &    0.73 & $\pm$ 0.25 &  2	 &    0.45 & $\pm$ 0.14 &  5	 &    0.37 & $\pm$ 0.01 &  1	 &    0.39 & $\pm$ 0.15 &  9	 &    0.20 & $\pm$ 0.13 &  2	 \\
   SY\,Nor & 54709.075 &    0.20 & $\pm$ 0.09 &  (58, 4) &    0.50 & $\pm$ 0.01 &  2	 &    0.36 & $\pm$ 0.11 &  5	 &    0.37 & $\pm$ 0.09 &  1	 &    0.35 & $\pm$ 0.14 & 11	 &    0.18 & $\pm$ 0.15 &  3	 \\
   TW\,Nor & 54666.127 &    0.27 & $\pm$ 0.10 &  (69, 7) &    0.58 & $\pm$ 0.11 &  2	 &    0.24 & $\pm$ 0.07 &  2	 &    0.58 & $\pm$ 0.60 &  1	 &    0.25 & $\pm$ 0.10 & 10	 &    0.05 & $\pm$ 0.06 &  2	 \\
 V340\,Nor & 54873.376 &    0.07 & $\pm$ 0.07 &  (47, 4) &    0.40 & $\pm$ 0.05 &  1	 &    0.37 & $\pm$ 0.08 &  1	 &    0.12 & $\pm$ 0.11 &  1	 &    0.30 & $\pm$ 0.21 &  9	 &    0.19 & $\pm$ 0.18 &  3	 \\
   CS\,Ori & 54845.085 & $-$0.25 & $\pm$ 0.06 &  (68, 6) &    0.11 & $\pm$ 0.06 &  2	 & $-$0.27 & $\pm$ 0.14 &  4	 & $-$0.25 & $\pm$ 0.12 &  1	 & $-$0.10 & $\pm$ 0.13 & 11	 & $-$0.16 & $\pm$ 0.04 &  2	 \\
   GQ\,Ori & 54845.082 &    0.20 & $\pm$ 0.08 &  (92,14) &	   & \ldots	& \ldots &	   & \ldots	& \ldots &	   & \ldots	& \ldots &	   & \ldots	& \ldots &	   & \ldots	& \ldots \\
   RS\,Ori & 54845.100 &    0.11 & $\pm$ 0.09 &  (71, 5) &    0.12 & $\pm$ 0.16 &  2	 &    0.11 & $\pm$ 0.22 &  5	 &    0.41 & $\pm$ 0.17 &  1	 &    0.30 & $\pm$ 0.20 & 11	 &    0.22 & $\pm$ 0.19 &  4	 \\
   AQ\,Pup & 54839.075 &    0.06 & $\pm$ 0.05 &  (14, 2) &    0.36 & $\pm$ 0.05 &  1	 &	   & \ldots	& \ldots &    0.06 & $\pm$ 0.04 &  1	 &    0.25 & $\pm$ 0.22 &  6	 &    0.05 & $\pm$ 0.00 &  2	 \\
   BC\,Pup & 54839.147 & $-$0.31 & $\pm$ 0.07 &  (57, 3) &    0.20 & $\pm$ 0.05 &  2	 & $-$0.07 & $\pm$ 0.22 &  6	 & $-$0.14 & $\pm$ 0.21 &  1	 & $-$0.13 & $\pm$ 0.17 & 12	 & $-$0.18 & $\pm$ 0.06 &  3	 \\
   BM\,Pup & 54839.086 & $-$0.07 & $\pm$ 0.08 &  (61, 7) &    0.17 & $\pm$ 0.15 &  2	 & $-$0.02 & $\pm$ 0.18 &  5	 &	   & \ldots	& \ldots &    0.04 & $\pm$ 0.18 & 11	 & $-$0.01 & $\pm$ 0.03 &  3	 \\
   BN\,Pup & 54839.109 &    0.03 & $\pm$ 0.05 &  (69, 4) &    0.22 & $\pm$ 0.02 &  2	 &    0.16 & $\pm$ 0.10 &  6	 &    0.23 & $\pm$ 0.07 &  1	 &    0.16 & $\pm$ 0.14 & 14	 &    0.09 & $\pm$ 0.17 &  3	 \\
   CK\,Pup & 54839.113 & $-$0.15 & $\pm$ 0.08 &  (72, 4) &    0.17 & $\pm$ 0.06 &  2	 & $-$0.12 & $\pm$ 0.20 &  5	 &    0.00 & $\pm$ 0.08 &  1	 &    0.00 & $\pm$ 0.21 & 12	 & $-$0.10 & $\pm$ 0.06 &  2	 \\
   CK\,Pup & 54839.173 & $-$0.12 & $\pm$ 0.08 &  (78,11) &    0.16 & $\pm$ 0.01 &  2	 & $-$0.05 & $\pm$ 0.29 &  6	 & $-$0.02 & $\pm$ 0.02 &  1	 &    0.02 & $\pm$ 0.20 & 12	 & $-$0.05 & $\pm$ 0.06 &  1	 \\
   HW\,Pup & 54792.249 & $-$0.22 & $\pm$ 0.09 &  (70, 3) & $-$0.02 & $\pm$ 0.01 &  2	 & $-$0.12 & $\pm$ 0.12 &  6	 & $-$0.06 & $\pm$ 0.14 &  1	 & $-$0.05 & $\pm$ 0.24 & 13	 & $-$0.11 & $\pm$ 0.12 &  3	 \\
   LS\,Pup & 54839.081 & $-$0.12 & $\pm$ 0.11 &  (18, 1) &	   & \ldots	& \ldots &    0.42 & $\pm$ 0.08 &  1	 &    0.10 & $\pm$ 0.10 &  1	 &    0.04 & $\pm$ 0.09 &  7	 &    0.09 & $\pm$ 0.06 &  1	 \\
   VW\,Pup & 54832.331 & $-$0.14 & $\pm$ 0.06 &  (50, 4) &    0.11 & $\pm$ 0.03 &  2	 & $-$0.04 & $\pm$ 0.17 &  5	 & $-$0.33 & $\pm$ 0.21 &  1	 & $-$0.03 & $\pm$ 0.14 & 12	 & $-$0.08 & $\pm$ 0.09 &  3	 \\
   VZ\,Pup & 54839.096 & $-$0.01 & $\pm$ 0.04 &  (27, 2) &	   & \ldots	& \ldots &    0.30 & $\pm$ 0.39 &  3	 &    0.01 & $\pm$ 0.16 &  1	 &    0.25 & $\pm$ 0.16 & 11	 &    0.11 & $\pm$ 0.01 &  2	 \\
   WW\,Pup & 54839.091 &    0.13 & $\pm$ 0.16 &  (18, 1) & $-$0.30 & $\pm$ 0.11 &  2	 & $-$0.38 & $\pm$ 0.24 &  6	 & $-$0.34 & $\pm$ 0.29 &  1	 & $-$0.36 & $\pm$ 0.22 & 14	 & $-$0.70 & $\pm$ 0.04 &  2	 \\
   WY\,Pup & 54839.100 & $-$0.10 & $\pm$ 0.08 &  (49, 6) &    0.21 & $\pm$ 0.05 &  1	 & $-$0.14 & $\pm$ 0.29 &  3	 &    0.05 & $\pm$ 0.08 &  1	 & $-$0.01 & $\pm$ 0.10 & 11	 &    0.01 & $\pm$ 0.05 &  2	 \\
\hline\noalign{\smallskip}
\multicolumn{20}{r}{\it {\footnotesize continued on next page}} \\
\end{tabular}}
\tablefoot{Column 3 lists the weighted mean and standard deviation of the
\ion{Fe}{i} and \ion{Fe}{ii} abundances derived by \citet{Genovalietal2014}.
Column 4 lists the respective number ($N_{\rm L}$) of iron lines used. The
other $N_{\rm L}$ values indicate the number of lines used for the other
elements to derive their abundances. For these elements, the quoted errors
represent either the dispersion around the mean if two or more lines were
measured, or the mean dispersion computed for the eleven calibrating stars
if only one line was available.}
\end{table*}
\addtocounter{table}{-1}
\begin{table*}[p]
\centering
\caption[]{continued.}
{\scriptsize 
\begin{tabular}{rc r@{ }l c r@{ }l c r@{ }l c r@{ }l c r@{ }l c r@{ }l c}
\noalign{\smallskip}\hline\hline\noalign{\smallskip}
Name & MJD &
\multicolumn{2}{c}{[Fe/H]} & \parbox[c]{1.0cm}{\centering $N_{\rm L}$ (\ion{Fe}{i},\ion{Fe}{ii})} &
\multicolumn{2}{c}{[Na/H]} & $N_{\rm L}$ &
\multicolumn{2}{c}{[Al/H]} & $N_{\rm L}$ &
\multicolumn{2}{c}{[Mg/H]} & $N_{\rm L}$ &
\multicolumn{2}{c}{[Si/H]} & $N_{\rm L}$ &
\multicolumn{2}{c}{[Ca/H]} & $N_{\rm L}$ \\
\noalign{\smallskip}\hline\noalign{\smallskip}
   WZ\,Pup & 54839.104 & $-$0.07 & $\pm$ 0.06 &  (72, 7) &    0.09  & $\pm$ 0.01 &  2     & $-$0.11 & $\pm$ 0.07 &  4	  &    0.14 & $\pm$ 0.09 &  1	  &    0.07 & $\pm$ 0.17 & 13	  &    0.04 & $\pm$ 0.07 &  3	  \\
    X\,Pup & 54839.070 &    0.02 & $\pm$ 0.08 &  (15, 2) &    0.42  & $\pm$ 0.05 &  1     &    0.35 & $\pm$ 0.08 &  1	  & $-$0.01 & $\pm$ 0.01 &  1	  & $-$0.01 & $\pm$ 0.21 &  4	  &    0.09 & $\pm$ 0.02 &  2	  \\
   KQ\,Sco & 56139.021 &    0.26 & $\pm$ 0.15 &  (32, 0) &    0.69  & $\pm$ 0.05 &  1     &    0.59 & $\pm$ 0.08 &  1	  &	    & \ldots	 & \ldots &    0.18 & $\pm$ 0.06 &  1	  &    0.16 & $\pm$ 0.06 &  1	  \\
   KQ\,Sco & 54873.379 &    0.52 & $\pm$ 0.08 &  (51, 4) &          & \ldots     & \ldots &         & \ldots	 & \ldots &	    & \ldots	 & \ldots &    0.17 & $\pm$ 0.24 &  2	  &    0.23 & $\pm$ 0.06 &  1	  \\
   KQ\,Sco & 56152.097 &    0.30 & $\pm$ 0.21 &  (16, 1) &    0.72  & $\pm$ 0.05 &  2     &    0.38 & $\pm$ 0.02 &  2	  &	    & \ldots	 & \ldots &    0.47 & $\pm$ 0.06 &  1	  &    0.15 & $\pm$ 0.11 &  2	  \\
   KQ\,Sco & 56163.004 &    0.22 & $\pm$ 0.27 &  (20, 1) &          & \ldots     & \ldots &         & \ldots	 & \ldots &	    & \ldots	 & \ldots &    0.23 & $\pm$ 0.06 &  1	  &	    & \ldots	 & \ldots \\
   KQ\,Sco & 56166.004 &    0.21 & $\pm$ 0.28 &  (15, 1) &    0.70  & $\pm$ 0.05 &  1     &         & \ldots	 & \ldots &	    & \ldots	 & \ldots &    0.24 & $\pm$ 0.06 &  1	  &	    & \ldots	 & \ldots \\
   RY\,Sco & 56140.187 &    0.06 & $\pm$ 0.01 &  (74, 2) &    0.31  & $\pm$ 0.01 &  2     &    0.14 & $\pm$ 0.04 &  2	  &    0.00 & $\pm$ 0.07 &  1	  &    0.19 & $\pm$ 0.13 &  3	  & $-$0.01 & $\pm$ 0.14 &  7	  \\
   RY\,Sco & 54599.412 &    0.06 & $\pm$ 0.02 &  (34, 5) &    0.46  & $\pm$ 0.06 &  2     &    0.11 & $\pm$ 0.10 &  6	  &    0.25 & $\pm$ 0.14 &  1	  &    0.19 & $\pm$ 0.11 & 12	  &    0.02 & $\pm$ 0.10 &  3	  \\
   RY\,Sco & 56152.143 &    0.01 & $\pm$ 0.03 &  (75, 3) &    0.32  & $\pm$ 0.06 &  2     &    0.22 & $\pm$ 0.01 &  2	  & $-$0.01 & $\pm$ 0.07 &  1	  &    0.15 & $\pm$ 0.18 &  3	  & $-$0.06 & $\pm$ 0.12 &  7	  \\
   RY\,Sco & 56162.170 & $-$0.03 & $\pm$ 0.05 &  (66, 2) &    0.29  & $\pm$ 0.00 &  2     &    0.14 & $\pm$ 0.04 &  2	  &    0.01 & $\pm$ 0.07 &  1	  &    0.12 & $\pm$ 0.09 &  3	  & $-$0.02 & $\pm$ 0.14 &  7	  \\
   RY\,Sco & 56167.085 & $-$0.04 & $\pm$ 0.08 &  (45, 2) &    0.40  & $\pm$ 0.16 &  2     &    0.22 & $\pm$ 0.12 &  2	  &    0.00 & $\pm$ 0.07 &  1	  &    0.09 & $\pm$ 0.08 &  3	  & $-$0.17 & $\pm$ 0.11 &  3	  \\
 V470\,Sco & 54708.073 &    0.16 & $\pm$ 0.06 &  (66, 4) &    0.55  & $\pm$ 0.11 &  2     &    0.26 & $\pm$ 0.13 &  5	  &    0.18 & $\pm$ 0.01 &  1	  &    0.23 & $\pm$ 0.09 &  8	  &    0.08 & $\pm$ 0.10 &  3	  \\
 V500\,Sco & 56140.191 & $-$0.01 & $\pm$ 0.05 &  (86, 3) &    0.16  & $\pm$ 0.04 &  2     &    0.13 & $\pm$ 0.03 &  2	  & $-$0.18 & $\pm$ 0.07 &  1	  &    0.00 & $\pm$ 0.16 &  3	  & $-$0.17 & $\pm$ 0.09 &  7	  \\
 V500\,Sco & 56152.092 &    0.00 & $\pm$ 0.10 &  (67, 3) &    0.17  & $\pm$ 0.02 &  2     &    0.03 & $\pm$ 0.04 &  2	  & $-$0.04 & $\pm$ 0.07 &  1	  &    0.03 & $\pm$ 0.08 &  3	  & $-$0.10 & $\pm$ 0.11 &  7	  \\
 V500\,Sco & 56162.998 & $-$0.03 & $\pm$ 0.12 &  (53, 3) &    0.26  & $\pm$ 0.08 &  2     &    0.17 & $\pm$ 0.03 &  2	  & $-$0.05 & $\pm$ 0.07 &  1	  &    0.04 & $\pm$ 0.07 &  3	  & $-$0.09 & $\pm$ 0.07 &  5	  \\
 V500\,Sco & 56167.077 & $-$0.11 & $\pm$ 0.07 &  (97, 5) &    0.16  & $\pm$ 0.05 &  2     &    0.14 & $\pm$ 0.02 &  2	  & $-$0.19 & $\pm$ 0.07 &  1	  &    0.01 & $\pm$ 0.23 &  3	  & $-$0.21 & $\pm$ 0.12 &  7	  \\
   EV\,Sct & 54708.086 &    0.09 & $\pm$ 0.07 &  (57, 3) &    0.25  & $\pm$ 0.05 &  1     &    0.56 & $\pm$ 0.08 &  1	  &    0.06 & $\pm$ 0.14 &  1	  &    0.26 & $\pm$ 0.21 &  9	  &    0.10 & $\pm$ 0.18 &  2	  \\
   RU\,Sct & 54906.414 &    0.16 & $\pm$ 0.05 &  (95, 7) &    0.43  & $\pm$ 0.11 &  2     &    0.24 & $\pm$ 0.15 &  6	  &    0.31 & $\pm$ 0.18 &  1	  &    0.35 & $\pm$ 0.19 & 12	  &    0.21 & $\pm$ 0.25 &  3	  \\
   RU\,Sct & 54923.375 &    0.09 & $\pm$ 0.07 &  (45, 5) &    0.35  & $\pm$ 0.01 &  2     &    0.17 & $\pm$ 0.08 &  5	  &    0.31 & $\pm$ 0.12 &  1	  &    0.25 & $\pm$ 0.12 & 10	  &    0.14 & $\pm$ 0.22 &  3	  \\
   UZ\,Sct & 56137.160 &    0.28 & $\pm$ 0.12 &  (17, 4) &          & \ldots     & \ldots &         & \ldots	 & \ldots &	    & \ldots	 & \ldots &    0.32 & $\pm$ 0.06 &  1	  &    0.14 & $\pm$ 0.06 &  1	  \\
   UZ\,Sct & 54906.400 &    0.36 & $\pm$ 0.10 &  (34, 5) &    0.83  & $\pm$ 0.08 &  2     &    0.66 & $\pm$ 0.25 &  5	  &    0.36 & $\pm$ 0.21 &  1	  &    0.58 & $\pm$ 0.15 & 10	  &    0.42 & $\pm$ 0.20 &  3	  \\
   UZ\,Sct & 54923.366 &    0.45 & $\pm$ 0.07 &  (63, 7) &    0.92  & $\pm$ 0.08 &  2     &    0.55 & $\pm$ 0.12 &  4	  &    0.31 & $\pm$ 0.21 &  1	  &    0.48 & $\pm$ 0.13 & 11	  &    0.28 & $\pm$ 0.16 &  2	  \\
   UZ\,Sct & 56152.064 &    0.25 & $\pm$ 0.28 &  ( 8, 0) &          & \ldots     & \ldots &         & \ldots	 & \ldots &	    & \ldots	 & \ldots &    0.41 & $\pm$ 0.06 &  1	  &    0.09 & $\pm$ 0.06 &  1	  \\
   UZ\,Sct & 56160.167 &    0.36 & $\pm$ 0.10 &  (47, 2) &    0.72  & $\pm$ 0.03 &  2     &         & \ldots	 & \ldots &    0.31 & $\pm$ 0.07 &  1	  &    0.40 & $\pm$ 0.05 &  3	  &    0.22 & $\pm$ 0.10 &  5	  \\
   UZ\,Sct & 56175.049 &    0.31 & $\pm$ 0.21 &  (36, 2) &    0.67  & $\pm$ 0.13 &  2     &         & \ldots	 & \ldots &    0.30 & $\pm$ 0.07 &  1	  &    0.48 & $\pm$ 0.13 &  2	  &    0.19 & $\pm$ 0.05 &  4	  \\
 V367\,Sct & 56137.147 &    0.13 & $\pm$ 0.07 &  (72, 3) &    0.28  & $\pm$ 0.01 &  2     &    0.33 & $\pm$ 0.08 &  1	  & $-$0.02 & $\pm$ 0.07 &  1	  &    0.17 & $\pm$ 0.19 &  3	  &    0.04 & $\pm$ 0.18 &  7	  \\
 V367\,Sct & 54709.128 & $-$0.04 & $\pm$ 0.04 &  (56, 4) &    0.26  & $\pm$ 0.02 &  2     &    0.19 & $\pm$ 0.26 &  6	  &    0.01 & $\pm$ 0.29 &  1	  &    0.14 & $\pm$ 0.10 & 10	  & $-$0.06 & $\pm$ 0.07 &  3	  \\
 V367\,Sct & 56175.105 &    0.14 & $\pm$ 0.06 &  (50, 3) &    0.36  & $\pm$ 0.05 &  1     &    0.37 & $\pm$ 0.08 &  1	  &    0.00 & $\pm$ 0.07 &  1	  &    0.18 & $\pm$ 0.19 &  2	  &    0.07 & $\pm$ 0.15 &  5	  \\
 V367\,Sct & 56184.000 &    0.03 & $\pm$ 0.07 &  (84, 4) &    0.31  & $\pm$ 0.11 &  2     &    0.44 & $\pm$ 0.08 &  1	  &    0.00 & $\pm$ 0.07 &  1	  &    0.15 & $\pm$ 0.24 &  2	  & $-$0.02 & $\pm$ 0.10 &  7	  \\
    X\,Sct & 54709.122 &    0.12 & $\pm$ 0.09 &  (72, 9) &    0.41  & $\pm$ 0.25 &  2     &    0.34 & $\pm$ 0.26 &  5	  &    0.12 & $\pm$ 0.12 &  1	  &    0.32 & $\pm$ 0.27 & 11	  &    0.14 & $\pm$ 0.14 &  3	  \\
    Z\,Sct & 56137.123 &    0.10 & $\pm$ 0.16 &  (20, 0) &    0.77  & $\pm$ 0.05 &  1     &    0.58 & $\pm$ 0.08 &  1	  & $-$0.21 & $\pm$ 0.07 &  1	  &	    & \ldots	 & \ldots &    0.13 & $\pm$ 0.01 &  2	  \\
    Z\,Sct & 54678.090 &    0.18 & $\pm$ 0.09 &  (41, 3) &    0.68  & $\pm$ 0.05 &  1     &    0.41 & $\pm$ 0.08 &  1	  & $-$0.08 & $\pm$ 0.23 &  1	  &    0.07 & $\pm$ 0.14 &  4	  &    0.06 & $\pm$ 0.06 &  1	  \\
    Z\,Sct & 56152.073 &    0.11 & $\pm$ 0.02 &  (49, 2) &    0.49  & $\pm$ 0.05 &  1     &    0.40 & $\pm$ 0.08 &  1	  & $-$0.06 & $\pm$ 0.07 &  1	  &    0.18 & $\pm$ 0.14 &  3	  & $-$0.05 & $\pm$ 0.07 &  4	  \\
    Z\,Sct & 56159.186 &    0.26 & $\pm$ 0.08 &  (45, 2) &    0.74  & $\pm$ 0.14 &  2     &    0.53 & $\pm$ 0.08 &  1	  &    0.20 & $\pm$ 0.07 &  1	  &    0.38 & $\pm$ 0.12 &  3	  &    0.08 & $\pm$ 0.04 &  2	  \\
    Z\,Sct & 56175.038 &    0.00 & $\pm$ 0.30 &  (25, 0) &    0.55  & $\pm$ 0.05 &  1     &    0.44 & $\pm$ 0.08 &  1	  &	    & \ldots	 & \ldots &    0.03 & $\pm$ 0.06 &  1	  & $-$0.08 & $\pm$ 0.03 &  2	  \\
   AA\,Ser & 54708.040 &    0.38 & $\pm$ 0.20 &  (24, 1) &    0.98  & $\pm$ 0.05 &  1     &    0.50 & $\pm$ 0.08 &  1	  &    0.28 & $\pm$ 0.07 &  1	  &    0.47 & $\pm$ 0.10 &  6	  &    0.16 & $\pm$ 0.06 &  1	  \\
   CR\,Ser & 54709.116 &    0.12 & $\pm$ 0.08 &  (53, 5) &    0.61  & $\pm$ 0.29 &  2     &    0.29 & $\pm$ 0.19 &  6	  &    0.32 & $\pm$ 0.22 &  1	  &    0.28 & $\pm$ 0.21 & 10	  &    0.11 & $\pm$ 0.10 &  3	  \\
   AV\,Sgr & 56136.169 &    0.40 & $\pm$ 0.15 &  (29, 2) &    0.85  & $\pm$ 0.00 &  2     &    0.54 & $\pm$ 0.08 &  1	  &	    & \ldots	 & \ldots &    0.48 & $\pm$ 0.06 &  1	  &    0.22 & $\pm$ 0.01 &  2	  \\
   AV\,Sgr & 56136.192 &    0.44 & $\pm$ 0.15 &  (31, 2) &    0.93  & $\pm$ 0.03 &  2     &    0.59 & $\pm$ 0.08 &  1	  &	    & \ldots	 & \ldots &    0.50 & $\pm$ 0.06 &  1	  &    0.24 & $\pm$ 0.07 &  2	  \\
   AV\,Sgr & 54923.348 &    0.53 & $\pm$ 0.17 &  (16, 2) &    0.90  & $\pm$ 0.05 &  1     &    0.03 & $\pm$ 0.08 &  1	  &	    & \ldots	 & \ldots &    0.35 & $\pm$ 0.02 &  2	  &    0.05 & $\pm$ 0.06 &  1	  \\
   AV\,Sgr & 56152.082 &    0.42 & $\pm$ 0.17 &  (24, 2) &    0.77  & $\pm$ 0.01 &  2     &    0.61 & $\pm$ 0.04 &  2	  &	    & \ldots	 & \ldots &    0.53 & $\pm$ 0.06 &  1	  &    0.17 & $\pm$ 0.02 &  2	  \\
   AV\,Sgr & 56168.049 &    0.30 & $\pm$ 0.22 &  (19, 1) &    0.88  & $\pm$ 0.05 &  1     &    0.75 & $\pm$ 0.13 &  2	  &	    & \ldots	 & \ldots &    0.49 & $\pm$ 0.06 &  1	  &    0.21 & $\pm$ 0.04 &  2	  \\
   AY\,Sgr & 54599.398 &    0.11 & $\pm$ 0.06 &  (58, 5) &    0.32  & $\pm$ 0.06 &  2     &    0.19 & $\pm$ 0.17 &  6	  &    0.20 & $\pm$ 0.06 &  1	  &    0.23 & $\pm$ 0.15 & 14	  &    0.11 & $\pm$ 0.10 &  3	  \\
V1954\,Sgr & 54599.389 &    0.24 & $\pm$ 0.10 &  (61, 4) &    0.62  & $\pm$ 0.06 &  2     &    0.27 & $\pm$ 0.14 &  5	  &    0.17 & $\pm$ 0.29 &  1	  &    0.47 & $\pm$ 0.22 & 13	  &    0.25 & $\pm$ 0.08 &  3	  \\
 V773\,Sgr & 54669.207 &    0.11 & $\pm$ 0.06 &  (58, 8) &    0.30  & $\pm$ 0.05 &  1     &    0.05 & $\pm$ 0.08 &  1	  &    0.20 & $\pm$ 0.00 &  1	  &    0.30 & $\pm$ 0.17 & 10	  &    0.14 & $\pm$ 0.02 &  2	  \\
   VY\,Sgr & 56160.179 &    0.27 & $\pm$ 0.25 &  (14, 1) &    0.73  & $\pm$ 0.05 &  1     &    0.71 & $\pm$ 0.08 &  1	  &	    & \ldots	 & \ldots &    0.17 & $\pm$ 0.06 &  1	  &    0.08 & $\pm$ 0.06 &  1	  \\
   VY\,Sgr & 54923.356 &    0.42 & $\pm$ 0.14 &  (30, 6) &    0.94  & $\pm$ 0.09 &  2     &    0.54 & $\pm$ 0.11 &  6	  &    0.53 & $\pm$ 0.13 &  1	  &    0.54 & $\pm$ 0.16 & 11	  &    0.32 & $\pm$ 0.10 &  2	  \\
   VY\,Sgr & 56162.162 &    0.32 & $\pm$ 0.27 &  (17, 1) &    0.85  & $\pm$ 0.05 &  1     &    0.72 & $\pm$ 0.08 &  1	  &	    & \ldots	 & \ldots &    0.06 & $\pm$ 0.06 &  1	  &    0.09 & $\pm$ 0.11 &  2	  \\
   VY\,Sgr & 56168.062 &    0.31 & $\pm$ 0.05 &  (51, 2) &    0.61  & $\pm$ 0.06 &  2     &         & \ldots	 & \ldots &    0.19 & $\pm$ 0.07 &  1	  &    0.34 & $\pm$ 0.08 &  3	  &    0.22 & $\pm$ 0.08 &  5	  \\
   WZ\,Sgr & 56132.190 &    0.18 & $\pm$ 0.08 &  (56, 2) &    0.56  & $\pm$ 0.12 &  2     &    0.51 & $\pm$ 0.08 &  1	  &    0.04 & $\pm$ 0.07 &  1	  &    0.24 & $\pm$ 0.11 &  3	  &    0.08 & $\pm$ 0.11 &  5	  \\
   WZ\,Sgr & 54599.395 &    0.35 & $\pm$ 0.08 &  (42, 2) &    0.78  & $\pm$ 0.05 &  1     &    0.72 & $\pm$ 0.07 &  2	  &	    & \ldots	 & \ldots &    0.48 & $\pm$ 0.28 &  3	  &    0.23 & $\pm$ 0.06 &  1	  \\
   WZ\,Sgr & 56136.213 &    0.24 & $\pm$ 0.01 &  (48, 2) &    0.41  & $\pm$ 0.05 &  2     &         & \ldots	 & \ldots &    0.03 & $\pm$ 0.07 &  1	  &    0.31 & $\pm$ 0.07 &  3	  &    0.18 & $\pm$ 0.07 &  5	  \\
   WZ\,Sgr & 56152.044 &    0.28 & $\pm$ 0.12 &  (28, 2) &    0.56  & $\pm$ 0.05 &  1     &    0.56 & $\pm$ 0.08 &  1	  &    0.00 & $\pm$ 0.07 &  1	  &    0.13 & $\pm$ 0.06 &  1	  &    0.13 & $\pm$ 0.01 &  2	  \\
   WZ\,Sgr & 56159.125 &    0.37 & $\pm$ 0.06 &  (44, 2) &    0.65  & $\pm$ 0.16 &  2     &    0.48 & $\pm$ 0.33 &  2	  & $-$0.02 & $\pm$ 0.07 &  1	  &    0.36 & $\pm$ 0.07 &  3	  &    0.23 & $\pm$ 0.09 &  5	  \\
   XX\,Sgr & 56054.234 & $-$0.01 & $\pm$ 0.10 & (100, 5) &    0.22  & $\pm$ 0.05 &  2     &    0.25 & $\pm$ 0.01 &  2	  & $-$0.05 & $\pm$ 0.07 &  1	  &    0.04 & $\pm$ 0.13 &  3	  & $-$0.06 & $\pm$ 0.16 &  7	  \\
   XX\,Sgr & 54599.404 & $-$0.07 & $\pm$ 0.07 &  (59, 4) &    0.29  & $\pm$ 0.03 &  2     &    0.02 & $\pm$ 0.15 &  5	  &    0.06 & $\pm$ 0.05 &  1	  &    0.23 & $\pm$ 0.21 & 13	  & $-$0.03 & $\pm$ 0.09 &  2	  \\
   XX\,Sgr & 56136.223 & $-$0.05 & $\pm$ 0.05 & (101, 6) &    0.20  & $\pm$ 0.08 &  2     &    0.21 & $\pm$ 0.08 &  1	  & $-$0.10 & $\pm$ 0.07 &  1	  &    0.07 & $\pm$ 0.13 &  3	  & $-$0.06 & $\pm$ 0.11 &  6	  \\
   XX\,Sgr & 56152.047 &    0.05 & $\pm$ 0.03 &  (43, 3) &    0.35  & $\pm$ 0.11 &  2     &    0.41 & $\pm$ 0.01 &  2	  &    0.08 & $\pm$ 0.07 &  1	  &    0.14 & $\pm$ 0.14 &  3	  &    0.17 & $\pm$ 0.20 &  6	  \\
   XX\,Sgr & 56159.128 & $-$0.02 & $\pm$ 0.12 &  (64, 4) &    0.40  & $\pm$ 0.29 &  2     &    0.30 & $\pm$ 0.01 &  2	  & $-$0.04 & $\pm$ 0.07 &  1	  &    0.08 & $\pm$ 0.05 &  3	  & $-$0.02 & $\pm$ 0.09 &  5	  \\
   EZ\,Vel & 54759.348 & $-$0.17 & $\pm$ 0.15 &  (23, 1) &    0.17  & $\pm$ 0.28 &  2     &    0.01 & $\pm$ 0.18 &  4	  & $-$0.03 & $\pm$ 0.28 &  1	  &	    & \ldots	 & \ldots & $-$0.08 & $\pm$ 0.16 &  2	  \\
\hline
\end{tabular}}
\end{table*}

\clearpage
\begin{table*}[p]
\centering
\caption[]{Mean abundances of Fe, Na, Al, and $\alpha$ elements for our
           sample of classical Cepheids.}
\label{table_ab_alpha_mean}
{\scriptsize 
\begin{tabular}{rc r@{ }l r@{ }l r@{ }l r@{ }l r@{ }l r@{ }l r@{ }l r}
\noalign{\smallskip}\hline\hline\noalign{\smallskip}
Name &
\parbox[c]{0.6cm}{\centering \logP\ [days]} &
\multicolumn{2}{c}{\parbox[c]{0.4cm}{\centering \RG\ [pc]}} &
\multicolumn{2}{c}{[Fe/H]} &
\multicolumn{2}{c}{[Na/H]} &
\multicolumn{2}{c}{[Al/H]} &
\multicolumn{2}{c}{[Mg/H]} &
\multicolumn{2}{c}{[Si/H]} &
\multicolumn{2}{c}{[Ca/H]} & $N_{\rm S}$ \\
\noalign{\smallskip}\hline\noalign{\smallskip}
V340\,Ara  & 1.3183 &  4657 & $\pm$ 427 &    0.33 & $\pm$ 0.09 &    0.80  & $\pm$ 0.03 &    0.61 & $\pm$ 0.10 &    0.34 & $\pm$ 0.00 &    0.47 & $\pm$ 0.08 &	 0.24 & $\pm$ 0.07 &  6     \\ 
  QZ\,Nor  & 0.5782 &  6283 & $\pm$ 447 &    0.21 & $\pm$ 0.06 &    0.56  & $\pm$ 0.03 &    0.30 & $\pm$ 0.18 &    0.39 & $\pm$ 0.14 &    0.41 & $\pm$ 0.17 &	 0.20 & $\pm$ 0.15 &  2     \\ 
  SY\,Nor  & 1.1019 &  6286 & $\pm$ 446 &    0.23 & $\pm$ 0.07 &    0.61  & $\pm$ 0.13 &    0.41 & $\pm$ 0.12 &    0.37 & $\pm$ 0.05 &    0.37 & $\pm$ 0.15 &	 0.19 & $\pm$ 0.14 &  2     \\ 
  CK\,Pup  & 0.8703 & 13357 & $\pm$ 423 & $-$0.13 & $\pm$ 0.06 &    0.17  & $\pm$ 0.03 & $-$0.08 & $\pm$ 0.24 & $-$0.02 & $\pm$ 0.02 &    0.02 & $\pm$ 0.20 & $-$0.08 & $\pm$ 0.03 &  2     \\ 
  KQ\,Sco  & 1.4577 &  5948 & $\pm$ 451 &    0.52 & $\pm$ 0.08 &    0.70  & $\pm$ 0.02 &    0.48 & $\pm$ 0.01 & 	& \ldots     &    0.26 & $\pm$ 0.05 &	 0.18 & $\pm$ 0.04 &  5     \\ 
  RY\,Sco  & 1.3078 &  6663 & $\pm$ 453 &    0.01 & $\pm$ 0.06 &    0.36  & $\pm$ 0.06 &    0.17 & $\pm$ 0.06 &    0.08 & $\pm$ 0.05 &    0.15 & $\pm$ 0.12 & $-$0.05 & $\pm$ 0.12 &  5     \\ 
V500\,Sco  & 0.9693 &  6590 & $\pm$ 453 & $-$0.07 & $\pm$ 0.08 &    0.19  & $\pm$ 0.05 &    0.12 & $\pm$ 0.03 & $-$0.12 & $\pm$ 0.00 &    0.03 & $\pm$ 0.13 & $-$0.14 & $\pm$ 0.10 &  4     \\ 
  RU\,Sct  & 1.2945 &  6361 & $\pm$ 449 &    0.14 & $\pm$ 0.04 &    0.39  & $\pm$ 0.06 &    0.21 & $\pm$ 0.11 &    0.31 & $\pm$ 0.15 &    0.30 & $\pm$ 0.15 &	 0.17 & $\pm$ 0.23 &  2     \\ 
  UZ\,Sct  & 1.1686 &  5309 & $\pm$ 448 &    0.33 & $\pm$ 0.08 &    0.79  & $\pm$ 0.08 &    0.60 & $\pm$ 0.18 &    0.32 & $\pm$ 0.10 &    0.45 & $\pm$ 0.08 &	 0.22 & $\pm$ 0.09 &  6     \\ 
V367\,Sct  & 0.7989 &  6332 & $\pm$ 451 &    0.05 & $\pm$ 0.08 &    0.30  & $\pm$ 0.04 &    0.33 & $\pm$ 0.07 & $-$0.01 & $\pm$ 0.14 &    0.16 & $\pm$ 0.18 &	 0.01 & $\pm$ 0.12 &  4     \\ 
   Z\,Sct  & 1.1106 &  5733 & $\pm$ 445 &    0.12 & $\pm$ 0.09 &    0.71  & $\pm$ 0.05 &    0.51 & $\pm$ 0.05 &    0.14 & $\pm$ 0.08 &    0.33 & $\pm$ 0.09 &	 0.13 & $\pm$ 0.06 &  5     \\ 
  AV\,Sgr  & 1.1879 &  5980 & $\pm$ 455 &    0.35 & $\pm$ 0.17 &    0.87  & $\pm$ 0.01 &    0.50 & $\pm$ 0.03 & 	& \ldots     &    0.47 & $\pm$ 0.00 &	 0.18 & $\pm$ 0.03 &  5     \\ 
  VY\,Sgr  & 1.1322 &  5862 & $\pm$ 453 &    0.33 & $\pm$ 0.12 &    0.78  & $\pm$ 0.04 &    0.66 & $\pm$ 0.04 &    0.36 & $\pm$ 0.07 &    0.28 & $\pm$ 0.06 &	 0.18 & $\pm$ 0.07 &  4     \\ 
  WZ\,Sgr  & 1.3394 &  6326 & $\pm$ 453 &    0.28 & $\pm$ 0.08 &    0.59  & $\pm$ 0.07 &    0.57 & $\pm$ 0.10 &    0.02 & $\pm$ 0.00 &    0.30 & $\pm$ 0.11 &	 0.17 & $\pm$ 0.06 &  5     \\ 
  XX\,Sgr  & 0.8078 &  6706 & $\pm$ 453 & $-$0.01 & $\pm$ 0.06 &    0.29  & $\pm$ 0.11 &    0.24 & $\pm$ 0.04 & $-$0.01 & $\pm$ 0.01 &    0.11 & $\pm$ 0.13 &	 0.00 & $\pm$ 0.13 &  5     \\ 
\noalign{\smallskip}\hline\noalign{\smallskip}
   AS\,Aur & 0.5017 & 12244 & $\pm$ 469 &    0.00 & $\pm$ 0.08 &    0.13  & $\pm$ 0.00 &    0.12 & $\pm$ 0.40 &    0.12 & $\pm$ 0.18 &    0.13 & $\pm$ 0.25 & $-$0.02 & $\pm$ 0.17 &  1     \\
   KN\,Cen & 1.5321 &  6498 & $\pm$ 417 &    0.55 & $\pm$ 0.12 &    0.69  & $\pm$ 0.04 &    0.73 & $\pm$ 0.17 &    0.32 & $\pm$ 0.17 &    0.65 & $\pm$ 0.24 &	 0.32 & $\pm$ 0.26 &  1     \\
   MZ\,Cen & 1.0151 &  6501 & $\pm$ 391 &    0.27 & $\pm$ 0.10 &    0.59  & $\pm$ 0.00 &    0.36 & $\pm$ 0.18 &    0.21 & $\pm$ 0.36 &    0.33 & $\pm$ 0.19 &	 0.19 & $\pm$ 0.12 &  1     \\
   OO\,Cen & 1.1099 &  6025 & $\pm$ 389 &    0.20 & $\pm$ 0.06 &    0.55  & $\pm$ 0.00 &    0.24 & $\pm$ 0.19 &    0.30 & $\pm$ 0.10 &    0.40 & $\pm$ 0.23 &	 0.26 & $\pm$ 0.12 &  1     \\
   TX\,Cen & 1.2328 &  6070 & $\pm$ 419 &    0.44 & $\pm$ 0.12 &    0.73  & $\pm$ 0.00 &    0.73 & $\pm$ 0.16 &    0.37 & $\pm$ 0.16 &    0.53 & $\pm$ 0.27 &	 0.47 & $\pm$ 0.11 &  1     \\
 V339\,Cen & 0.9762 &  6917 & $\pm$ 446 &    0.06 & $\pm$ 0.03 &    0.53  & $\pm$ 0.23 &    0.26 & $\pm$ 0.11 &    0.24 & $\pm$ 0.19 &    0.19 & $\pm$ 0.18 &	 0.02 & $\pm$ 0.07 &  1     \\
   VW\,Cen & 1.1771 &  6417 & $\pm$ 405 &    0.41 & $\pm$ 0.08 &    0.79  & $\pm$ 0.00 &    0.92 & $\pm$ 0.00 & 	& \ldots     &    0.49 & $\pm$ 0.20 &	 0.22 & $\pm$ 0.01 &  1     \\
   AO\,CMa & 0.7646 & 10430 & $\pm$ 433 &    0.01 & $\pm$ 0.06 &    0.16  & $\pm$ 0.03 &    0.10 & $\pm$ 0.15 &    0.13 & $\pm$ 0.02 &    0.12 & $\pm$ 0.14 &	 0.08 & $\pm$ 0.10 &  1     \\
   RW\,CMa & 0.7581 & 10057 & $\pm$ 445 & $-$0.07 & $\pm$ 0.08 &    0.22  & $\pm$ 0.08 &	 & \ldots     & 	& \ldots     &    0.05 & $\pm$ 0.04 &	 0.03 & $\pm$ 0.08 &  1     \\
   SS\,CMa & 1.0921 &  9829 & $\pm$ 439 &    0.06 & $\pm$ 0.04 &    0.32  & $\pm$ 0.04 &    0.17 & $\pm$ 0.15 &    0.12 & $\pm$ 0.17 &    0.26 & $\pm$ 0.18 &	 0.22 & $\pm$ 0.14 &  1     \\
   TV\,CMa & 0.6693 &  9575 & $\pm$ 447 &    0.01 & $\pm$ 0.07 &    0.25  & $\pm$ 0.02 &    0.15 & $\pm$ 0.12 &    0.19 & $\pm$ 0.07 &    0.18 & $\pm$ 0.14 &	 0.08 & $\pm$ 0.09 &  1     \\
   TW\,CMa & 0.8448 &  9788 & $\pm$ 445 &    0.04 & $\pm$ 0.09 &	  & \ldots     &    0.17 & $\pm$ 0.23 & 	& \ldots     &    0.27 & $\pm$ 0.16 &	 0.27 & $\pm$ 0.31 &  1     \\
   AA\,Gem & 1.0532 & 11454 & $\pm$ 459 & $-$0.08 & $\pm$ 0.05 &    0.22  & $\pm$ 0.03 & $-$0.01 & $\pm$ 0.15 &    0.42 & $\pm$ 0.20 &         & \ldots     &	 0.10 & $\pm$ 0.22 &  1     \\
   AD\,Gem & 0.5784 & 10662 & $\pm$ 455 & $-$0.14 & $\pm$ 0.06 &    0.11  & $\pm$ 0.13 & $-$0.23 & $\pm$ 0.23 & $-$0.02 & $\pm$ 0.12 &    0.03 & $\pm$ 0.19 & $-$0.06 & $\pm$ 0.13 &  1     \\
   BB\,Gem & 0.3633 & 11199 & $\pm$ 460 & $-$0.09 & $\pm$ 0.04 &	  & \ldots     &	 & \ldots     & 	& \ldots     &         & \ldots     &	      & \ldots     & \ldots \\
   BW\,Gem & 0.3633 & 11302 & $\pm$ 463 & $-$0.22 & $\pm$ 0.09 & $-$0.02  & $\pm$ 0.00 & $-$0.15 & $\pm$ 0.19 & $-$0.07 & $\pm$ 0.04 & $-$0.13 & $\pm$ 0.14 & $-$0.12 & $\pm$ 0.12 &  1     \\
   DX\,Gem & 0.4966 & 11407 & $\pm$ 473 & $-$0.01 & $\pm$ 0.09 &    0.12  & $\pm$ 0.00 &    0.10 & $\pm$ 0.16 & $-$0.03 & $\pm$ 0.15 &    0.06 & $\pm$ 0.11 & $-$0.03 & $\pm$ 0.06 &  1     \\
   RZ\,Gem & 0.7427 &  9973 & $\pm$ 454 & $-$0.16 & $\pm$ 0.03 &    0.19  & $\pm$ 0.16 &	 & \ldots     &    0.03 & $\pm$ 0.00 &    0.07 & $\pm$ 0.15 & $-$0.12 & $\pm$ 0.06 &  1     \\
   BE\,Mon & 0.4322 &  9609 & $\pm$ 452 &    0.05 & $\pm$ 0.09 &    0.34  & $\pm$ 0.13 &    0.13 & $\pm$ 0.24 &    0.33 & $\pm$ 0.22 &    0.14 & $\pm$ 0.16 &	 0.08 & $\pm$ 0.12 &  1     \\
   CV\,Mon & 0.7307 &  9362 & $\pm$ 452 &    0.09 & $\pm$ 0.09 &    0.31  & $\pm$ 0.20 &    0.15 & $\pm$ 0.16 &    0.22 & $\pm$ 0.04 &    0.24 & $\pm$ 0.27 &	 0.18 & $\pm$ 0.24 &  1     \\
   FT\,Mon & 0.6843 & 14344 & $\pm$ 468 & $-$0.13 & $\pm$ 0.08 &    0.15  & $\pm$ 0.19 & $-$0.10 & $\pm$ 0.17 & $-$0.08 & $\pm$ 0.01 &         & \ldots     &	 0.02 & $\pm$ 0.01 &  1     \\
   SV\,Mon & 1.1828 & 10070 & $\pm$ 453 &    0.12 & $\pm$ 0.08 &    0.63  & $\pm$ 0.04 &    0.16 & $\pm$ 0.06 &    0.30 & $\pm$ 0.17 &    0.21 & $\pm$ 0.16 &	 0.24 & $\pm$ 0.20 &  1     \\
   TW\,Mon & 0.8511 & 13059 & $\pm$ 457 & $-$0.13 & $\pm$ 0.07 &    0.16  & $\pm$ 0.06 & $-$0.03 & $\pm$ 0.31 & $-$0.04 & $\pm$ 0.07 & $-$0.02 & $\pm$ 0.14 &	 0.02 & $\pm$ 0.12 &  1     \\
   TX\,Mon & 0.9396 & 11790 & $\pm$ 452 & $-$0.03 & $\pm$ 0.05 &    0.23  & $\pm$ 0.06 & $-$0.08 & $\pm$ 0.11 & $-$0.04 & $\pm$ 0.08 &    0.06 & $\pm$ 0.13 &	 0.04 & $\pm$ 0.10 &  1     \\
   TY\,Mon & 0.6045 & 11180 & $\pm$ 451 &    0.02 & $\pm$ 0.08 &    0.17  & $\pm$ 0.09 & $-$0.19 & $\pm$ 0.07 &    0.03 & $\pm$ 0.11 &    0.14 & $\pm$ 0.23 &	 0.12 & $\pm$ 0.15 &  1     \\
   TZ\,Mon & 0.8709 & 11183 & $\pm$ 451 & $-$0.02 & $\pm$ 0.07 &    0.16  & $\pm$ 0.01 &    0.02 & $\pm$ 0.08 &    0.11 & $\pm$ 0.05 &    0.03 & $\pm$ 0.12 &	 0.04 & $\pm$ 0.18 &  1     \\
 V465\,Mon & 0.4335 & 11037 & $\pm$ 450 & $-$0.07 & $\pm$ 0.07 &    0.26  & $\pm$ 0.08 &    0.06 & $\pm$ 0.17 & $-$0.08 & $\pm$ 0.17 &    0.12 & $\pm$ 0.18 &	 0.04 & $\pm$ 0.08 &  1     \\
 V495\,Mon & 0.6124 & 12098 & $\pm$ 453 & $-$0.13 & $\pm$ 0.07 &    0.04  & $\pm$ 0.08 & $-$0.06 & $\pm$ 0.30 & 	& \ldots     & $-$0.01 & $\pm$ 0.18 & $-$0.04 & $\pm$ 0.12 &  1     \\
 V508\,Mon & 0.6163 & 10714 & $\pm$ 452 & $-$0.04 & $\pm$ 0.10 &    0.12  & $\pm$ 0.01 &    0.04 & $\pm$ 0.28 &    0.01 & $\pm$ 0.05 &    0.04 & $\pm$ 0.20 &	 0.04 & $\pm$ 0.21 &  1     \\
 V510\,Mon & 0.8637 & 12550 & $\pm$ 456 & $-$0.16 & $\pm$ 0.06 & $-$0.05  & $\pm$ 0.03 & $-$0.16 & $\pm$ 0.15 & $-$0.04 & $\pm$ 0.06 & $-$0.09 & $\pm$ 0.15 & $-$0.15 & $\pm$ 0.03 &  1     \\
   XX\,Mon & 0.7369 & 11854 & $\pm$ 451 &    0.01 & $\pm$ 0.08 &    0.51  & $\pm$ 0.37 &    0.05 & $\pm$ 0.22 &    0.21 & $\pm$ 0.07 &    0.17 & $\pm$ 0.23 &	 0.23 & $\pm$ 0.06 &  1     \\
   GU\,Nor & 0.5382 &  6663 & $\pm$ 450 &    0.08 & $\pm$ 0.06 &    0.32  & $\pm$ 0.04 &    0.17 & $\pm$ 0.10 &    0.18 & $\pm$ 0.19 &    0.28 & $\pm$ 0.15 &	 0.09 & $\pm$ 0.04 &  1     \\
   IQ\,Nor & 0.9159 &  6691 & $\pm$ 448 &    0.22 & $\pm$ 0.07 &    0.57  & $\pm$ 0.07 &    0.39 & $\pm$ 0.13 &    0.36 & $\pm$ 0.13 &    0.38 & $\pm$ 0.16 &	 0.25 & $\pm$ 0.06 &  1     \\
   RS\,Nor & 0.7923 &  6385 & $\pm$ 449 &    0.18 & $\pm$ 0.08 &    0.57  & $\pm$ 0.07 &    0.39 & $\pm$ 0.13 &    0.36 & $\pm$ 0.13 &    0.38 & $\pm$ 0.16 &	 0.25 & $\pm$ 0.06 &  1     \\
   TW\,Nor & 1.0329 &  6160 & $\pm$ 447 &    0.27 & $\pm$ 0.10 &    0.58  & $\pm$ 0.11 &    0.24 & $\pm$ 0.07 &    0.58 & $\pm$ 0.60 &    0.25 & $\pm$ 0.10 &	 0.05 & $\pm$ 0.06 &  1     \\
 V340\,Nor & 1.0526 &  6483 & $\pm$ 449 &    0.07 & $\pm$ 0.07 &    0.40  & $\pm$ 0.00 &    0.37 & $\pm$ 0.00 &    0.12 & $\pm$ 0.11 &    0.30 & $\pm$ 0.21 &	 0.19 & $\pm$ 0.18 &  1     \\
   CS\,Ori & 0.5899 & 11701 & $\pm$ 458 & $-$0.25 & $\pm$ 0.06 &    0.11  & $\pm$ 0.06 & $-$0.27 & $\pm$ 0.14 & $-$0.25 & $\pm$ 0.12 & $-$0.10 & $\pm$ 0.13 & $-$0.16 & $\pm$ 0.04 &  1     \\
   GQ\,Ori & 0.9353 & 10129 & $\pm$ 453 &    0.20 & $\pm$ 0.08 &	  & \ldots     &	 & \ldots     & 	& \ldots     &         & \ldots     &	      & \ldots     & \ldots \\
   RS\,Ori & 0.8789 &  9470 & $\pm$ 453 &    0.11 & $\pm$ 0.09 &    0.12  & $\pm$ 0.16 &    0.11 & $\pm$ 0.22 &    0.41 & $\pm$ 0.17 &    0.30 & $\pm$ 0.20 &	 0.22 & $\pm$ 0.19 &  1     \\
   AQ\,Pup & 1.4786 &  9472 & $\pm$ 436 &    0.06 & $\pm$ 0.05 &    0.36  & $\pm$ 0.00 &	 & \ldots     &    0.06 & $\pm$ 0.04 &    0.25 & $\pm$ 0.22 &	 0.05 & $\pm$ 0.00 &  1     \\
   BC\,Pup & 0.5495 & 12763 & $\pm$ 426 & $-$0.31 & $\pm$ 0.07 &    0.20  & $\pm$ 0.05 & $-$0.07 & $\pm$ 0.22 & $-$0.14 & $\pm$ 0.21 & $-$0.13 & $\pm$ 0.17 & $-$0.18 & $\pm$ 0.06 &  1     \\
   BM\,Pup & 0.8572 &  9981 & $\pm$ 435 & $-$0.07 & $\pm$ 0.08 &    0.17  & $\pm$ 0.15 & $-$0.02 & $\pm$ 0.18 & 	& \ldots     &    0.04 & $\pm$ 0.18 & $-$0.01 & $\pm$ 0.03 &  1     \\
   BN\,Pup & 1.1359 &  9930 & $\pm$ 428 &    0.03 & $\pm$ 0.05 &    0.22  & $\pm$ 0.02 &    0.16 & $\pm$ 0.10 &    0.23 & $\pm$ 0.07 &    0.16 & $\pm$ 0.14 &	 0.09 & $\pm$ 0.17 &  1     \\
   HW\,Pup & 1.1289 & 13554 & $\pm$ 436 & $-$0.22 & $\pm$ 0.09 & $-$0.02  & $\pm$ 0.01 & $-$0.12 & $\pm$ 0.12 & $-$0.06 & $\pm$ 0.14 & $-$0.05 & $\pm$ 0.24 & $-$0.11 & $\pm$ 0.12 &  1     \\
   LS\,Pup & 1.1506 & 10610 & $\pm$ 423 & $-$0.12 & $\pm$ 0.11 &	  & \ldots     &    0.42 & $\pm$ 0.00 &    0.10 & $\pm$ 0.10 &    0.04 & $\pm$ 0.09 &	 0.09 & $\pm$ 0.00 &  1     \\
   VW\,Pup & 0.6320 & 10175 & $\pm$ 443 & $-$0.14 & $\pm$ 0.06 &    0.11  & $\pm$ 0.03 & $-$0.04 & $\pm$ 0.17 & $-$0.33 & $\pm$ 0.21 & $-$0.03 & $\pm$ 0.14 & $-$0.08 & $\pm$ 0.09 &  1     \\
   VZ\,Pup & 1.3649 & 10867 & $\pm$ 425 & $-$0.01 & $\pm$ 0.04 &	  & \ldots     &    0.30 & $\pm$ 0.39 &    0.01 & $\pm$ 0.16 &    0.25 & $\pm$ 0.16 &	 0.11 & $\pm$ 0.01 &  1     \\
   WW\,Pup & 0.7417 & 10382 & $\pm$ 436 &    0.13 & $\pm$ 0.16 & $-$0.30  & $\pm$ 0.11 & $-$0.38 & $\pm$ 0.24 & $-$0.34 & $\pm$ 0.29 & $-$0.36 & $\pm$ 0.22 & $-$0.70 & $\pm$ 0.04 &  1     \\
   WY\,Pup & 0.7202 & 10549 & $\pm$ 430 & $-$0.10 & $\pm$ 0.08 &    0.21  & $\pm$ 0.00 & $-$0.14 & $\pm$ 0.29 &    0.05 & $\pm$ 0.08 & $-$0.01 & $\pm$ 0.10 &	 0.01 & $\pm$ 0.05 &  1     \\
   WZ\,Pup & 0.7013 & 10123 & $\pm$ 437 & $-$0.07 & $\pm$ 0.06 &    0.09  & $\pm$ 0.01 & $-$0.11 & $\pm$ 0.07 &    0.14 & $\pm$ 0.09 &    0.07 & $\pm$ 0.17 &	 0.04 & $\pm$ 0.07 &  1     \\
    X\,Pup & 1.4143 &  9788 & $\pm$ 441 &    0.02 & $\pm$ 0.08 &    0.42  & $\pm$ 0.00 &    0.35 & $\pm$ 0.00 & $-$0.01 & $\pm$ 0.01 & $-$0.01 & $\pm$ 0.21 &	 0.09 & $\pm$ 0.02 &  1     \\
 V470\,Sco & 1.2112 &  6461 & $\pm$ 454 &    0.16 & $\pm$ 0.06 &    0.55  & $\pm$ 0.11 &    0.26 & $\pm$ 0.13 &    0.18 & $\pm$ 0.01 &    0.23 & $\pm$ 0.09 &	 0.08 & $\pm$ 0.10 &  1     \\
   EV\,Sct & 0.4901 &  6135 & $\pm$ 449 &    0.09 & $\pm$ 0.07 &    0.25  & $\pm$ 0.00 &    0.56 & $\pm$ 0.00 &    0.06 & $\pm$ 0.14 &    0.26 & $\pm$ 0.21 &	 0.10 & $\pm$ 0.18 &  1     \\
    X\,Sct & 0.6230 &  6464 & $\pm$ 452 &    0.12 & $\pm$ 0.09 &    0.41  & $\pm$ 0.25 &    0.34 & $\pm$ 0.26 &    0.12 & $\pm$ 0.12 &    0.32 & $\pm$ 0.27 &	 0.14 & $\pm$ 0.14 &  1     \\
   AA\,Ser & 1.2340 &  5572 & $\pm$ 437 &    0.38 & $\pm$ 0.20 &    0.98  & $\pm$ 0.00 &    0.50 & $\pm$ 0.00 &    0.28 & $\pm$ 0.00 &    0.47 & $\pm$ 0.10 &	 0.16 & $\pm$ 0.00 &  1     \\
   CR\,Ser & 0.7244 &  6510 & $\pm$ 452 &    0.12 & $\pm$ 0.08 &    0.61  & $\pm$ 0.29 &    0.29 & $\pm$ 0.19 &    0.32 & $\pm$ 0.22 &    0.28 & $\pm$ 0.21 &	 0.11 & $\pm$ 0.10 &  1     \\
   AY\,Sgr & 0.8175 &  6429 & $\pm$ 452 &    0.11 & $\pm$ 0.06 &    0.32  & $\pm$ 0.06 &    0.19 & $\pm$ 0.17 &    0.20 & $\pm$ 0.06 &    0.23 & $\pm$ 0.15 &	 0.11 & $\pm$ 0.10 &  1     \\
V1954\,Sgr & 0.7909 &  5687 & $\pm$ 456 &    0.24 & $\pm$ 0.10 &    0.62  & $\pm$ 0.06 &    0.27 & $\pm$ 0.14 &    0.17 & $\pm$ 0.29 &    0.47 & $\pm$ 0.22 &	 0.25 & $\pm$ 0.08 &  1     \\
 V773\,Sgr & 0.7596 &  6595 & $\pm$ 454 &    0.11 & $\pm$ 0.06 &    0.30  & $\pm$ 0.00 &    0.05 & $\pm$ 0.00 &    0.20 & $\pm$ 0.00 &    0.30 & $\pm$ 0.17 &	 0.14 & $\pm$ 0.02 &  1     \\
   EZ\,Vel & 1.5383 & 12119 & $\pm$ 358 & $-$0.17 & $\pm$ 0.15 &    0.17  & $\pm$ 0.28 &    0.01 & $\pm$ 0.18 & $-$0.03 & $\pm$ 0.28 &         & \ldots     & $-$0.08 & $\pm$ 0.16 &  1     \\
\hline
\end{tabular}}
\tablefoot{The weighted (in the case of iron) or the arithmetic (for the
other elements) mean abundances of the stars with multiple spectra
(Table~\ref{table_ab_alpha_spec}) are listed first. Columns 2 and 3 shows
the logarithmic of the pulsation period and the Galactocentric distance
(\RG), respectively. The $N_{\rm S}$ values indicate the number of spectra
available for each star.}
\end{table*}

\clearpage
\begin{table*}[p]
\centering
\caption[]{Abundance difference of stars in common among the current sample
           and other data sets.}
\label{table_diff_ab}
\begin{tabular}{cc r@{ }l c}
\noalign{\smallskip}\hline\hline\noalign{\smallskip}
\parbox[c]{1.6cm}{\centering Abundance ratio} &
Data sets$^1$ &
\multicolumn{2}{c}{\parbox[c]{1.6cm}{\centering Zero-point difference}} &
$N_{\rm Common}$ \\
\noalign{\smallskip}\hline\noalign{\smallskip}
 {[Fe/H]} & LII--G14  & $-$0.05 & $\pm$ 0.11 & 45 \\
 {[Fe/H]} & LIII--G14 &    0.03 & $\pm$ 0.08 & 33 \\
 {[Fe/H]} & LII--LEM  &    0.08 & $\pm$ 0.12 & 51 \\
 {[Fe/H]} & LIII--YON &    0.34 & $\pm$ 0.20 & 20 \\[0.1cm]
 {[Na/H]} & LII--TS   & $-$0.13 & $\pm$ 0.14 & 38 \\
 {[Na/H]} & LIII--TS  & $-$0.10 & $\pm$ 0.13 & 34 \\
 {[Na/H]} & LII--LEM  & $-$0.11 & $\pm$ 0.17 & 36 \\[0.1cm]
 {[Al/H]} & LII--TS   & $-$0.08 & $\pm$ 0.16 & 36 \\
 {[Al/H]} & LIII--TS  & $-$0.03 & $\pm$ 0.14 & 33 \\
 {[Al/H]} & LII--LEM  & $-$0.03 & $\pm$ 0.16 & 41 \\[0.1cm]
 {[Mg/H]} & LII--TS   & $-$0.23 & $\pm$ 0.24 & 26 \\
 {[Mg/H]} & LIII--TS  & $-$0.10 & $\pm$ 0.17 & 30 \\
 {[Mg/H]} & LII--LEM  & $-$0.27 & $\pm$ 0.24 & 35 \\
 {[Mg/H]} & LIII--YON &    0.08 & $\pm$ 0.15 & 16 \\[0.1cm]
 {[Si/H]} & LII--TS   & $-$0.11 & $\pm$ 0.12 & 41 \\
 {[Si/H]} & LIII--TS  & $-$0.06 & $\pm$ 0.11 & 33 \\
 {[Si/H]} & LII--LEM  & $-$0.06 & $\pm$ 0.11 & 55 \\
 {[Si/H]} & LIII--YON &    0.12 & $\pm$ 0.08 & 18 \\[0.1cm]
 {[Ca/H]} & LII--TS   & $-$0.11 & $\pm$ 0.17 & 42 \\
 {[Ca/H]} & LIII--TS  & $-$0.08 & $\pm$ 0.15 & 32 \\
 {[Ca/H]} & LII--LEM  & $-$0.06 & $\pm$ 0.17 & 54 \\
 {[Ca/H]} & LIII--YON &    0.14 & $\pm$ 0.11 & 19 \\
\hline\noalign{\smallskip}
\end{tabular}
\tablefoot{\tablefoottext{1}{G14: \citet{Genovalietal2014}; TS: this study;
LII: \citet{Lucketal2011}; LIII: \citet{LuckLambert2011};
LEM: \citet{Lemasleetal2013}; YON: \citet{Yongetal2006}.
The quoted errors represent the dispersion around the mean.}}
\end{table*}

\clearpage
\begin{table*}[p]
\centering
\caption[]{Slopes and zero-points of the abundance gradients as a function
           of the Galactocentric distance and of the pulsation period.}
\label{table_slopes}
{\footnotesize 
\begin{tabular}{c r@{ }l r@{ }l cc r@{ }l r@{ }l r@{ }l r@{ }l}
\noalign{\smallskip}\hline\hline\noalign{\smallskip}
\parbox[c]{1.6cm}{\centering Abundance ratio} &
\multicolumn{2}{c}{\parbox[c]{1.4cm}{\centering Slope$^a$}} &
\multicolumn{2}{c}{\parbox[c]{1.6cm}{\centering Zero-point [dex]}} &
\parbox[c]{0.8cm}{\centering $\sigma$ [dex]} &
$N$ &
\multicolumn{2}{c}{\parbox[c]{1.4cm}{\centering Slope$^a$ (TS)}} &
\multicolumn{2}{c}{\parbox[c]{1.7cm}{\centering Slope$^a$ (LEM)}} &
\multicolumn{2}{c}{\parbox[c]{1.5cm}{\centering Slope$^a$ (LII)}} &
\multicolumn{2}{c}{\parbox[c]{1.6cm}{\centering Slope$^a$ (LIII)}} \\
\noalign{\smallskip}\hline\noalign{\smallskip}
\multicolumn{15}{c}{as a function of \RG} \\
\noalign{\smallskip}\hline\noalign{\smallskip}
 {[Na/H]}  & $-$0.052 & $\pm$ 0.003 &	 0.79 & $\pm$ 0.03 & 0.14 & 428 & $-$0.072 & $\pm$ 0.007 & $-$0.066 & $\pm$ 0.015 & $-$0.044 & $\pm$ 0.004 & $-$0.047 & $\pm$ 0.003 \\
 {[Al/H]}  & $-$0.055 & $\pm$ 0.003 &	 0.64 & $\pm$ 0.03 & 0.13 & 426 & $-$0.073 & $\pm$ 0.007 & $-$0.046 & $\pm$ 0.013 & $-$0.053 & $\pm$ 0.004 & $-$0.049 & $\pm$ 0.003 \\
 {[Mg/H]}  & $-$0.045 & $\pm$ 0.004 &	 0.56 & $\pm$ 0.03 & 0.17 & 417 & $-$0.039 & $\pm$ 0.007 & $-$0.050 & $\pm$ 0.013 & $-$0.048 & $\pm$ 0.006 & $-$0.048 & $\pm$ 0.004 \\
 {[Si/H]}  & $-$0.049 & $\pm$ 0.002 &	 0.59 & $\pm$ 0.02 & 0.09 & 432 & $-$0.055 & $\pm$ 0.005 & $-$0.068 & $\pm$ 0.009 & $-$0.049 & $\pm$ 0.003 & $-$0.048 & $\pm$ 0.002 \\
 {[Ca/H]}  & $-$0.039 & $\pm$ 0.002 &	 0.42 & $\pm$ 0.02 & 0.11 & 434 & $-$0.029 & $\pm$ 0.005 & $-$0.044 & $\pm$ 0.012 & $-$0.040 & $\pm$ 0.004 & $-$0.041 & $\pm$ 0.003 \\[0.1cm]
 {[Na/Fe]} &    0.007 & $\pm$ 0.002 &	 0.23 & $\pm$ 0.02 & 0.11 & 428 & $-$0.015 & $\pm$ 0.005 & $-$0.026 & $\pm$ 0.011 &    0.011 & $\pm$ 0.003 &	0.015 & $\pm$ 0.002 \\
 {[Al/Fe]} &    0.001 & $\pm$ 0.002 &	 0.09 & $\pm$ 0.02 & 0.10 & 425 & $-$0.020 & $\pm$ 0.006 &    0.001 & $\pm$ 0.012 &    0.004 & $\pm$ 0.002 &	0.012 & $\pm$ 0.002 \\
 {[Mg/Fe]} &    0.013 & $\pm$ 0.003 &	 0.01 & $\pm$ 0.03 & 0.15 & 415 &    0.015 & $\pm$ 0.006 &    0.003 & $\pm$ 0.010 &    0.008 & $\pm$ 0.005 &	0.010 & $\pm$ 0.003 \\
 {[Si/Fe]} &    0.009 & $\pm$ 0.001 &	 0.04 & $\pm$ 0.01 & 0.06 & 432 &    0.000 & $\pm$ 0.003 & $-$0.011 & $\pm$ 0.007 &    0.006 & $\pm$ 0.002 &	0.012 & $\pm$ 0.001 \\
 {[Ca/Fe]} &    0.018 & $\pm$ 0.002 & $-$0.12 & $\pm$ 0.02 & 0.10 & 433 &    0.028 & $\pm$ 0.004 &    0.010 & $\pm$ 0.010 &    0.017 & $\pm$ 0.002 &	0.020 & $\pm$ 0.002 \\
\noalign{\smallskip}\hline\noalign{\smallskip}
\multicolumn{15}{c}{as a function of \logP} \\
\noalign{\smallskip}\hline\noalign{\smallskip}
 {[Na/H]}  & 0.20 & $\pm$ 0.03 &    0.16 & $\pm$ 0.02 & 0.17 & 427 & 0.47 & $\pm$ 0.09 & 0.23 & $\pm$ 0.08 & 0.18 & $\pm$ 0.03 & 0.10 & $\pm$ 0.03 \\
 {[Al/H]}  & 0.25 & $\pm$ 0.03 & $-$0.06 & $\pm$ 0.02 & 0.17 & 427 & 0.49 & $\pm$ 0.09 & 0.19 & $\pm$ 0.07 & 0.23 & $\pm$ 0.03 & 0.13 & $\pm$ 0.03 \\
 {[Mg/H]}  & 0.13 & $\pm$ 0.03 &    0.04 & $\pm$ 0.03 & 0.19 & 417 & 0.18 & $\pm$ 0.07 & 0.24 & $\pm$ 0.09 & 0.09 & $\pm$ 0.04 & 0.07 & $\pm$ 0.04 \\
 {[Si/H]}  & 0.16 & $\pm$ 0.02 &    0.02 & $\pm$ 0.02 & 0.14 & 435 & 0.28 & $\pm$ 0.06 & 0.17 & $\pm$ 0.07 & 0.14 & $\pm$ 0.03 & 0.06 & $\pm$ 0.02 \\
 {[Ca/H]}  & 0.09 & $\pm$ 0.02 &    0.00 & $\pm$ 0.02 & 0.13 & 432 & 0.16 & $\pm$ 0.05 & 0.05 & $\pm$ 0.07 & 0.05 & $\pm$ 0.03 & 0.07 & $\pm$ 0.03 \\[0.1cm]
 {[Al/Fe]} &    0.05 & $\pm$ 0.02 & 0.07 & $\pm$ 0.02 & 0.10 & 423 &	0.15 & $\pm$ 0.05 &    0.03 & $\pm$ 0.07 &    0.04 & $\pm$ 0.02 &    0.06 & $\pm$ 0.02 \\
 {[Si/Fe]} & $-$0.04 & $\pm$ 0.01 & 0.15 & $\pm$ 0.01 & 0.07 & 431 & $-$0.04 & $\pm$ 0.03 & $-$0.02 & $\pm$ 0.04 & $-$0.05 & $\pm$ 0.01 & $-$0.04 & $\pm$ 0.01 \\
 {[Ca/Fe]} & $-$0.11 & $\pm$ 0.02 & 0.13 & $\pm$ 0.02 & 0.11 & 436 & $-$0.15 & $\pm$ 0.04 & $-$0.10 & $\pm$ 0.05 & $-$0.04 & $\pm$ 0.02 & $-$0.10 & $\pm$ 0.02 \\
\hline
\end{tabular}
}
\tablefoot{\tablefoottext{a}{In units of \dexkpc\ if in function of \RG, and
dex per logarithmic day if in function of \logP}.
Columns from 2 to 5 shows the results for all the different samples fitted
together. We also list the standard deviation ($\sigma$) of the residuals
and the number of data points ($N$) used in the fit. The slopes using only
the stars of our sample (TS: this study) and of previous studies (LEM, LII,
and LIII) are shown for comparison.}
\end{table*}

\clearpage
\begin{table*}[p]
\centering
\caption[]{Galactic Cepheids for which the abundances of Na, Al, and
           $\alpha$-elements was available in the literature.}
\label{table_liter}
{\scriptsize 
\begin{tabular}{r r@{}l r@{}l c r@{}l r@{}l c r@{}l r@{}l c r@{}l r@{}l c r@{}l r@{}l c}
\noalign{\smallskip}\hline\hline\noalign{\smallskip}
Name &
\multicolumn{2}{c}{[Na/H]$_{\rm lit}$} & \multicolumn{2}{c}{[Na/H]} & Ref. &
\multicolumn{2}{c}{[Al/H]$_{\rm lit}$} & \multicolumn{2}{c}{[Al/H]} & Ref. &
\multicolumn{2}{c}{[Mg/H]$_{\rm lit}$} & \multicolumn{2}{c}{[Mg/H]} & Ref. &
\multicolumn{2}{c}{[Si/H]$_{\rm lit}$} & \multicolumn{2}{c}{[Si/H]} & Ref. &
\multicolumn{2}{c}{[Ca/H]$_{\rm lit}$} & \multicolumn{2}{c}{[Ca/H]} & Ref. \\
\noalign{\smallskip}\hline\noalign{\smallskip}
    T\,Ant & $-$0&.04	 &    0&.06    & LIII	& $-$0&.09    & $-$0&.06    & LIII   & $-$0&.20    & $-$0&.10    & LIII   & $-$0&.09 & $-$0&.03 & LIII & $-$0&.18 & $-$0&.10 & LIII \\
   BC\,Aql & $-$0&.26	 & $-$0&.16    & LIII	& $-$0&.48    & $-$0&.45    & LIII   & $-$0&.06    &	0&.04    & LIII   & $-$0&.10 & $-$0&.04 & LIII &    0&.01 &    0&.09 & LIII \\
   EV\,Aql &	0&.16	 &    0&.26    & LIII	&    0&.10    &    0&.13    & LIII   &    0&.03    &	0&.13    & LIII   &    0&.05 &    0&.11 & LIII & $-$0&.15 & $-$0&.07 & LIII \\
   FF\,Aql &	0&.23	 &    0&.36    &  LII	&    0&.12    &    0&.20    &  LII   & $-$0&.26    & $-$0&.03    &  LII   &    0&.05 &    0&.16 &  LII & $-$0&.01 &    0&.10 &  LII \\
   FM\,Aql &	0&.28	 &    0&.38    & LIII	&    0&.27    &    0&.30    & LIII   &    0&.28    &	0&.38    & LIII   &    0&.19 &    0&.25 & LIII &    0&.04 &    0&.12 & LIII \\
   FN\,Aql &	0&.19	 &    0&.29    & LIII	&    0&.01    &    0&.04    & LIII   & $-$0&.10    &	0&.00    & LIII   & $-$0&.03 &    0&.03 & LIII & $-$0&.13 & $-$0&.05 & LIII \\
   KL\,Aql &	0&.46	 &    0&.56    & LIII	&    0&.32    &    0&.35    & LIII   &    0&.28    &	0&.38    & LIII   &    0&.28 &    0&.34 & LIII &    0&.13 &    0&.21 & LIII \\
   SZ\,Aql &	0&.25	 &    0&.38    &  LII	&    0&.31    &    0&.39    &  LII   & $-$0&.08    &	0&.15    &  LII   &    0&.16 &    0&.27 &  LII &    0&.10 &    0&.21 &  LII \\
   TT\,Aql &	0&.37	 &    0&.47    & LIII	&    0&.24    &    0&.27    & LIII   &    0&.27    &	0&.37    & LIII   &    0&.24 &    0&.30 & LIII &    0&.02 &    0&.10 & LIII \\
    U\,Aql &	0&.32	 &    0&.42    & LIII	&    0&.27    &    0&.30    & LIII   &    0&.09    &	0&.19    & LIII   &    0&.16 &    0&.22 & LIII & $-$0&.01 &    0&.07 & LIII \\
V1162\,Aql &	0&.13	 &    0&.26    &  LII	&    0&.13    &    0&.21    &  LII   & $-$0&.19    &	0&.04    &  LII   &    0&.06 &    0&.17 &  LII & $-$0&.03 &    0&.08 &  LII \\
V1344\,Aql &	0&.21	 &    0&.31    & LIII	&    0&.21    &    0&.24    & LIII   &    0&.03    &	0&.13    & LIII   &    0&.12 &    0&.18 & LIII & $-$0&.04 &    0&.04 & LIII \\
V1359\,Aql & $-$0&.19	 & $-$0&.09    & LIII	&    0&.01    &    0&.04    & LIII   & $-$0&.03    &	0&.07    & LIII   &    0&.32 &    0&.38 & LIII & $-$0&.58 & $-$0&.50 & LIII \\
 V336\,Aql &	0&.26	 &    0&.36    & LIII	&    0&.25    &    0&.28    & LIII   &    0&.11    &	0&.21    & LIII   &    0&.18 &    0&.24 & LIII &    0&.04 &    0&.12 & LIII \\
 V493\,Aql &	0&.32	 &    0&.42    & LIII	&    0&.01    &    0&.04    & LIII   &    0&.14    &	0&.24    & LIII   &    0&.07 &    0&.13 & LIII &    0&.01 &    0&.09 & LIII \\
 V496\,Aql &	0&.24	 &    0&.37    &  LII	&    0&.10    &    0&.18    &  LII   & $-$0&.12    &	0&.11    &  LII   &    0&.11 &    0&.22 &  LII & $-$0&.03 &    0&.08 &  LII \\
 V526\,Aql &	0&.63	 &    0&.73    & LIII	&    0&.58    &    0&.61    & LIII   &    0&.22    &	0&.32    & LIII   &    0&.38 &    0&.44 & LIII &    0&.16 &    0&.24 & LIII \\
 V600\,Aql &	0&.30	 &    0&.43    &  LII	&    0&.15    &    0&.23    &  LII   &    0&.20    &	0&.43    &  LII   &    0&.08 &    0&.19 &  LII &    0&.06 &    0&.17 &  LII \\
 V733\,Aql &	0&.19	 &    0&.32    &  LII	&    0&.08    &    0&.16    &  LII   &    0&.22    &	0&.45    &  LII   &    0&.09 &    0&.20 &  LII & $-$0&.02 &    0&.09 &  LII \\
 V916\,Aql &	0&.55	 &    0&.65    & LIII	&    0&.63    &    0&.66    & LIII   &    0&.41    &	0&.51    & LIII   &    0&.39 &    0&.45 & LIII &    0&.22 &    0&.30 & LIII \\
$\eta$\,Aql &	0&.20	 &    0&.33    &  LII	&    0&.15    &    0&.23    &  LII   & $-$0&.05    &	0&.18    &  LII   &    0&.12 &    0&.23 &  LII & $-$0&.02 &    0&.09 &  LII \\
 V340\,Ara &	0&.80	 &    0&.80    &   TS	&    0&.61    &    0&.61    &   TS   &    0&.34    &	0&.34    &   TS   &    0&.47 &    0&.47 &   TS &    0&.24 &    0&.24 &   TS \\
   AN\,Aur &	0&.07	 &    0&.17    & LIII	& $-$0&.08    & $-$0&.05    & LIII   & $-$0&.11    & $-$0&.01    & LIII   & $-$0&.09 & $-$0&.03 & LIII & $-$0&.19 & $-$0&.11 & LIII \\
   AO\,Aur & $-$0&.10	 & $-$0&.08    &  LEM	& $-$0&.11    & $-$0&.07    &  LEM   & $-$0&.08    & $-$0&.12    &  LEM   & $-$0&.15 & $-$0&.10 &  LEM & $-$0&.18 & $-$0&.13 &  LEM \\
   AS\,Aur &	0&.13	 &    0&.13    &   TS	&    0&.12    &    0&.12    &   TS   &    0&.12    &	0&.12    &   TS   &    0&.13 &    0&.13 &   TS & $-$0&.02 & $-$0&.02 &   TS \\
   AX\,Aur &	0&.17	 &    0&.19    &  LEM	& $-$0&.09    & $-$0&.05    &  LEM   & $-$0&.01    & $-$0&.05    &  LEM   & $-$0&.02 &    0&.03 &  LEM & $-$0&.04 &    0&.01 &  LEM \\
   BK\,Aur &	0&.51	 &    0&.61    & LIII	& $-$0&.03    &    0&.01    &  LEM   &    0&.32    &	0&.28    &  LEM   &    0&.16 &    0&.21 &  LEM &    0&.05 &    0&.10 &  LEM \\
   CO\,Aur &	0&.19	 &    0&.29    & LIII	&    0&.00    &    0&.03    & LIII   & $-$0&.12    & $-$0&.02    & LIII   &    0&.07 &    0&.13 & LIII & $-$0&.01 &    0&.07 & LIII \\
   CY\,Aur &	0&.06	 &    0&.16    & LIII	& $-$0&.04    & $-$0&.01    & LIII   & $-$0&.16    & $-$0&.06    & LIII   & $-$0&.10 & $-$0&.04 & LIII & $-$0&.20 & $-$0&.12 & LIII \\
   ER\,Aur & $-$0&.01	 &    0&.09    & LIII	& $-$0&.17    & $-$0&.14    & LIII   & $-$0&.31    & $-$0&.21    & LIII   & $-$0&.18 & $-$0&.12 & LIII & $-$0&.29 & $-$0&.21 & LIII \\
   EW\,Aur & $-$0&.27	 & $-$0&.17    & LIII	& $-$0&.20    & $-$0&.17    & LIII   & $-$0&.40    & $-$0&.30    & LIII   & $-$0&.31 & $-$0&.25 & LIII & $-$0&.44 & $-$0&.36 & LIII \\
   FF\,Aur & $-$0&.52	 & $-$0&.42    & LIII	& $-$0&.64    & $-$0&.61    & LIII   & $-$0&.37    & $-$0&.27    & LIII   & $-$0&.33 & $-$0&.27 & LIII & $-$0&.60 & $-$0&.52 & LIII \\
   GT\,Aur &	0&.01	 &    0&.11    & LIII	& $-$0&.02    &    0&.01    & LIII   & $-$0&.04    &	0&.06    & LIII   &    0&.05 &    0&.11 & LIII &    0&.00 &    0&.08 & LIII \\
   GV\,Aur &	0&.04	 &    0&.14    & LIII	& $-$0&.03    & $-$0&.00    & LIII   & $-$0&.09    &	0&.01    & LIII   & $-$0&.05 &    0&.01 & LIII & $-$0&.09 & $-$0&.01 & LIII \\
   IN\,Aur & $-$0&.02	 &    0&.08    & LIII	& $-$0&.13    & $-$0&.10    & LIII   & $-$0&.18    & $-$0&.08    & LIII   & $-$0&.18 & $-$0&.12 & LIII & $-$0&.11 & $-$0&.03 & LIII \\
   RT\,Aur &	0&.37	 &    0&.47    & LIII	&    0&.14    &    0&.17    & LIII   &    0&.17    &	0&.27    & LIII   &    0&.19 &    0&.25 & LIII &    0&.09 &    0&.17 & LIII \\
   RX\,Aur &	0&.27	 &    0&.37    & LIII	&    0&.10    &    0&.13    & LIII   &    0&.22    &	0&.32    & LIII   &    0&.15 &    0&.21 & LIII &    0&.06 &    0&.14 & LIII \\
   SY\,Aur &	0&.35	 &    0&.37    &  LEM	&    0&.01    &    0&.05    &  LEM   &    0&.16    &	0&.12    &  LEM   &    0&.12 &    0&.17 &  LEM &    0&.06 &    0&.11 &  LEM \\
 V335\,Aur &	0&.05	 &    0&.15    & LIII	& $-$0&.12    & $-$0&.09    & LIII   & $-$0&.16    & $-$0&.06    & LIII   & $-$0&.08 & $-$0&.02 & LIII & $-$0&.22 & $-$0&.14 & LIII \\
 V637\,Aur & $-$0&.04	 &    0&.06    & LIII	& $-$0&.19    & $-$0&.16    & LIII   & $-$0&.18    & $-$0&.08    & LIII   & $-$0&.13 & $-$0&.07 & LIII & $-$0&.24 & $-$0&.16 & LIII \\
    Y\,Aur &	0&.27	 &    0&.29    &  LEM	& $-$0&.11    & $-$0&.07    &  LEM   & $-$0&.01    & $-$0&.05    &  LEM   & $-$0&.01 &    0&.04 &  LEM & $-$0&.17 & $-$0&.12 &  LEM \\
   YZ\,Aur &	0&.22	 &    0&.24    &  LEM	& $-$0&.14    & $-$0&.11    & LIII   & $-$0&.20    & $-$0&.24    &  LEM   & $-$0&.13 & $-$0&.08 &  LEM & $-$0&.14 & $-$0&.09 &  LEM \\
   AO\,CMa &	0&.16	 &    0&.16    &   TS	&    0&.10    &    0&.10    &   TS   &    0&.13    &	0&.13    &   TS   &    0&.12 &    0&.12 &   TS &    0&.08 &    0&.08 &   TS \\
   RW\,CMa &	0&.22	 &    0&.22    &   TS	&     &\ldots &     &\ldots & \ldots &     &\ldots &     &\ldots & \ldots &    0&.05 &    0&.05 &   TS &    0&.03 &    0&.03 &   TS \\
   RY\,CMa &	0&.12	 &    0&.14    &  LEM	&    0&.12    &    0&.16    &  LEM   &    0&.07    &	0&.03    &  LEM   &    0&.07 &    0&.12 &  LEM &    0&.02 &    0&.07 &  LEM \\
   RZ\,CMa &	0&.06	 &    0&.08    &  LEM	& $-$0&.12    & $-$0&.04    &  LII   &    0&.12    &	0&.08    &  LEM   &    0&.06 &    0&.17 &  LII &    0&.12 &    0&.17 &  LEM \\
   SS\,CMa &	0&.32	 &    0&.32    &   TS	&    0&.17    &    0&.17    &   TS   &    0&.12    &	0&.12    &   TS   &    0&.26 &    0&.26 &   TS &    0&.22 &    0&.22 &   TS \\
   TV\,CMa &	0&.25	 &    0&.25    &   TS	&    0&.15    &    0&.15    &   TS   &    0&.19    &	0&.19    &   TS   &    0&.18 &    0&.18 &   TS &    0&.08 &    0&.08 &   TS \\
   TW\,CMa & $-$0&.03	 &    0&.10    &  LII	&    0&.17    &    0&.17    &   TS   & $-$0&.23    & $-$0&.27    &  LEM   &    0&.27 &    0&.27 &   TS &    0&.27 &    0&.27 &   TS \\
   VZ\,CMa &	0&.16	 &    0&.29    &  LII	&     &\ldots &     &\ldots & \ldots &     &\ldots &     &\ldots & \ldots &    0&.00 &    0&.11 &  LII & $-$0&.06 &    0&.05 &  LII \\
   XZ\,CMa &	 &\ldots &     &\ldots & \ldots &     &\ldots &     &\ldots & \ldots &     &\ldots &     &\ldots & \ldots & $-$0&.20 & $-$0&.02 &  YON & $-$0&.43 & $-$0&.21 &  YON \\
   AB\,Cam &	0&.09	 &    0&.19    & LIII	& $-$0&.05    & $-$0&.02    & LIII   & $-$0&.11    & $-$0&.01    & LIII   & $-$0&.05 &    0&.01 & LIII & $-$0&.15 & $-$0&.07 & LIII \\
   AC\,Cam &	0&.18	 &    0&.28    & LIII	& $-$0&.01    &    0&.02    & LIII   & $-$0&.15    & $-$0&.05    & LIII   & $-$0&.02 &    0&.04 & LIII & $-$0&.03 &    0&.05 & LIII \\
   AD\,Cam &	0&.07	 &    0&.17    & LIII	& $-$0&.07    & $-$0&.04    & LIII   & $-$0&.06    &	0&.04    & LIII   & $-$0&.15 & $-$0&.09 & LIII & $-$0&.20 & $-$0&.12 & LIII \\
   AM\,Cam &	0&.08	 &    0&.18    & LIII	& $-$0&.10    & $-$0&.07    & LIII   & $-$0&.01    &	0&.09    & LIII   & $-$0&.02 &    0&.04 & LIII & $-$0&.01 &    0&.07 & LIII \\
   CK\,Cam &	0&.19	 &    0&.29    & LIII	&    0&.10    &    0&.13    & LIII   &    0&.15    &	0&.25    & LIII   &    0&.12 &    0&.18 & LIII &    0&.07 &    0&.15 & LIII \\
   LO\,Cam &	0&.12	 &    0&.22    & LIII	& $-$0&.14    & $-$0&.11    & LIII   & $-$0&.06    &	0&.04    & LIII   & $-$0&.07 & $-$0&.01 & LIII & $-$0&.19 & $-$0&.11 & LIII \\
   MN\,Cam &	0&.18	 &    0&.28    & LIII	&    0&.04    &    0&.07    & LIII   &    0&.01    &	0&.11    & LIII   &    0&.02 &    0&.08 & LIII & $-$0&.08 & $-$0&.00 & LIII \\
   MQ\,Cam &	0&.18	 &    0&.28    & LIII	& $-$0&.01    &    0&.02    & LIII   &    0&.00    &	0&.10    & LIII   & $-$0&.09 & $-$0&.03 & LIII & $-$0&.18 & $-$0&.10 & LIII \\
   OX\,Cam &	0&.02	 &    0&.12    & LIII	& $-$0&.06    & $-$0&.03    & LIII   & $-$0&.16    & $-$0&.06    & LIII   & $-$0&.07 & $-$0&.01 & LIII & $-$0&.26 & $-$0&.18 & LIII \\
   PV\,Cam & $-$0&.06	 &    0&.04    & LIII	& $-$0&.13    & $-$0&.10    & LIII   & $-$0&.19    & $-$0&.09    & LIII   & $-$0&.09 & $-$0&.03 & LIII & $-$0&.15 & $-$0&.07 & LIII \\
   QS\,Cam & $-$0&.07	 &    0&.03    & LIII	&    0&.07    &    0&.10    & LIII   &    0&.29    &	0&.39    & LIII   &    0&.07 &    0&.13 & LIII & $-$0&.07 &    0&.01 & LIII \\
   RW\,Cam &	0&.07	 &    0&.17    & LIII	&    0&.33    &    0&.36    & LIII   &    0&.13    &	0&.23    & LIII   &    0&.13 &    0&.19 & LIII & $-$0&.28 & $-$0&.20 & LIII \\
   RX\,Cam &	0&.22	 &    0&.32    & LIII	&    0&.16    &    0&.19    & LIII   &    0&.20    &	0&.30    & LIII   &    0&.11 &    0&.17 & LIII &    0&.03 &    0&.11 & LIII \\
   TV\,Cam &	0&.19	 &    0&.29    & LIII	& $-$0&.03    & $-$0&.00    & LIII   & $-$0&.07    &	0&.03    & LIII   &    0&.02 &    0&.08 & LIII & $-$0&.05 &    0&.03 & LIII \\
 V359\,Cam &	0&.08	 &    0&.18    & LIII	& $-$0&.12    & $-$0&.09    & LIII   & $-$0&.02    &	0&.08    & LIII   & $-$0&.07 & $-$0&.01 & LIII & $-$0&.07 &    0&.01 & LIII \\
   AQ\,Car &	0&.15	 &    0&.25    & LIII	&    0&.28    &    0&.32    &  LEM   &    0&.38    &	0&.34    &  LEM   &    0&.18 &    0&.23 &  LEM &    0&.36 &    0&.41 &  LEM \\
   CN\,Car &	0&.42	 &    0&.52    & LIII	&    0&.10    &    0&.13    & LIII   &    0&.18    &	0&.28    & LIII   &    0&.20 &    0&.26 & LIII &    0&.14 &    0&.22 & LIII \\
   CY\,Car &	0&.25	 &    0&.35    & LIII	&    0&.17    &    0&.20    & LIII   &    0&.08    &	0&.18    & LIII   &    0&.13 &    0&.19 & LIII &    0&.05 &    0&.13 & LIII \\
   DY\,Car &	0&.28	 &    0&.38    & LIII	&    0&.08    &    0&.11    & LIII   &    0&.11    &	0&.21    & LIII   &    0&.09 &    0&.15 & LIII & $-$0&.05 &    0&.03 & LIII \\
   ER\,Car &	0&.22	 &    0&.32    & LIII	&    0&.17    &    0&.20    & LIII   &    0&.21    &	0&.31    & LIII   &    0&.13 &    0&.19 & LIII &    0&.05 &    0&.13 & LIII \\
   FI\,Car &	0&.21	 &    0&.31    & LIII	&    0&.28    &    0&.31    & LIII   &    0&.24    &	0&.34    & LIII   &    0&.28 &    0&.34 & LIII &    0&.14 &    0&.22 & LIII \\
   FR\,Car &	0&.14	 &    0&.24    & LIII	&    0&.11    &    0&.14    & LIII   &    0&.00    &	0&.10    & LIII   &    0&.08 &    0&.14 & LIII & $-$0&.08 & $-$0&.00 & LIII \\
   GH\,Car &	0&.41	 &    0&.51    & LIII	&    0&.15    &    0&.18    & LIII   &    0&.23    &	0&.33    & LIII   &    0&.21 &    0&.27 & LIII &    0&.11 &    0&.19 & LIII \\
   GX\,Car &	0&.19	 &    0&.29    & LIII	&    0&.15    &    0&.18    & LIII   &    0&.33    &	0&.43    & LIII   &    0&.17 &    0&.23 & LIII &    0&.07 &    0&.15 & LIII \\
\hline\noalign{\smallskip}
\multicolumn{26}{r}{\it {\footnotesize continued on next page}} \\
\end{tabular}}
\tablefoot{The columns first give the original abundance estimate available
           in the literature and then the abundances rescaled according to
	   the zero-point differences listed in Table~\ref{table_diff_ab}.
	   The priority was given in the following order: we first adopt the
	   abundances provided by our group, this study (TS) and LEM, and
	   then those provided by the other studies, LIII, LII, and YON.}
\end{table*}
\addtocounter{table}{-1}
\begin{table*}[p]
\centering
\caption[]{continued.}
{\scriptsize 
\begin{tabular}{r r@{}l r@{}l c r@{}l r@{}l c r@{}l r@{}l c r@{}l r@{}l c r@{}l r@{}l c}
\noalign{\smallskip}\hline\hline\noalign{\smallskip}
Name &
\multicolumn{2}{c}{[Na/H]$_{\rm lit}$} & \multicolumn{2}{c}{[Na/H]} & Ref. &
\multicolumn{2}{c}{[Al/H]$_{\rm lit}$} & \multicolumn{2}{c}{[Al/H]} & Ref. &
\multicolumn{2}{c}{[Mg/H]$_{\rm lit}$} & \multicolumn{2}{c}{[Mg/H]} & Ref. &
\multicolumn{2}{c}{[Si/H]$_{\rm lit}$} & \multicolumn{2}{c}{[Si/H]} & Ref. &
\multicolumn{2}{c}{[Ca/H]$_{\rm lit}$} & \multicolumn{2}{c}{[Ca/H]} & Ref. \\
\noalign{\smallskip}\hline\noalign{\smallskip}
   HQ\,Car &	 &\ldots &     &\ldots & \ldots &     &\ldots &     &\ldots & \ldots &     &\ldots &     &\ldots & \ldots &    0&.03 &    0&.21 &  YON & $-$0&.45 & $-$0&.23 &  YON \\
   HW\,Car &	0&.24	 &    0&.34    & LIII	&    0&.09    &    0&.12    & LIII   & $-$0&.04    &	0&.06    & LIII   &    0&.05 &    0&.11 & LIII & $-$0&.08 & $-$0&.00 & LIII \\
   IO\,Car &	0&.15	 &    0&.25    & LIII	&    0&.05    &    0&.08    & LIII   & $-$0&.09    &	0&.01    & LIII   &    0&.04 &    0&.10 & LIII & $-$0&.17 & $-$0&.09 & LIII \\
   IT\,Car &	0&.36	 &    0&.46    & LIII	&    0&.25    &    0&.28    & LIII   &    0&.25    &	0&.35    & LIII   &    0&.19 &    0&.25 & LIII &    0&.13 &    0&.21 & LIII \\
    L\,Car & $-$0&.01	 &    0&.09    & LIII	&    0&.11    &    0&.15    &  LEM   &    0&.01    &	0&.11    & LIII   &    0&.20 &    0&.25 &  LEM &    0&.00 &    0&.05 &  LEM \\
   SX\,Car &	0&.29	 &    0&.39    & LIII	&    0&.02    &    0&.05    & LIII   &    0&.24    &	0&.34    & LIII   &    0&.14 &    0&.20 & LIII &    0&.01 &    0&.09 & LIII \\
    U\,Car &	0&.43	 &    0&.53    & LIII	&    0&.41    &    0&.44    & LIII   &    0&.22    &	0&.45    &  LII   &    0&.23 &    0&.29 & LIII &    0&.10 &    0&.18 & LIII \\
   UW\,Car &	0&.18	 &    0&.28    & LIII	&    0&.05    &    0&.08    & LIII   &    0&.26    &	0&.36    & LIII   &    0&.14 &    0&.20 & LIII &    0&.09 &    0&.17 & LIII \\
   UX\,Car &	0&.27	 &    0&.37    & LIII	&    0&.32    &    0&.36    &  LEM   & $-$0&.06    & $-$0&.10    &  LEM   &    0&.05 &    0&.10 &  LEM &    0&.14 &    0&.19 &  LEM \\
   UY\,Car &	0&.23	 &    0&.33    & LIII	&    0&.12    &    0&.15    & LIII   &    0&.46    &	0&.56    & LIII   &    0&.19 &    0&.25 & LIII &    0&.17 &    0&.25 & LIII \\
   UZ\,Car &	0&.20	 &    0&.30    & LIII	&    0&.17    &    0&.20    & LIII   &    0&.38    &	0&.48    & LIII   &    0&.21 &    0&.27 & LIII &    0&.19 &    0&.27 & LIII \\
    V\,Car &	0&.20	 &    0&.30    & LIII	&    0&.23    &    0&.27    &  LEM   &    0&.31    &	0&.27    &  LEM   &    0&.20 &    0&.25 &  LEM &    0&.31 &    0&.36 &  LEM \\
 V397\,Car &	0&.22	 &    0&.32    & LIII	&    0&.21    &    0&.25    &  LEM   &    0&.14    &	0&.24    & LIII   &    0&.20 &    0&.25 &  LEM &    0&.09 &    0&.14 &  LEM \\
   VY\,Car &	0&.29	 &    0&.31    &  LEM	&    0&.71    &    0&.74    & LIII   &    0&.14    &	0&.10    &  LEM   &    0&.49 &    0&.54 &  LEM &    0&.16 &    0&.21 &  LEM \\
   WW\,Car &	0&.20	 &    0&.30    & LIII	&    0&.12    &    0&.15    & LIII   &    0&.04    &	0&.14    & LIII   &    0&.05 &    0&.11 & LIII &    0&.00 &    0&.08 & LIII \\
   WZ\,Car &	0&.43	 &    0&.53    & LIII	&    0&.19    &    0&.22    & LIII   &    0&.45    &	0&.55    & LIII   &    0&.10 &    0&.16 & LIII &    0&.06 &    0&.14 & LIII \\
   XX\,Car &	0&.36	 &    0&.46    & LIII	&    0&.31    &    0&.34    & LIII   &    0&.17    &	0&.27    & LIII   &    0&.19 &    0&.25 & LIII &    0&.08 &    0&.16 & LIII \\
   XY\,Car &	0&.20	 &    0&.30    & LIII	&    0&.16    &    0&.19    & LIII   &    0&.11    &	0&.21    & LIII   &    0&.13 &    0&.19 & LIII &    0&.01 &    0&.09 & LIII \\
   XZ\,Car &	0&.44	 &    0&.54    & LIII	&    0&.29    &    0&.32    & LIII   &    0&.28    &	0&.38    & LIII   &    0&.27 &    0&.33 & LIII &    0&.22 &    0&.30 & LIII \\
   YZ\,Car &	0&.16	 &    0&.26    & LIII	&    0&.12    &    0&.15    & LIII   &    0&.15    &	0&.25    & LIII   &    0&.07 &    0&.13 & LIII & $-$0&.03 &    0&.05 & LIII \\
   AP\,Cas &	0&.24	 &    0&.34    & LIII	&    0&.11    &    0&.14    & LIII   & $-$0&.02    &	0&.08    & LIII   &    0&.05 &    0&.11 & LIII & $-$0&.06 &    0&.02 & LIII \\
   AS\,Cas &	0&.12	 &    0&.22    & LIII	& $-$0&.10    & $-$0&.07    & LIII   &    0&.38    &	0&.48    & LIII   &    0&.07 &    0&.13 & LIII &    0&.11 &    0&.19 & LIII \\
   AW\,Cas &	0&.06	 &    0&.16    & LIII	&    0&.02    &    0&.05    & LIII   &    0&.05    &	0&.15    & LIII   &    0&.03 &    0&.09 & LIII & $-$0&.04 &    0&.04 & LIII \\
   AY\,Cas &	0&.13	 &    0&.23    & LIII	&    0&.07    &    0&.10    & LIII   &    0&.09    &	0&.19    & LIII   &    0&.08 &    0&.14 & LIII &    0&.03 &    0&.11 & LIII \\
   BF\,Cas &	0&.11	 &    0&.21    & LIII	& $-$0&.04    & $-$0&.01    & LIII   & $-$0&.13    & $-$0&.03    & LIII   &    0&.10 &    0&.16 & LIII & $-$0&.02 &    0&.06 & LIII \\
   BP\,Cas &	0&.24	 &    0&.34    & LIII	&    0&.06    &    0&.09    & LIII   &    0&.03    &	0&.13    & LIII   &    0&.09 &    0&.15 & LIII &    0&.02 &    0&.10 & LIII \\
   BV\,Cas &	0&.09	 &    0&.19    & LIII	&    0&.12    &    0&.15    & LIII   &    0&.05    &	0&.15    & LIII   &    0&.06 &    0&.12 & LIII & $-$0&.02 &    0&.06 & LIII \\
   BY\,Cas &	0&.26	 &    0&.36    & LIII	&    0&.11    &    0&.14    & LIII   &    0&.18    &	0&.28    & LIII   &    0&.21 &    0&.27 & LIII &    0&.09 &    0&.17 & LIII \\
   CD\,Cas &	0&.29	 &    0&.39    & LIII	&    0&.11    &    0&.14    & LIII   &     &\ldots &     &\ldots & \ldots &    0&.21 &    0&.27 & LIII &    0&.06 &    0&.14 & LIII \\
   CF\,Cas &	0&.20	 &    0&.30    & LIII	&    0&.16    &    0&.19    & LIII   &    0&.10    &	0&.20    & LIII   &    0&.06 &    0&.12 & LIII &    0&.01 &    0&.09 & LIII \\
   CG\,Cas &	0&.28	 &    0&.38    & LIII	&    0&.03    &    0&.06    & LIII   &    0&.21    &	0&.31    & LIII   &    0&.13 &    0&.19 & LIII &    0&.04 &    0&.12 & LIII \\
   CH\,Cas &	0&.28	 &    0&.41    &  LII	&     &\ldots &     &\ldots & \ldots &     &\ldots &     &\ldots & \ldots &    0&.14 &    0&.25 &  LII & $-$0&.04 &    0&.07 &  LII \\
   CT\,Cas &	0&.26	 &    0&.36    & LIII	&    0&.04    &    0&.07    & LIII   & $-$0&.08    &	0&.02    & LIII   &    0&.02 &    0&.08 & LIII & $-$0&.01 &    0&.07 & LIII \\
   CY\,Cas &	 &\ldots &     &\ldots & \ldots &    0&.08    &    0&.16    &  LII   &     &\ldots &     &\ldots & \ldots &    0&.11 &    0&.22 &  LII & $-$0&.01 &    0&.10 &  LII \\
   CZ\,Cas &	0&.16	 &    0&.26    & LIII	&    0&.09    &    0&.12    & LIII   &    0&.09    &	0&.19    & LIII   &    0&.09 &    0&.15 & LIII &    0&.00 &    0&.08 & LIII \\
   DD\,Cas &	0&.30	 &    0&.43    &  LII	&    0&.10    &    0&.18    &  LII   &    0&.20    &	0&.43    &  LII   &    0&.10 &    0&.21 &  LII &    0&.00 &    0&.11 &  LII \\
   DF\,Cas &	0&.16	 &    0&.29    &  LII	&     &\ldots &     &\ldots & \ldots & $-$0&.33    & $-$0&.10    &  LII   &    0&.03 &    0&.14 &  LII & $-$0&.18 & $-$0&.07 &  LII \\
   DL\,Cas &	0&.11	 &    0&.24    &  LII	&    0&.14    &    0&.22    &  LII   &    0&.11    &	0&.34    &  LII   &    0&.02 &    0&.13 &  LII &    0&.01 &    0&.12 &  LII \\
   DW\,Cas &	0&.25	 &    0&.35    & LIII	&    0&.15    &    0&.18    & LIII   &    0&.23    &	0&.33    & LIII   &    0&.12 &    0&.18 & LIII &    0&.04 &    0&.12 & LIII \\
   EX\,Cas &	0&.28	 &    0&.38    & LIII	&    0&.13    &    0&.16    & LIII   &    0&.01    &	0&.11    & LIII   &    0&.07 &    0&.13 & LIII &    0&.15 &    0&.23 & LIII \\
   FM\,Cas &	0&.17	 &    0&.30    &  LII	&    0&.30    &    0&.38    &  LII   &    0&.20    &	0&.43    &  LII   &    0&.09 &    0&.20 &  LII &    0&.14 &    0&.25 &  LII \\
   FO\,Cas &	0&.07	 &    0&.17    & LIII	& $-$0&.12    & $-$0&.09    & LIII   & $-$0&.34    & $-$0&.24    & LIII   & $-$0&.27 & $-$0&.21 & LIII & $-$0&.46 & $-$0&.38 & LIII \\
   FW\,Cas & $-$0&.03	 &    0&.07    & LIII	&    0&.02    &    0&.05    & LIII   & $-$0&.16    & $-$0&.06    & LIII   & $-$0&.07 & $-$0&.01 & LIII & $-$0&.05 &    0&.03 & LIII \\
   GL\,Cas &	0&.32	 &    0&.42    & LIII	&    0&.19    &    0&.22    & LIII   &    0&.37    &	0&.47    & LIII   &    0&.07 &    0&.13 & LIII &    0&.15 &    0&.23 & LIII \\
   GM\,Cas &	0&.15	 &    0&.25    & LIII	& $-$0&.18    & $-$0&.15    & LIII   & $-$0&.18    & $-$0&.08    & LIII   & $-$0&.10 & $-$0&.04 & LIII & $-$0&.10 & $-$0&.02 & LIII \\
   GO\,Cas &	0&.33	 &    0&.43    & LIII	&    0&.09    &    0&.12    & LIII   &    0&.01    &	0&.11    & LIII   &    0&.19 &    0&.25 & LIII &    0&.13 &    0&.21 & LIII \\
   HK\,Cas &	0&.81	 &    0&.91    & LIII	&    0&.72    &    0&.75    & LIII   &    0&.59    &	0&.69    & LIII   &    0&.52 &    0&.58 & LIII &    0&.33 &    0&.41 & LIII \\
   IO\,Cas &	0&.09	 &    0&.19    & LIII	& $-$0&.05    & $-$0&.02    & LIII   & $-$0&.12    & $-$0&.02    & LIII   & $-$0&.17 & $-$0&.11 & LIII & $-$0&.24 & $-$0&.16 & LIII \\
   KK\,Cas &	0&.34	 &    0&.44    & LIII	&    0&.17    &    0&.20    & LIII   &    0&.04    &	0&.14    & LIII   &    0&.15 &    0&.21 & LIII &    0&.14 &    0&.22 & LIII \\
   LT\,Cas & $-$0&.09	 &    0&.01    & LIII	& $-$0&.23    & $-$0&.20    & LIII   & $-$0&.29    & $-$0&.19    & LIII   & $-$0&.28 & $-$0&.22 & LIII & $-$0&.35 & $-$0&.27 & LIII \\
   NP\,Cas &	0&.07	 &    0&.17    & LIII	&    0&.01    &    0&.04    & LIII   &    0&.01    &	0&.11    & LIII   &    0&.07 &    0&.13 & LIII &    0&.04 &    0&.12 & LIII \\
   NY\,Cas & $-$0&.12	 & $-$0&.02    & LIII	&    0&.09    &    0&.12    & LIII   & $-$0&.40    & $-$0&.30    & LIII   & $-$0&.21 & $-$0&.15 & LIII & $-$0&.40 & $-$0&.32 & LIII \\
   OP\,Cas &	0&.49	 &    0&.59    & LIII	&    0&.27    &    0&.30    & LIII   &    0&.14    &	0&.24    & LIII   &    0&.19 &    0&.25 & LIII &    0&.15 &    0&.23 & LIII \\
   OZ\,Cas &	0&.07	 &    0&.17    & LIII	&    0&.03    &    0&.06    & LIII   & $-$0&.07    &	0&.03    & LIII   &    0&.13 &    0&.19 & LIII &    0&.07 &    0&.15 & LIII \\
   PW\,Cas &	0&.12	 &    0&.22    & LIII	& $-$0&.06    & $-$0&.03    & LIII   & $-$0&.04    &	0&.06    & LIII   & $-$0&.03 &    0&.03 & LIII & $-$0&.14 & $-$0&.06 & LIII \\
   RS\,Cas &	0&.36	 &    0&.46    & LIII	&    0&.10    &    0&.13    & LIII   &    0&.11    &	0&.21    & LIII   &    0&.19 &    0&.25 & LIII &    0&.13 &    0&.21 & LIII \\
   RW\,Cas &	 &\ldots &     &\ldots & \ldots &    0&.05    &    0&.13    &  LII   &     &\ldots &     &\ldots & \ldots &    0&.13 &    0&.24 &  LII &    0&.28 &    0&.39 &  LII \\
   RY\,Cas &	0&.26	 &    0&.39    &  LII	&    0&.15    &    0&.23    &  LII   &     &\ldots &     &\ldots & \ldots &    0&.22 &    0&.33 &  LII & $-$0&.07 &    0&.04 &  LII \\
   SU\,Cas &	0&.26	 &    0&.39    &  LII	&    0&.10    &    0&.18    &  LII   & $-$0&.22    &	0&.01    &  LII   &    0&.09 &    0&.20 &  LII &    0&.07 &    0&.18 &  LII \\
   SW\,Cas &	0&.26	 &    0&.39    &  LII	&    0&.22    &    0&.30    &  LII   &     &\ldots &     &\ldots & \ldots &    0&.12 &    0&.23 &  LII &    0&.04 &    0&.15 &  LII \\
   SY\,Cas &	0&.16	 &    0&.29    &  LII	&    0&.15    &    0&.23    &  LII   &    0&.32    &	0&.55    &  LII   &    0&.07 &    0&.18 &  LII &    0&.10 &    0&.21 &  LII \\
   SZ\,Cas &	0&.34	 &    0&.44    & LIII	&    0&.08    &    0&.11    & LIII   &    0&.20    &	0&.30    & LIII   &    0&.14 &    0&.20 & LIII &    0&.07 &    0&.15 & LIII \\
   TU\,Cas &	0&.15	 &    0&.28    &  LII	&    0&.14    &    0&.22    &  LII   & $-$0&.19    &	0&.04    &  LII   &    0&.10 &    0&.21 &  LII & $-$0&.02 &    0&.09 &  LII \\
   UZ\,Cas &	0&.22	 &    0&.32    & LIII	&    0&.01    &    0&.04    & LIII   & $-$0&.23    & $-$0&.13    & LIII   & $-$0&.02 &    0&.04 & LIII & $-$0&.03 &    0&.05 & LIII \\
V1017\,Cas &	0&.05	 &    0&.15    & LIII	& $-$0&.14    & $-$0&.11    & LIII   & $-$0&.21    & $-$0&.11    & LIII   & $-$0&.16 & $-$0&.10 & LIII & $-$0&.25 & $-$0&.17 & LIII \\
V1019\,Cas &	0&.22	 &    0&.32    & LIII	&    0&.08    &    0&.11    & LIII   &    0&.12    &	0&.22    & LIII   &    0&.11 &    0&.17 & LIII &    0&.08 &    0&.16 & LIII \\
V1020\,Cas &	0&.29	 &    0&.39    & LIII	&    0&.25    &    0&.28    & LIII   &    0&.20    &	0&.30    & LIII   &    0&.17 &    0&.23 & LIII &    0&.09 &    0&.17 & LIII \\
V1100\,Cas &	0&.11	 &    0&.21    & LIII	& $-$0&.28    & $-$0&.25    & LIII   &    0&.04    &	0&.14    & LIII   & $-$0&.06 & $-$0&.00 & LIII & $-$0&.16 & $-$0&.08 & LIII \\
V1154\,Cas &	0&.10	 &    0&.20    & LIII	& $-$0&.07    & $-$0&.04    & LIII   &    0&.10    &	0&.20    & LIII   & $-$0&.09 & $-$0&.03 & LIII & $-$0&.27 & $-$0&.19 & LIII \\
V1206\,Cas &	0&.10	 &    0&.20    & LIII	&    0&.03    &    0&.06    & LIII   &    0&.08    &	0&.18    & LIII   & $-$0&.01 &    0&.05 & LIII & $-$0&.15 & $-$0&.07 & LIII \\
 V342\,Cas &	0&.17	 &    0&.27    & LIII	&    0&.09    &    0&.12    & LIII   & $-$0&.08    &	0&.02    & LIII   &    0&.08 &    0&.14 & LIII &    0&.04 &    0&.12 & LIII \\
 V379\,Cas &	0&.21	 &    0&.34    &  LII	&    0&.23    &    0&.31    &  LII   &     &\ldots &     &\ldots & \ldots &    0&.19 &    0&.30 &  LII &    0&.18 &    0&.29 &  LII \\
 V395\,Cas &	0&.28	 &    0&.38    & LIII	&    0&.16    &    0&.19    & LIII   &    0&.07    &	0&.17    & LIII   &    0&.11 &    0&.17 & LIII &    0&.01 &    0&.09 & LIII \\
 V407\,Cas &	0&.36	 &    0&.46    & LIII	&    0&.20    &    0&.23    & LIII   & $-$0&.01    &	0&.09    & LIII   &    0&.11 &    0&.17 & LIII &    0&.06 &    0&.14 & LIII \\
 V556\,Cas &	0&.24	 &    0&.34    & LIII	&    0&.06    &    0&.09    & LIII   &    0&.05    &	0&.15    & LIII   &    0&.02 &    0&.08 & LIII & $-$0&.05 &    0&.03 & LIII \\
\hline\noalign{\smallskip}
\multicolumn{26}{r}{\it {\footnotesize continued on next page}} \\
\end{tabular}}
\end{table*}
\addtocounter{table}{-1}
\begin{table*}[p]
\centering
\caption[]{continued.}
{\scriptsize 
\begin{tabular}{r r@{}l r@{}l c r@{}l r@{}l c r@{}l r@{}l c r@{}l r@{}l c r@{}l r@{}l c}
\noalign{\smallskip}\hline\hline\noalign{\smallskip}
Name &
\multicolumn{2}{c}{[Na/H]$_{\rm lit}$} & \multicolumn{2}{c}{[Na/H]} & Ref. &
\multicolumn{2}{c}{[Al/H]$_{\rm lit}$} & \multicolumn{2}{c}{[Al/H]} & Ref. &
\multicolumn{2}{c}{[Mg/H]$_{\rm lit}$} & \multicolumn{2}{c}{[Mg/H]} & Ref. &
\multicolumn{2}{c}{[Si/H]$_{\rm lit}$} & \multicolumn{2}{c}{[Si/H]} & Ref. &
\multicolumn{2}{c}{[Ca/H]$_{\rm lit}$} & \multicolumn{2}{c}{[Ca/H]} & Ref. \\
\noalign{\smallskip}\hline\noalign{\smallskip}
 V636\,Cas &	0&.28	 &    0&.41    &  LII	&    0&.12    &    0&.20    &  LII   & $-$0&.14    &	0&.09    &  LII   &    0&.06 &    0&.17 &  LII &    0&.06 &    0&.17 &  LII \\
   VV\,Cas &	0&.15	 &    0&.25    & LIII	&    0&.00    &    0&.03    & LIII   & $-$0&.15    & $-$0&.05    & LIII   &    0&.04 &    0&.10 & LIII & $-$0&.09 & $-$0&.01 & LIII \\
   VW\,Cas &	0&.20	 &    0&.30    & LIII	&    0&.21    &    0&.24    & LIII   &    0&.24    &	0&.34    & LIII   &    0&.14 &    0&.20 & LIII &    0&.10 &    0&.18 & LIII \\
   XY\,Cas &	0&.37	 &    0&.47    & LIII	&    0&.12    &    0&.15    & LIII   &    0&.16    &	0&.26    & LIII   &    0&.14 &    0&.20 & LIII &    0&.10 &    0&.18 & LIII \\
   AY\,Cen &	0&.26	 &    0&.36    & LIII	&    0&.12    &    0&.15    & LIII   & $-$0&.01    &	0&.09    & LIII   &    0&.08 &    0&.14 & LIII &    0&.00 &    0&.08 & LIII \\
   AZ\,Cen &	0&.24	 &    0&.34    & LIII	&    0&.03    &    0&.06    & LIII   &    0&.02    &	0&.12    & LIII   &    0&.10 &    0&.16 & LIII & $-$0&.03 &    0&.05 & LIII \\
   BB\,Cen &	0&.30	 &    0&.40    & LIII	&    0&.18    &    0&.21    & LIII   &    0&.11    &	0&.21    & LIII   &    0&.19 &    0&.25 & LIII &    0&.01 &    0&.09 & LIII \\
   KK\,Cen &	0&.52	 &    0&.62    & LIII	&    0&.27    &    0&.30    & LIII   &    0&.18    &	0&.28    & LIII   &    0&.24 &    0&.30 & LIII &    0&.16 &    0&.24 & LIII \\
   KN\,Cen &	0&.69	 &    0&.69    &   TS	&    0&.73    &    0&.73    &   TS   &    0&.32    &	0&.32    &   TS   &    0&.65 &    0&.65 &   TS &    0&.32 &    0&.32 &   TS \\
   MZ\,Cen &	0&.59	 &    0&.59    &   TS	&    0&.36    &    0&.36    &   TS   &    0&.21    &	0&.21    &   TS   &    0&.33 &    0&.33 &   TS &    0&.19 &    0&.19 &   TS \\
   OO\,Cen &	0&.55	 &    0&.55    &   TS	&    0&.24    &    0&.24    &   TS   &    0&.30    &	0&.30    &   TS   &    0&.40 &    0&.40 &   TS &    0&.26 &    0&.26 &   TS \\
   QY\,Cen &	0&.41	 &    0&.51    & LIII	&    0&.21    &    0&.24    & LIII   & $-$0&.17    & $-$0&.07    & LIII   &    0&.24 &    0&.30 & LIII &    0&.11 &    0&.19 & LIII \\
   TX\,Cen &	0&.73	 &    0&.73    &   TS	&    0&.73    &    0&.73    &   TS   &    0&.37    &	0&.37    &   TS   &    0&.53 &    0&.53 &   TS &    0&.47 &    0&.47 &   TS \\
    V\,Cen &	0&.28	 &    0&.38    & LIII	&    0&.10    &    0&.13    & LIII   &    0&.02    &	0&.12    & LIII   &    0&.09 &    0&.15 & LIII & $-$0&.07 &    0&.01 & LIII \\
 V339\,Cen &	0&.53	 &    0&.53    &   TS	&    0&.26    &    0&.26    &   TS   &    0&.24    &	0&.24    &   TS   &    0&.19 &    0&.19 &   TS &    0&.02 &    0&.02 &   TS \\
 V378\,Cen &	0&.25	 &    0&.35    & LIII	&    0&.06    &    0&.09    & LIII   & $-$0&.05    &	0&.05    & LIII   &    0&.07 &    0&.13 & LIII & $-$0&.09 & $-$0&.01 & LIII \\
 V381\,Cen &	0&.20	 &    0&.30    & LIII	&    0&.08    &    0&.11    & LIII   &    0&.00    &	0&.10    & LIII   &    0&.08 &    0&.14 & LIII & $-$0&.03 &    0&.05 & LIII \\
 V419\,Cen &	0&.36	 &    0&.46    & LIII	&    0&.33    &    0&.36    & LIII   &    0&.42    &	0&.52    & LIII   &    0&.20 &    0&.26 & LIII &    0&.16 &    0&.24 & LIII \\
 V496\,Cen &	0&.36	 &    0&.46    & LIII	&    0&.13    &    0&.16    & LIII   &    0&.12    &	0&.22    & LIII   &    0&.14 &    0&.20 & LIII &    0&.06 &    0&.14 & LIII \\
 V659\,Cen &	0&.19	 &    0&.29    & LIII	&    0&.14    &    0&.17    & LIII   &    0&.00    &	0&.10    & LIII   &    0&.09 &    0&.15 & LIII & $-$0&.05 &    0&.03 & LIII \\
 V737\,Cen &	0&.25	 &    0&.35    & LIII	&    0&.16    &    0&.19    & LIII   &    0&.01    &	0&.11    & LIII   &    0&.11 &    0&.17 & LIII & $-$0&.04 &    0&.04 & LIII \\
   VW\,Cen &	0&.79	 &    0&.79    &   TS	&    0&.92    &    0&.92    &   TS   &     &\ldots &     &\ldots & \ldots &    0&.49 &    0&.49 &   TS &    0&.22 &    0&.22 &   TS \\
   XX\,Cen &	0&.24	 &    0&.34    & LIII	&    0&.21    &    0&.24    & LIII   &    0&.10    &	0&.20    & LIII   &    0&.17 &    0&.23 & LIII &    0&.01 &    0&.09 & LIII \\
   AK\,Cep &	0&.08	 &    0&.18    & LIII	&    0&.08    &    0&.11    & LIII   &    0&.21    &	0&.31    & LIII   &    0&.12 &    0&.18 & LIII &    0&.03 &    0&.11 & LIII \\
   CN\,Cep &	0&.29	 &    0&.39    & LIII	& $-$0&.15    & $-$0&.12    & LIII   &    0&.12    &	0&.22    & LIII   &    0&.06 &    0&.12 & LIII & $-$0&.03 &    0&.05 & LIII \\
   CP\,Cep &	0&.13	 &    0&.26    &  LII	&    0&.07    &    0&.15    &  LII   &    0&.10    &	0&.33    &  LII   &    0&.10 &    0&.21 &  LII &    0&.14 &    0&.25 &  LII \\
   CR\,Cep &	0&.09	 &    0&.22    &  LII	&    0&.19    &    0&.27    &  LII   &    0&.39    &	0&.62    &  LII   &    0&.02 &    0&.13 &  LII &    0&.02 &    0&.13 &  LII \\
   DR\,Cep &	0&.02	 &    0&.12    & LIII	& $-$0&.06    & $-$0&.03    & LIII   & $-$0&.36    & $-$0&.26    & LIII   & $-$0&.10 & $-$0&.04 & LIII & $-$0&.47 & $-$0&.39 & LIII \\
   IR\,Cep &	0&.36	 &    0&.49    &  LII	&    0&.24    &    0&.32    &  LII   & $-$0&.08    &	0&.15    &  LII   &    0&.19 &    0&.30 &  LII &    0&.08 &    0&.19 &  LII \\
   IY\,Cep &	0&.19	 &    0&.29    & LIII	&    0&.14    &    0&.17    & LIII   &    0&.01    &	0&.11    & LIII   &    0&.07 &    0&.13 & LIII & $-$0&.02 &    0&.06 & LIII \\
   MU\,Cep &	0&.20	 &    0&.30    & LIII	& $-$0&.03    & $-$0&.00    & LIII   &    0&.46    &	0&.56    & LIII   &    0&.11 &    0&.17 & LIII & $-$0&.01 &    0&.07 & LIII \\
 V901\,Cep &	0&.12	 &    0&.22    & LIII	& $-$0&.03    & $-$0&.00    & LIII   &    0&.16    &	0&.26    & LIII   &    0&.07 &    0&.13 & LIII &    0&.04 &    0&.12 & LIII \\
 V911\,Cep &	0&.35	 &    0&.45    & LIII	& $-$0&.11    & $-$0&.08    & LIII   &    0&.24    &	0&.34    & LIII   &    0&.09 &    0&.15 & LIII &    0&.02 &    0&.10 & LIII \\
$\delta$\,Cep &	0&.28	 &    0&.38    & LIII	&    0&.19    &    0&.22    & LIII   &    0&.10    &	0&.20    & LIII   &    0&.14 &    0&.20 & LIII &    0&.01 &    0&.09 & LIII \\
   AV\,Cir &	0&.21	 &    0&.31    & LIII	&    0&.13    &    0&.16    & LIII   &    0&.09    &	0&.19    & LIII   &    0&.15 &    0&.21 & LIII & $-$0&.01 &    0&.07 & LIII \\
   AX\,Cir &	0&.17	 &    0&.27    & LIII	&    0&.03    &    0&.06    & LIII   & $-$0&.08    &	0&.02    & LIII   &    0&.01 &    0&.07 & LIII & $-$0&.12 & $-$0&.04 & LIII \\
   BP\,Cir &	0&.12	 &    0&.22    & LIII	& $-$0&.05    & $-$0&.02    & LIII   &    0&.11    &	0&.21    & LIII   &    0&.10 &    0&.16 & LIII & $-$0&.06 &    0&.02 & LIII \\
   AD\,Cru &	0&.31	 &    0&.41    & LIII	&    0&.07    &    0&.10    & LIII   &    0&.24    &	0&.34    & LIII   &    0&.14 &    0&.20 & LIII &    0&.09 &    0&.17 & LIII \\
   AG\,Cru &	0&.31	 &    0&.41    & LIII	& $-$0&.03    & $-$0&.00    & LIII   &    0&.14    &	0&.24    & LIII   &    0&.14 &    0&.20 & LIII &    0&.04 &    0&.12 & LIII \\
   BG\,Cru & $-$0&.13	 & $-$0&.03    & LIII	&    0&.13    &    0&.16    & LIII   &    0&.30    &	0&.40    & LIII   & $-$0&.07 & $-$0&.01 & LIII & $-$0&.16 & $-$0&.08 & LIII \\
    R\,Cru &	0&.23	 &    0&.33    & LIII	&    0&.14    &    0&.17    & LIII   &    0&.09    &	0&.19    & LIII   &    0&.14 &    0&.20 & LIII &    0&.05 &    0&.13 & LIII \\
    S\,Cru &	0&.23	 &    0&.33    & LIII	&    0&.10    &    0&.13    & LIII   &    0&.15    &	0&.25    & LIII   &    0&.16 &    0&.22 & LIII &    0&.07 &    0&.15 & LIII \\
    T\,Cru &	0&.22	 &    0&.32    & LIII	&    0&.17    &    0&.20    & LIII   &    0&.02    &	0&.12    & LIII   &    0&.11 &    0&.17 & LIII & $-$0&.01 &    0&.07 & LIII \\
   VW\,Cru &	0&.29	 &    0&.39    & LIII	&    0&.18    &    0&.21    & LIII   &    0&.07    &	0&.17    & LIII   &    0&.15 &    0&.21 & LIII &    0&.01 &    0&.09 & LIII \\
    X\,Cru &	0&.36	 &    0&.46    & LIII	&    0&.22    &    0&.25    & LIII   &    0&.18    &	0&.28    & LIII   &    0&.16 &    0&.22 & LIII &    0&.08 &    0&.16 & LIII \\
   BZ\,Cyg &	0&.38	 &    0&.51    &  LII	&    0&.24    &    0&.32    &  LII   &     &\ldots &     &\ldots & \ldots &    0&.29 &    0&.40 &  LII &    0&.17 &    0&.28 &  LII \\
   CD\,Cyg &	0&.34	 &    0&.44    & LIII	&    0&.34    &    0&.37    & LIII   &    0&.18    &	0&.28    & LIII   &    0&.27 &    0&.33 & LIII &    0&.22 &    0&.30 & LIII \\
   DT\,Cyg &	0&.28	 &    0&.41    &  LII	&    0&.13    &    0&.21    &  LII   & $-$0&.05    &	0&.18    &  LII   &    0&.12 &    0&.23 &  LII &    0&.09 &    0&.20 &  LII \\
   EP\,Cyg &	0&.12	 &    0&.22    & LIII	&    0&.01    &    0&.04    & LIII   & $-$0&.07    &	0&.03    & LIII   & $-$0&.01 &    0&.05 & LIII & $-$0&.01 &    0&.07 & LIII \\
   EU\,Cyg &	0&.01	 &    0&.11    & LIII	&    0&.07    &    0&.10    & LIII   & $-$0&.10    &	0&.00    & LIII   & $-$0&.04 &    0&.02 & LIII & $-$0&.17 & $-$0&.09 & LIII \\
   EX\,Cyg &	0&.44	 &    0&.54    & LIII	&    0&.21    &    0&.24    & LIII   & $-$0&.13    & $-$0&.03    & LIII   &    0&.20 &    0&.26 & LIII &    0&.01 &    0&.09 & LIII \\
   EZ\,Cyg &	0&.60	 &    0&.70    & LIII	&    0&.44    &    0&.47    & LIII   &    0&.45    &	0&.55    & LIII   &    0&.36 &    0&.42 & LIII &    0&.25 &    0&.33 & LIII \\
   GH\,Cyg &	0&.24	 &    0&.34    & LIII	&    0&.14    &    0&.17    & LIII   & $-$0&.02    &	0&.08    & LIII   &    0&.17 &    0&.23 & LIII &    0&.01 &    0&.09 & LIII \\
   GI\,Cyg &	0&.34	 &    0&.44    & LIII	&    0&.32    &    0&.35    & LIII   &    0&.49    &	0&.59    & LIII   &    0&.22 &    0&.28 & LIII &    0&.21 &    0&.29 & LIII \\
   GL\,Cyg &	0&.16	 &    0&.26    & LIII	&    0&.07    &    0&.10    & LIII   & $-$0&.14    & $-$0&.04    & LIII   &    0&.04 &    0&.10 & LIII &    0&.00 &    0&.08 & LIII \\
   IY\,Cyg &	0&.08	 &    0&.18    & LIII	&    0&.03    &    0&.06    & LIII   & $-$0&.11    & $-$0&.01    & LIII   & $-$0&.10 & $-$0&.04 & LIII & $-$0&.23 & $-$0&.15 & LIII \\
   KX\,Cyg &	0&.05	 &    0&.15    & LIII	&    0&.18    &    0&.21    & LIII   &    0&.09    &	0&.19    & LIII   &    0&.15 &    0&.21 & LIII & $-$0&.05 &    0&.03 & LIII \\
   MW\,Cyg &	0&.15	 &    0&.28    &  LII	&    0&.18    &    0&.26    &  LII   &    0&.08    &	0&.31    &  LII   &    0&.06 &    0&.17 &  LII & $-$0&.01 &    0&.10 &  LII \\
   SU\,Cyg &	0&.21	 &    0&.34    &  LII	&    0&.14    &    0&.22    &  LII   & $-$0&.16    &	0&.07    &  LII   &    0&.04 &    0&.15 &  LII &    0&.03 &    0&.14 &  LII \\
   SZ\,Cyg &	0&.44	 &    0&.57    &  LII	&    0&.22    &    0&.30    &  LII   &    0&.03    &	0&.26    &  LII   &    0&.25 &    0&.36 &  LII &    0&.19 &    0&.30 &  LII \\
   TX\,Cyg &	0&.32	 &    0&.45    &  LII	&    0&.21    &    0&.29    &  LII   &     &\ldots &     &\ldots & \ldots &    0&.22 &    0&.33 &  LII &    0&.15 &    0&.26 &  LII \\
V1020\,Cyg &	0&.44	 &    0&.54    & LIII	&    0&.20    &    0&.23    & LIII   & $-$0&.08    &	0&.02    & LIII   &    0&.18 &    0&.24 & LIII &    0&.12 &    0&.20 & LIII \\
V1025\,Cyg &	0&.24	 &    0&.34    & LIII	&    0&.25    &    0&.28    & LIII   &    0&.16    &	0&.26    & LIII   &    0&.20 &    0&.26 & LIII &    0&.06 &    0&.14 & LIII \\
V1033\,Cyg & $-$0&.01	 &    0&.09    & LIII	&    0&.14    &    0&.17    & LIII   &    0&.12    &	0&.22    & LIII   &    0&.14 &    0&.20 & LIII &    0&.06 &    0&.14 & LIII \\
V1046\,Cyg &	0&.37	 &    0&.47    & LIII	&    0&.24    &    0&.27    & LIII   &    0&.28    &	0&.38    & LIII   &    0&.23 &    0&.29 & LIII &    0&.08 &    0&.16 & LIII \\
V1154\,Cyg &	0&.11	 &    0&.24    &  LII	&    0&.09    &    0&.17    &  LII   &    0&.53    &	0&.76    &  LII   &    0&.04 &    0&.15 &  LII &    0&.13 &    0&.24 &  LII \\
V1334\,Cyg &	0&.19	 &    0&.32    &  LII	&    0&.17    &    0&.25    &  LII   & $-$0&.29    & $-$0&.06    &  LII   &    0&.06 &    0&.17 &  LII & $-$0&.02 &    0&.09 &  LII \\
V1364\,Cyg &	0&.42	 &    0&.52    & LIII	&    0&.38    &    0&.41    & LIII   &    0&.28    &	0&.38    & LIII   &    0&.31 &    0&.37 & LIII &    0&.10 &    0&.18 & LIII \\
V1397\,Cyg &	0&.17	 &    0&.27    & LIII	&    0&.15    &    0&.18    & LIII   &    0&.30    &	0&.40    & LIII   &    0&.11 &    0&.17 & LIII &    0&.01 &    0&.09 & LIII \\
V1726\,Cyg &	0&.29	 &    0&.42    &  LII	&    0&.07    &    0&.15    &  LII   & $-$0&.12    &	0&.11    &  LII   &    0&.11 &    0&.22 &  LII & $-$0&.16 & $-$0&.05 &  LII \\
 V347\,Cyg &	0&.36	 &    0&.46    & LIII	&    0&.14    &    0&.17    & LIII   &    0&.30    &	0&.40    & LIII   &    0&.22 &    0&.28 & LIII &    0&.12 &    0&.20 & LIII \\
 V356\,Cyg &	0&.33	 &    0&.43    & LIII	&    0&.22    &    0&.25    & LIII   &    0&.02    &	0&.12    & LIII   &    0&.19 &    0&.25 & LIII &    0&.09 &    0&.17 & LIII \\
 V386\,Cyg &	0&.23	 &    0&.36    &  LII	&    0&.15    &    0&.23    &  LII   &     &\ldots &     &\ldots & \ldots &    0&.14 &    0&.25 &  LII &    0&.02 &    0&.13 &  LII \\
 V396\,Cyg &	0&.30	 &    0&.40    & LIII	&    0&.32    &    0&.35    & LIII   &    0&.19    &	0&.29    & LIII   &    0&.17 &    0&.23 & LIII &    0&.29 &    0&.37 & LIII \\
 V402\,Cyg &	0&.40	 &    0&.53    &  LII	&    0&.19    &    0&.27    &  LII   &    0&.15    &	0&.38    &  LII   &    0&.08 &    0&.19 &  LII &    0&.12 &    0&.23 &  LII \\
\hline\noalign{\smallskip}
\multicolumn{26}{r}{\it {\footnotesize continued on next page}} \\
\end{tabular}}
\end{table*}
\addtocounter{table}{-1}
\begin{table*}[p]
\centering
\caption[]{continued.}
{\scriptsize 
\begin{tabular}{r r@{}l r@{}l c r@{}l r@{}l c r@{}l r@{}l c r@{}l r@{}l c r@{}l r@{}l c}
\noalign{\smallskip}\hline\hline\noalign{\smallskip}
Name &
\multicolumn{2}{c}{[Na/H]$_{\rm lit}$} & \multicolumn{2}{c}{[Na/H]} & Ref. &
\multicolumn{2}{c}{[Al/H]$_{\rm lit}$} & \multicolumn{2}{c}{[Al/H]} & Ref. &
\multicolumn{2}{c}{[Mg/H]$_{\rm lit}$} & \multicolumn{2}{c}{[Mg/H]} & Ref. &
\multicolumn{2}{c}{[Si/H]$_{\rm lit}$} & \multicolumn{2}{c}{[Si/H]} & Ref. &
\multicolumn{2}{c}{[Ca/H]$_{\rm lit}$} & \multicolumn{2}{c}{[Ca/H]} & Ref. \\
\noalign{\smallskip}\hline\noalign{\smallskip}
 V438\,Cyg &	0&.11	 &    0&.21    & LIII	&    0&.17    &    0&.20    & LIII   & $-$0&.27    & $-$0&.17    & LIII   &    0&.14 &    0&.20 & LIII & $-$0&.03 &    0&.05 & LIII \\
 V459\,Cyg &	0&.38	 &    0&.48    & LIII	&    0&.28    &    0&.31    & LIII   &    0&.26    &	0&.36    & LIII   &    0&.24 &    0&.30 & LIII &    0&.10 &    0&.18 & LIII \\
 V492\,Cyg &	0&.32	 &    0&.42    & LIII	&    0&.23    &    0&.26    & LIII   &    0&.18    &	0&.28    & LIII   &    0&.17 &    0&.23 & LIII &    0&.05 &    0&.13 & LIII \\
 V495\,Cyg &	0&.44	 &    0&.54    & LIII	&    0&.20    &    0&.23    & LIII   &     &\ldots &     &\ldots & \ldots &    0&.26 &    0&.32 & LIII &    0&.11 &    0&.19 & LIII \\
 V514\,Cyg &	0&.32	 &    0&.42    & LIII	&    0&.13    &    0&.16    & LIII   &    0&.18    &	0&.28    & LIII   &    0&.17 &    0&.23 & LIII &    0&.07 &    0&.15 & LIII \\
 V520\,Cyg &	0&.26	 &    0&.36    & LIII	&    0&.24    &    0&.27    & LIII   & $-$0&.05    &	0&.05    & LIII   &    0&.17 &    0&.23 & LIII &    0&.10 &    0&.18 & LIII \\
 V532\,Cyg &	0&.26	 &    0&.39    &  LII	&    0&.16    &    0&.24    &  LII   &    0&.14    &	0&.37    &  LII   &    0&.15 &    0&.26 &  LII &    0&.12 &    0&.23 &  LII \\
 V538\,Cyg &	0&.25	 &    0&.35    & LIII	&    0&.14    &    0&.17    & LIII   & $-$0&.06    &	0&.04    & LIII   &    0&.13 &    0&.19 & LIII &    0&.03 &    0&.11 & LIII \\
 V547\,Cyg &	0&.26	 &    0&.36    & LIII	&    0&.24    &    0&.27    & LIII   &    0&.24    &	0&.34    & LIII   &    0&.16 &    0&.22 & LIII &    0&.05 &    0&.13 & LIII \\
 V609\,Cyg &	0&.32	 &    0&.42    & LIII	&    0&.21    &    0&.24    & LIII   &    0&.04    &	0&.14    & LIII   &    0&.20 &    0&.26 & LIII &    0&.02 &    0&.10 & LIII \\
 V621\,Cyg &	0&.31	 &    0&.41    & LIII	&    0&.11    &    0&.14    & LIII   &    0&.22    &	0&.32    & LIII   &    0&.15 &    0&.21 & LIII &    0&.02 &    0&.10 & LIII \\
 V924\,Cyg & $-$0&.04	 &    0&.09    &  LII	&    0&.09    &    0&.17    &  LII   & $-$0&.38    & $-$0&.15    &  LII   & $-$0&.04 &    0&.07 &  LII & $-$0&.21 & $-$0&.10 &  LII \\
   VX\,Cyg &	0&.41	 &    0&.54    &  LII	&    0&.13    &    0&.21    &  LII   & $-$0&.16    &	0&.07    &  LII   &    0&.18 &    0&.29 &  LII &    0&.20 &    0&.31 &  LII \\
   VY\,Cyg &	0&.23	 &    0&.36    &  LII	&    0&.15    &    0&.23    &  LII   &    0&.19    &	0&.42    &  LII   &    0&.06 &    0&.17 &  LII &    0&.06 &    0&.17 &  LII \\
   VZ\,Cyg &	0&.26	 &    0&.39    &  LII	&    0&.39    &    0&.47    &  LII   &    0&.21    &	0&.44    &  LII   &    0&.10 &    0&.21 &  LII &    0&.03 &    0&.14 &  LII \\
    X\,Cyg &	0&.25	 &    0&.38    &  LII	&    0&.25    &    0&.33    &  LII   &    0&.04    &	0&.27    &  LII   &    0&.09 &    0&.20 &  LII &    0&.09 &    0&.20 &  LII \\
   EK\,Del & $-$1&.62	 & $-$1&.52    & LIII	& $-$0&.75    & $-$0&.72    & LIII   & $-$1&.53    & $-$1&.43    & LIII   & $-$1&.17 & $-$1&.11 & LIII & $-$1&.24 & $-$1&.16 & LIII \\
$\beta$\,Dor &	0&.07	 &    0&.20    &  LII	&    0&.07    &    0&.15    &  LII   & $-$0&.30    & $-$0&.07    &  LII   &    0&.00 &    0&.11 &  LII & $-$0&.18 & $-$0&.07 &  LII \\
   AA\,Gem &	0&.22	 &    0&.22    &   TS	& $-$0&.01    & $-$0&.01    &   TS   &    0&.42    &	0&.42    &   TS   & $-$0&.24 & $-$0&.19 &  LEM &    0&.10 &    0&.10 &   TS \\
   AD\,Gem &	0&.11	 &    0&.11    &   TS	& $-$0&.23    & $-$0&.23    &   TS   & $-$0&.02    & $-$0&.02    &   TS   &    0&.03 &    0&.03 &   TS & $-$0&.06 & $-$0&.06 &   TS \\
   BB\,Gem &	0&.09	 &    0&.19    & LIII	&    0&.05    &    0&.08    & LIII   &    0&.02    &	0&.12    & LIII   &    0&.01 &    0&.07 & LIII & $-$0&.04 &    0&.04 & LIII \\
   BW\,Gem & $-$0&.02	 & $-$0&.02    &   TS	& $-$0&.15    & $-$0&.15    &   TS   & $-$0&.07    & $-$0&.07    &   TS   & $-$0&.13 & $-$0&.13 &   TS & $-$0&.12 & $-$0&.12 &   TS \\
   DX\,Gem &	0&.12	 &    0&.12    &   TS	&    0&.10    &    0&.10    &   TS   & $-$0&.03    & $-$0&.03    &   TS   &    0&.06 &    0&.06 &   TS & $-$0&.03 & $-$0&.03 &   TS \\
   RZ\,Gem &	0&.19	 &    0&.19    &   TS	&    0&.04    &    0&.08    &  LEM   &    0&.03    &	0&.03    &   TS   &    0&.07 &    0&.07 &   TS & $-$0&.12 & $-$0&.12 &   TS \\
    W\,Gem &	0&.03	 &    0&.13    & LIII	& $-$0&.08    & $-$0&.05    & LIII   & $-$0&.16    & $-$0&.06    & LIII   & $-$0&.07 & $-$0&.01 & LIII & $-$0&.14 & $-$0&.06 & LIII \\
 $\zeta$\,Gem &	0&.31	 &    0&.41    & LIII	&    0&.17    &    0&.20    & LIII   &    0&.06    &	0&.16    & LIII   &    0&.07 &    0&.13 & LIII & $-$0&.10 & $-$0&.02 & LIII \\
   BB\,Her &	0&.48	 &    0&.58    & LIII	&    0&.29    &    0&.32    & LIII   &    0&.23    &	0&.33    & LIII   &    0&.25 &    0&.31 & LIII &    0&.14 &    0&.22 & LIII \\
   BG\,Lac &	0&.31	 &    0&.41    & LIII	&    0&.19    &    0&.22    & LIII   &    0&.22    &	0&.32    & LIII   &    0&.12 &    0&.18 & LIII & $-$0&.04 &    0&.04 & LIII \\
   DF\,Lac &	0&.23	 &    0&.33    & LIII	&    0&.12    &    0&.15    & LIII   &    0&.06    &	0&.16    & LIII   &    0&.10 &    0&.16 & LIII &    0&.01 &    0&.09 & LIII \\
   FQ\,Lac &	0&.11	 &    0&.21    & LIII	& $-$0&.50    & $-$0&.47    & LIII   &    0&.42    &	0&.52    & LIII   & $-$0&.05 &    0&.01 & LIII &    0&.04 &    0&.12 & LIII \\
   RR\,Lac &	0&.21	 &    0&.31    & LIII	&    0&.18    &    0&.21    & LIII   & $-$0&.12    & $-$0&.02    & LIII   &    0&.07 &    0&.13 & LIII &    0&.02 &    0&.10 & LIII \\
    V\,Lac &	0&.32	 &    0&.42    & LIII	&    0&.11    &    0&.14    & LIII   &    0&.16    &	0&.26    & LIII   &    0&.19 &    0&.25 & LIII &    0&.09 &    0&.17 & LIII \\
 V411\,Lac &	0&.34	 &    0&.44    & LIII	&    0&.08    &    0&.11    & LIII   &    0&.08    &	0&.18    & LIII   &    0&.04 &    0&.10 & LIII & $-$0&.07 &    0&.01 & LIII \\
    X\,Lac &	0&.36	 &    0&.46    & LIII	&    0&.11    &    0&.14    & LIII   &    0&.12    &	0&.22    & LIII   &    0&.16 &    0&.22 & LIII &    0&.08 &    0&.16 & LIII \\
    Y\,Lac &	0&.15	 &    0&.25    & LIII	&    0&.09    &    0&.12    & LIII   &    0&.16    &	0&.26    & LIII   &    0&.14 &    0&.20 & LIII &    0&.13 &    0&.21 & LIII \\
    Z\,Lac &	0&.50	 &    0&.60    & LIII	&    0&.22    &    0&.25    & LIII   &    0&.19    &	0&.29    & LIII   &    0&.19 &    0&.25 & LIII &    0&.14 &    0&.22 & LIII \\
   GH\,Lup &	0&.25	 &    0&.35    & LIII	&    0&.23    &    0&.26    & LIII   &    0&.12    &	0&.22    & LIII   &    0&.15 &    0&.21 & LIII & $-$0&.01 &    0&.07 & LIII \\
 V473\,Lyr & $-$0&.01	 &    0&.09    & LIII	& $-$0&.02    &    0&.01    & LIII   & $-$0&.07    &	0&.03    & LIII   &    0&.00 &    0&.06 & LIII & $-$0&.10 & $-$0&.02 & LIII \\
   AA\,Mon &	0&.21	 &    0&.31    & LIII	& $-$0&.13    & $-$0&.10    & LIII   & $-$0&.19    & $-$0&.09    & LIII   & $-$0&.04 &    0&.02 & LIII & $-$0&.17 & $-$0&.09 & LIII \\
   AC\,Mon &	0&.19	 &    0&.29    & LIII	&    0&.02    &    0&.05    & LIII   & $-$0&.05    &	0&.05    & LIII   &    0&.03 &    0&.09 & LIII & $-$0&.08 & $-$0&.00 & LIII \\
   BE\,Mon &	0&.34	 &    0&.34    &   TS	&    0&.13    &    0&.13    &   TS   &    0&.33    &	0&.33    &   TS   &    0&.14 &    0&.14 &   TS &    0&.08 &    0&.08 &   TS \\
   BV\,Mon &	0&.08	 &    0&.18    & LIII	&    0&.02    &    0&.06    &  LEM   &    0&.14    &	0&.10    &  LEM   &    0&.01 &    0&.06 &  LEM & $-$0&.03 &    0&.02 &  LEM \\
   CS\,Mon &	0&.06	 &    0&.16    & LIII	& $-$0&.01    &    0&.02    & LIII   & $-$0&.07    &	0&.03    & LIII   & $-$0&.07 & $-$0&.01 & LIII & $-$0&.21 & $-$0&.13 & LIII \\
   CU\,Mon &	0&.03	 &    0&.13    & LIII	& $-$0&.20    & $-$0&.17    & LIII   & $-$0&.23    & $-$0&.13    & LIII   & $-$0&.16 & $-$0&.10 & LIII & $-$0&.19 & $-$0&.11 & LIII \\
   CV\,Mon &	0&.31	 &    0&.31    &   TS	&    0&.15    &    0&.15    &   TS   &    0&.22    &	0&.22    &   TS   &    0&.24 &    0&.24 &   TS &    0&.18 &    0&.18 &   TS \\
   EE\,Mon & $-$0&.27	 & $-$0&.17    & LIII	& $-$0&.65    & $-$0&.62    & LIII   & $-$0&.53    & $-$0&.43    & LIII   & $-$0&.38 & $-$0&.32 & LIII & $-$0&.48 & $-$0&.40 & LIII \\
   EK\,Mon &	0&.22	 &    0&.32    & LIII	&    0&.02    &    0&.06    &  LEM   &    0&.05    &	0&.01    &  LEM   & $-$0&.06 & $-$0&.01 &  LEM & $-$0&.04 &    0&.01 &  LEM \\
   FG\,Mon &	0&.04	 &    0&.14    & LIII	&    0&.01    &    0&.04    & LIII   &    0&.05    &	0&.15    & LIII   & $-$0&.06 & $-$0&.00 & LIII & $-$0&.10 & $-$0&.02 & LIII \\
   FI\,Mon & $-$0&.02	 &    0&.08    & LIII	& $-$0&.10    & $-$0&.07    & LIII   & $-$0&.24    & $-$0&.14    & LIII   &    0&.06 &    0&.12 & LIII & $-$0&.08 & $-$0&.00 & LIII \\
   FT\,Mon &	0&.15	 &    0&.15    &   TS	& $-$0&.10    & $-$0&.10    &   TS   & $-$0&.08    & $-$0&.08    &   TS   & $-$0&.15 & $-$0&.09 & LIII &    0&.02 &    0&.02 &   TS \\
   SV\,Mon &	0&.63	 &    0&.63    &   TS	&    0&.16    &    0&.16    &   TS   &    0&.30    &	0&.30    &   TS   &    0&.21 &    0&.21 &   TS &    0&.24 &    0&.24 &   TS \\
    T\,Mon &	0&.43	 &    0&.53    & LIII	&    0&.45    &    0&.48    & LIII   &    0&.39    &	0&.49    & LIII   &    0&.28 &    0&.34 & LIII &    0&.23 &    0&.31 & LIII \\
   TW\,Mon &	0&.16	 &    0&.16    &   TS	& $-$0&.03    & $-$0&.03    &   TS   & $-$0&.04    & $-$0&.04    &   TS   & $-$0&.02 & $-$0&.02 &   TS &    0&.02 &    0&.02 &   TS \\
   TX\,Mon &	0&.23	 &    0&.23    &   TS	& $-$0&.08    & $-$0&.08    &   TS   & $-$0&.04    & $-$0&.04    &   TS   &    0&.06 &    0&.06 &   TS &    0&.04 &    0&.04 &   TS \\
   TY\,Mon &	0&.17	 &    0&.17    &   TS	& $-$0&.19    & $-$0&.19    &   TS   &    0&.03    &	0&.03    &   TS   &    0&.14 &    0&.14 &   TS &    0&.12 &    0&.12 &   TS \\
   TZ\,Mon &	0&.16	 &    0&.16    &   TS	&    0&.02    &    0&.02    &   TS   &    0&.11    &	0&.11    &   TS   &    0&.03 &    0&.03 &   TS &    0&.04 &    0&.04 &   TS \\
   UY\,Mon & $-$0&.08	 & $-$0&.06    &  LEM	& $-$0&.25    & $-$0&.21    &  LEM   & $-$0&.23    & $-$0&.27    &  LEM   & $-$0&.22 & $-$0&.17 &  LEM & $-$0&.28 & $-$0&.23 &  LEM \\
 V446\,Mon &	0&.03	 &    0&.13    & LIII	& $-$0&.18    & $-$0&.15    & LIII   & $-$0&.35    & $-$0&.25    & LIII   & $-$0&.24 & $-$0&.18 & LIII & $-$0&.33 & $-$0&.25 & LIII \\
 V447\,Mon & $-$0&.07	 &    0&.03    & LIII	& $-$0&.16    & $-$0&.13    & LIII   & $-$0&.40    & $-$0&.30    & LIII   & $-$0&.26 & $-$0&.20 & LIII & $-$0&.35 & $-$0&.27 & LIII \\
 V465\,Mon &	0&.26	 &    0&.26    &   TS	&    0&.06    &    0&.06    &   TS   & $-$0&.08    & $-$0&.08    &   TS   &    0&.12 &    0&.12 &   TS &    0&.04 &    0&.04 &   TS \\
 V484\,Mon & $-$0&.10	 &    0&.00    & LIII	& $-$0&.01    &    0&.02    & LIII   & $-$0&.13    & $-$0&.03    & LIII   & $-$0&.10 & $-$0&.04 & LIII & $-$0&.13 & $-$0&.05 & LIII \\
 V495\,Mon &	0&.04	 &    0&.04    &   TS	& $-$0&.06    & $-$0&.06    &   TS   &    0&.23    &	0&.19    &  LEM   & $-$0&.01 & $-$0&.01 &   TS & $-$0&.04 & $-$0&.04 &   TS \\
 V504\,Mon &	0&.07	 &    0&.17    & LIII	&    0&.04    &    0&.07    & LIII   & $-$0&.02    &	0&.08    & LIII   & $-$0&.02 &    0&.04 & LIII & $-$0&.09 & $-$0&.01 & LIII \\
 V508\,Mon &	0&.12	 &    0&.12    &   TS	&    0&.04    &    0&.04    &   TS   &    0&.01    &	0&.01    &   TS   &    0&.04 &    0&.04 &   TS &    0&.04 &    0&.04 &   TS \\
 V510\,Mon & $-$0&.05	 & $-$0&.05    &   TS	& $-$0&.16    & $-$0&.16    &   TS   & $-$0&.04    & $-$0&.04    &   TS   & $-$0&.09 & $-$0&.09 &   TS & $-$0&.15 & $-$0&.15 &   TS \\
 V526\,Mon &	0&.05	 &    0&.15    & LIII	& $-$0&.21    & $-$0&.18    & LIII   & $-$0&.14    & $-$0&.04    & LIII   & $-$0&.06 & $-$0&.00 & LIII & $-$0&.16 & $-$0&.08 & LIII \\
 V911\,Mon &	0&.20	 &    0&.30    & LIII	&    0&.04    &    0&.07    & LIII   &    0&.16    &	0&.26    & LIII   &    0&.06 &    0&.12 & LIII & $-$0&.03 &    0&.05 & LIII \\
   VZ\,Mon &	0&.11	 &    0&.21    & LIII	& $-$0&.02    &    0&.01    & LIII   &    0&.01    &	0&.11    & LIII   & $-$0&.12 & $-$0&.06 & LIII & $-$0&.21 & $-$0&.13 & LIII \\
   WW\,Mon &	0&.13	 &    0&.15    &  LEM	& $-$0&.11    & $-$0&.08    & LIII   & $-$0&.09    & $-$0&.13    &  LEM   & $-$0&.02 &    0&.03 &  LEM & $-$0&.26 & $-$0&.21 &  LEM \\
   XX\,Mon &	0&.51	 &    0&.51    &   TS	&    0&.05    &    0&.05    &   TS   &    0&.21    &	0&.21    &   TS   &    0&.17 &    0&.17 &   TS &    0&.23 &    0&.23 &   TS \\
   YY\,Mon & $-$0&.19	 & $-$0&.09    & LIII	& $-$0&.23    & $-$0&.20    & LIII   & $-$0&.10    &	0&.00    & LIII   & $-$0&.38 & $-$0&.32 & LIII & $-$0&.36 & $-$0&.28 & LIII \\
    R\,Mus &	0&.37	 &    0&.47    & LIII	&    0&.18    &    0&.21    & LIII   &    0&.23    &	0&.33    & LIII   &    0&.14 &    0&.20 & LIII &    0&.03 &    0&.11 & LIII \\
   RT\,Mus &	0&.20	 &    0&.30    & LIII	&    0&.07    &    0&.10    & LIII   &    0&.04    &	0&.14    & LIII   &    0&.14 &    0&.20 & LIII &    0&.03 &    0&.11 & LIII \\
    S\,Mus &	0&.29	 &    0&.39    & LIII	&    0&.15    &    0&.18    & LIII   &    0&.03    &	0&.13    & LIII   &    0&.08 &    0&.14 & LIII & $-$0&.04 &    0&.04 & LIII \\
   TZ\,Mus &	0&.14	 &    0&.24    & LIII	&    0&.05    &    0&.08    & LIII   &    0&.43    &	0&.53    & LIII   &    0&.17 &    0&.23 & LIII &    0&.05 &    0&.13 & LIII \\
\hline\noalign{\smallskip}
\multicolumn{26}{r}{\it {\footnotesize continued on next page}} \\
\end{tabular}}
\end{table*}
\addtocounter{table}{-1}
\begin{table*}[p]
\centering
\caption[]{continued.}
{\scriptsize 
\begin{tabular}{r r@{}l r@{}l c r@{}l r@{}l c r@{}l r@{}l c r@{}l r@{}l c r@{}l r@{}l c}
\noalign{\smallskip}\hline\hline\noalign{\smallskip}
Name &
\multicolumn{2}{c}{[Na/H]$_{\rm lit}$} & \multicolumn{2}{c}{[Na/H]} & Ref. &
\multicolumn{2}{c}{[Al/H]$_{\rm lit}$} & \multicolumn{2}{c}{[Al/H]} & Ref. &
\multicolumn{2}{c}{[Mg/H]$_{\rm lit}$} & \multicolumn{2}{c}{[Mg/H]} & Ref. &
\multicolumn{2}{c}{[Si/H]$_{\rm lit}$} & \multicolumn{2}{c}{[Si/H]} & Ref. &
\multicolumn{2}{c}{[Ca/H]$_{\rm lit}$} & \multicolumn{2}{c}{[Ca/H]} & Ref. \\
\noalign{\smallskip}\hline\noalign{\smallskip}
   UU\,Mus &	0&.27	 &    0&.37    & LIII	&    0&.16    &    0&.19    & LIII   &    0&.19    &	0&.29    & LIII   &    0&.17 &    0&.23 & LIII &    0&.06 &    0&.14 & LIII \\
   GU\,Nor &	0&.32	 &    0&.32    &   TS	&    0&.17    &    0&.17    &   TS   &    0&.18    &	0&.18    &   TS   &    0&.28 &    0&.28 &   TS &    0&.09 &    0&.09 &   TS \\
   IQ\,Nor &	0&.57	 &    0&.57    &   TS	&    0&.39    &    0&.39    &   TS   &    0&.36    &	0&.36    &   TS   &    0&.38 &    0&.38 &   TS &    0&.25 &    0&.25 &   TS \\
   QZ\,Nor &	0&.56	 &    0&.56    &   TS	&    0&.30    &    0&.30    &   TS   &    0&.39    &	0&.39    &   TS   &    0&.41 &    0&.41 &   TS &    0&.20 &    0&.20 &   TS \\
   RS\,Nor &	0&.46	 &    0&.46    &   TS	&    0&.26    &    0&.26    &   TS   &    0&.26    &	0&.26    &   TS   &    0&.39 &    0&.39 &   TS &    0&.22 &    0&.22 &   TS \\
    S\,Nor &	0&.35	 &    0&.45    & LIII	&    0&.20    &    0&.23    & LIII   &    0&.14    &	0&.24    & LIII   &    0&.16 &    0&.22 & LIII &    0&.05 &    0&.13 & LIII \\
   SY\,Nor &	0&.61	 &    0&.61    &   TS	&    0&.41    &    0&.41    &   TS   &    0&.37    &	0&.37    &   TS   &    0&.37 &    0&.37 &   TS &    0&.19 &    0&.19 &   TS \\
   TW\,Nor &	0&.58	 &    0&.58    &   TS	&    0&.24    &    0&.24    &   TS   &    0&.58    &	0&.58    &   TS   &    0&.25 &    0&.25 &   TS &    0&.05 &    0&.05 &   TS \\
    U\,Nor &	0&.31	 &    0&.41    & LIII	&    0&.15    &    0&.18    & LIII   &    0&.08    &	0&.18    & LIII   &    0&.14 &    0&.20 & LIII & $-$0&.06 &    0&.02 & LIII \\
 V340\,Nor &	0&.40	 &    0&.40    &   TS	&    0&.37    &    0&.37    &   TS   &    0&.12    &	0&.12    &   TS   &    0&.30 &    0&.30 &   TS &    0&.19 &    0&.19 &   TS \\
   BF\,Oph &	0&.31	 &    0&.41    & LIII	&    0&.07    &    0&.10    & LIII   &    0&.05    &	0&.15    & LIII   &    0&.18 &    0&.24 & LIII &    0&.04 &    0&.12 & LIII \\
    Y\,Oph &	0&.07	 &    0&.20    &  LII	&    0&.14    &    0&.22    &  LII   & $-$0&.27    & $-$0&.04    &  LII   &    0&.02 &    0&.13 &  LII & $-$0&.07 &    0&.04 &  LII \\
   CR\,Ori &	0&.23	 &    0&.33    & LIII	& $-$0&.10    & $-$0&.07    & LIII   & $-$0&.32    & $-$0&.22    & LIII   & $-$0&.14 & $-$0&.08 & LIII & $-$0&.18 & $-$0&.10 & LIII \\
   CS\,Ori &	0&.11	 &    0&.11    &   TS	& $-$0&.27    & $-$0&.27    &   TS   & $-$0&.25    & $-$0&.25    &   TS   & $-$0&.10 & $-$0&.10 &   TS & $-$0&.16 & $-$0&.16 &   TS \\
   DF\,Ori &	0&.00	 &    0&.10    & LIII	& $-$0&.23    & $-$0&.20    & LIII   & $-$0&.25    & $-$0&.15    & LIII   & $-$0&.16 & $-$0&.10 & LIII & $-$0&.21 & $-$0&.13 & LIII \\
   GQ\,Ori &	0&.20	 &    0&.33    &  LII	&    0&.17    &    0&.25    &  LII   &    0&.21    &	0&.44    &  LII   &    0&.05 &    0&.16 &  LII & $-$0&.14 & $-$0&.03 &  LII \\
   RS\,Ori &	0&.12	 &    0&.12    &   TS	&    0&.11    &    0&.11    &   TS   &    0&.41    &	0&.41    &   TS   &    0&.30 &    0&.30 &   TS &    0&.22 &    0&.22 &   TS \\
   AS\,Per &	0&.32	 &    0&.42    & LIII	&    0&.17    &    0&.20    & LIII   &    0&.46    &	0&.56    & LIII   &    0&.17 &    0&.23 & LIII &    0&.11 &    0&.19 & LIII \\
   AW\,Per &	0&.39	 &    0&.49    & LIII	&    0&.20    &    0&.23    & LIII   &    0&.12    &	0&.22    & LIII   &    0&.04 &    0&.10 & LIII & $-$0&.17 & $-$0&.09 & LIII \\
   BM\,Per &	0&.19	 &    0&.29    & LIII	&    0&.19    &    0&.22    & LIII   &    0&.13    &	0&.23    & LIII   &    0&.15 &    0&.21 & LIII &    0&.01 &    0&.09 & LIII \\
   CI\,Per &	 &\ldots &     &\ldots & \ldots & $-$0&.34    & $-$0&.31    & LIII   &     &\ldots &     &\ldots & \ldots & $-$0&.30 & $-$0&.24 & LIII & $-$0&.26 & $-$0&.18 & LIII \\
   DW\,Per &	0&.12	 &    0&.22    & LIII	& $-$0&.17    & $-$0&.14    & LIII   &    0&.05    &	0&.15    & LIII   &    0&.10 &    0&.16 & LIII &    0&.03 &    0&.11 & LIII \\
   GP\,Per & $-$0&.43	 & $-$0&.33    & LIII	& $-$0&.34    & $-$0&.31    & LIII   & $-$0&.27    & $-$0&.17    & LIII   & $-$0&.39 & $-$0&.33 & LIII & $-$0&.70 & $-$0&.62 & LIII \\
   HQ\,Per & $-$0&.08	 &    0&.02    & LIII	& $-$0&.22    & $-$0&.19    & LIII   & $-$0&.21    & $-$0&.11    & LIII   & $-$0&.23 & $-$0&.17 & LIII & $-$0&.30 & $-$0&.22 & LIII \\
   HZ\,Per &	0&.03	 &    0&.13    & LIII	& $-$0&.10    & $-$0&.07    & LIII   & $-$0&.13    & $-$0&.03    & LIII   & $-$0&.17 & $-$0&.11 & LIII & $-$0&.16 & $-$0&.08 & LIII \\
   MM\,Per &	0&.07	 &    0&.17    & LIII	& $-$0&.02    &    0&.01    & LIII   & $-$0&.06    &	0&.04    & LIII   & $-$0&.03 &    0&.03 & LIII & $-$0&.06 &    0&.02 & LIII \\
   OT\,Per &	0&.25	 &    0&.35    & LIII	& $-$0&.08    & $-$0&.05    & LIII   & $-$0&.29    & $-$0&.19    & LIII   & $-$0&.05 &    0&.01 & LIII & $-$0&.07 &    0&.01 & LIII \\
   SV\,Per &	0&.17	 &    0&.27    & LIII	&    0&.06    &    0&.09    & LIII   & $-$0&.03    &	0&.07    & LIII   &    0&.05 &    0&.11 & LIII & $-$0&.09 & $-$0&.01 & LIII \\
   SX\,Per &	0&.10	 &    0&.20    & LIII	&    0&.08    &    0&.11    & LIII   & $-$0&.01    &	0&.09    & LIII   & $-$0&.01 &    0&.05 & LIII & $-$0&.03 &    0&.05 & LIII \\
   UX\,Per &	0&.22	 &    0&.32    & LIII	& $-$0&.07    & $-$0&.04    & LIII   &    0&.03    &	0&.13    & LIII   &    0&.05 &    0&.11 & LIII &    0&.04 &    0&.12 & LIII \\
   UY\,Per &	0&.26	 &    0&.36    & LIII	&    0&.21    &    0&.24    & LIII   &    0&.07    &	0&.17    & LIII   &    0&.15 &    0&.21 & LIII &    0&.05 &    0&.13 & LIII \\
 V440\,Per &	0&.07	 &    0&.20    &  LII	&    0&.02    &    0&.10    &  LII   & $-$0&.39    & $-$0&.16    &  LII   & $-$0&.02 &    0&.09 &  LII & $-$0&.18 & $-$0&.07 &  LII \\
 V891\,Per &	0&.23	 &    0&.33    & LIII	&    0&.06    &    0&.09    & LIII   &    0&.02    &	0&.12    & LIII   &    0&.05 &    0&.11 & LIII &    0&.00 &    0&.08 & LIII \\
   VX\,Per &	0&.35	 &    0&.45    & LIII	&    0&.25    &    0&.28    & LIII   &    0&.06    &	0&.16    & LIII   &    0&.13 &    0&.19 & LIII &    0&.06 &    0&.14 & LIII \\
   VY\,Per &	0&.27	 &    0&.37    & LIII	&    0&.09    &    0&.12    & LIII   &    0&.05    &	0&.15    & LIII   &    0&.07 &    0&.13 & LIII &    0&.03 &    0&.11 & LIII \\
   AD\,Pup &	0&.04	 &    0&.06    &  LEM	& $-$0&.09    & $-$0&.05    &  LEM   &    0&.02    & $-$0&.02    &  LEM   & $-$0&.10 & $-$0&.05 &  LEM & $-$0&.02 &    0&.03 &  LEM \\
   AP\,Pup &	0&.06	 &    0&.08    &  LEM	& $-$0&.11    & $-$0&.07    &  LEM   & $-$0&.05    & $-$0&.09    &  LEM   & $-$0&.09 & $-$0&.04 &  LEM & $-$0&.14 & $-$0&.09 &  LEM \\
   AQ\,Pup &	0&.36	 &    0&.36    &   TS	& $-$0&.02    &    0&.02    &  LEM   &    0&.06    &	0&.06    &   TS   &    0&.25 &    0&.25 &   TS &    0&.05 &    0&.05 &   TS \\
   AT\,Pup &	0&.29	 &    0&.31    &  LEM	& $-$0&.20    & $-$0&.16    &  LEM   &    0&.19    &	0&.15    &  LEM   & $-$0&.04 &    0&.01 &  LEM & $-$0&.15 & $-$0&.10 &  LEM \\
   BC\,Pup &	0&.20	 &    0&.20    &   TS	& $-$0&.07    & $-$0&.07    &   TS   & $-$0&.14    & $-$0&.14    &   TS   & $-$0&.13 & $-$0&.13 &   TS & $-$0&.18 & $-$0&.18 &   TS \\
   BM\,Pup &	0&.17	 &    0&.17    &   TS	& $-$0&.02    & $-$0&.02    &   TS   &     &\ldots &     &\ldots & \ldots &    0&.04 &    0&.04 &   TS & $-$0&.01 & $-$0&.01 &   TS \\
   BN\,Pup &	0&.22	 &    0&.22    &   TS	&    0&.16    &    0&.16    &   TS   &    0&.23    &	0&.23    &   TS   &    0&.16 &    0&.16 &   TS &    0&.09 &    0&.09 &   TS \\
   CE\,Pup &	0&.18	 &    0&.28    & LIII	&    0&.07    &    0&.10    & LIII   &    0&.31    &	0&.41    & LIII   &    0&.02 &    0&.08 & LIII & $-$0&.08 & $-$0&.00 & LIII \\
   CK\,Pup &	0&.17	 &    0&.17    &   TS	& $-$0&.08    & $-$0&.08    &   TS   & $-$0&.02    & $-$0&.02    &   TS   &    0&.02 &    0&.02 &   TS & $-$0&.08 & $-$0&.08 &   TS \\
   HW\,Pup & $-$0&.02	 & $-$0&.02    &   TS	& $-$0&.12    & $-$0&.12    &   TS   & $-$0&.06    & $-$0&.06    &   TS   & $-$0&.05 & $-$0&.05 &   TS & $-$0&.11 & $-$0&.11 &   TS \\
   LS\,Pup &	 &\ldots &     &\ldots & \ldots &    0&.42    &    0&.42    &   TS   &    0&.10    &	0&.10    &   TS   &    0&.04 &    0&.04 &   TS &    0&.09 &    0&.09 &   TS \\
   MY\,Pup &	0&.13	 &    0&.15    &  LEM	& $-$0&.18    & $-$0&.14    &  LEM   & $-$0&.18    & $-$0&.22    &  LEM   & $-$0&.06 & $-$0&.01 &  LEM & $-$0&.14 & $-$0&.09 &  LEM \\
   NT\,Pup &	0&.34	 &    0&.44    & LIII	&    0&.01    &    0&.04    & LIII   & $-$0&.15    & $-$0&.05    & LIII   & $-$0&.05 &    0&.01 & LIII & $-$0&.20 & $-$0&.12 & LIII \\
   RS\,Pup &	0&.73	 &    0&.75    &  LEM	&    0&.19    &    0&.23    &  LEM   &    0&.21    &	0&.17    &  LEM   &    0&.31 &    0&.36 &  LEM & $-$0&.03 &    0&.02 &  LEM \\
 V335\,Pup &	0&.17	 &    0&.27    & LIII	&    0&.03    &    0&.06    & LIII   & $-$0&.04    &	0&.06    & LIII   &    0&.10 &    0&.16 & LIII & $-$0&.04 &    0&.04 & LIII \\
   VW\,Pup &	0&.11	 &    0&.11    &   TS	& $-$0&.04    & $-$0&.04    &   TS   & $-$0&.33    & $-$0&.33    &   TS   & $-$0&.03 & $-$0&.03 &   TS & $-$0&.08 & $-$0&.08 &   TS \\
   VX\,Pup &	0&.16	 &    0&.26    & LIII	& $-$0&.16    & $-$0&.12    &  LEM   & $-$0&.11    & $-$0&.15    &  LEM   & $-$0&.12 & $-$0&.07 &  LEM & $-$0&.10 & $-$0&.05 &  LEM \\
   VZ\,Pup &	0&.06	 &    0&.08    &  LEM	&    0&.30    &    0&.30    &   TS   &    0&.01    &	0&.01    &   TS   &    0&.25 &    0&.25 &   TS &    0&.11 &    0&.11 &   TS \\
   WW\,Pup & $-$0&.30	 & $-$0&.30    &   TS	& $-$0&.38    & $-$0&.38    &   TS   & $-$0&.34    & $-$0&.34    &   TS   & $-$0&.36 & $-$0&.36 &   TS & $-$0&.70 & $-$0&.70 &   TS \\
   WX\,Pup &	0&.22	 &    0&.24    &  LEM	& $-$0&.02    &    0&.02    &  LEM   & $-$0&.15    & $-$0&.05    & LIII   & $-$0&.12 & $-$0&.07 &  LEM & $-$0&.04 &    0&.01 &  LEM \\
   WY\,Pup &	0&.21	 &    0&.21    &   TS	& $-$0&.14    & $-$0&.14    &   TS   &    0&.05    &	0&.05    &   TS   & $-$0&.01 & $-$0&.01 &   TS &    0&.01 &    0&.01 &   TS \\
   WZ\,Pup &	0&.09	 &    0&.09    &   TS	& $-$0&.11    & $-$0&.11    &   TS   &    0&.14    &	0&.14    &   TS   &    0&.07 &    0&.07 &   TS &    0&.04 &    0&.04 &   TS \\
    X\,Pup &	0&.42	 &    0&.42    &   TS	&    0&.35    &    0&.35    &   TS   & $-$0&.01    & $-$0&.01    &   TS   & $-$0&.01 & $-$0&.01 &   TS &    0&.09 &    0&.09 &   TS \\
   KQ\,Sco &	0&.70	 &    0&.70    &   TS	&    0&.48    &    0&.48    &   TS   &     &\ldots &     &\ldots & \ldots &    0&.26 &    0&.26 &   TS &    0&.18 &    0&.18 &   TS \\
   RV\,Sco &	0&.41	 &    0&.51    & LIII	&    0&.20    &    0&.23    & LIII   &    0&.01    &	0&.11    & LIII   &    0&.16 &    0&.22 & LIII &    0&.11 &    0&.19 & LIII \\
   RY\,Sco &	0&.36	 &    0&.36    &   TS	&    0&.17    &    0&.17    &   TS   &    0&.08    &	0&.08    &   TS   &    0&.15 &    0&.15 &   TS & $-$0&.05 & $-$0&.05 &   TS \\
 V470\,Sco &	0&.55	 &    0&.55    &   TS	&    0&.26    &    0&.26    &   TS   &    0&.18    &	0&.18    &   TS   &    0&.23 &    0&.23 &   TS &    0&.08 &    0&.08 &   TS \\
 V482\,Sco &	0&.38	 &    0&.48    & LIII	&    0&.23    &    0&.26    & LIII   &    0&.15    &	0&.25    & LIII   &    0&.19 &    0&.25 & LIII &    0&.08 &    0&.16 & LIII \\
 V500\,Sco &	0&.19	 &    0&.19    &   TS	&    0&.12    &    0&.12    &   TS   & $-$0&.12    & $-$0&.12    &   TS   &    0&.03 &    0&.03 &   TS & $-$0&.14 & $-$0&.14 &   TS \\
 V636\,Sco &	0&.31	 &    0&.41    & LIII	&    0&.29    &    0&.32    & LIII   &    0&.09    &	0&.19    & LIII   &    0&.11 &    0&.17 & LIII & $-$0&.02 &    0&.06 & LIII \\
 V950\,Sco &	0&.27	 &    0&.37    & LIII	&    0&.11    &    0&.14    & LIII   &    0&.06    &	0&.16    & LIII   &    0&.15 &    0&.21 & LIII & $-$0&.02 &    0&.06 & LIII \\
   BX\,Sct &	0&.46	 &    0&.56    & LIII	&    0&.27    &    0&.30    & LIII   &    0&.27    &	0&.37    & LIII   &    0&.21 &    0&.27 & LIII &    0&.15 &    0&.23 & LIII \\
   CK\,Sct &	0&.32	 &    0&.42    & LIII	&    0&.25    &    0&.28    & LIII   &    0&.10    &	0&.20    & LIII   &    0&.13 &    0&.19 & LIII &    0&.00 &    0&.08 & LIII \\
   CM\,Sct &	0&.33	 &    0&.43    & LIII	&    0&.19    &    0&.22    & LIII   &    0&.03    &	0&.13    & LIII   &    0&.15 &    0&.21 & LIII &    0&.04 &    0&.12 & LIII \\
   CN\,Sct &	0&.56	 &    0&.66    & LIII	&    0&.45    &    0&.48    & LIII   &    0&.32    &	0&.42    & LIII   &    0&.32 &    0&.38 & LIII &    0&.17 &    0&.25 & LIII \\
   EV\,Sct &	0&.25	 &    0&.25    &   TS	&    0&.56    &    0&.56    &   TS   &    0&.06    &	0&.06    &   TS   &    0&.26 &    0&.26 &   TS &    0&.10 &    0&.10 &   TS \\
   EW\,Sct &	0&.07	 &    0&.20    &  LII	&    0&.15    &    0&.23    &  LII   & $-$0&.10    &	0&.13    &  LII   &    0&.08 &    0&.19 &  LII & $-$0&.01 &    0&.10 &  LII \\
   RU\,Sct &	0&.41	 &    0&.41    &   TS	&    0&.23    &    0&.23    &   TS   &    0&.42    &	0&.42    &   TS   &    0&.32 &    0&.32 &   TS &    0&.16 &    0&.16 &   TS \\
   SS\,Sct &	0&.15	 &    0&.25    & LIII	&    0&.18    &    0&.21    & LIII   &    0&.24    &	0&.34    & LIII   &    0&.14 &    0&.20 & LIII &    0&.06 &    0&.14 & LIII \\
   TY\,Sct &	0&.45	 &    0&.55    & LIII	&    0&.40    &    0&.43    & LIII   &    0&.29    &	0&.39    & LIII   &    0&.26 &    0&.32 & LIII &    0&.07 &    0&.15 & LIII \\
\hline\noalign{\smallskip}
\multicolumn{26}{r}{\it {\footnotesize continued on next page}} \\
\end{tabular}}
\end{table*}
\addtocounter{table}{-1}
\begin{table*}[p]
\centering
\caption[]{continued.}
{\scriptsize 
\begin{tabular}{r r@{}l r@{}l c r@{}l r@{}l c r@{}l r@{}l c r@{}l r@{}l c r@{}l r@{}l c}
\noalign{\smallskip}\hline\hline\noalign{\smallskip}
Name &
\multicolumn{2}{c}{[Na/H]$_{\rm lit}$} & \multicolumn{2}{c}{[Na/H]} & Ref. &
\multicolumn{2}{c}{[Al/H]$_{\rm lit}$} & \multicolumn{2}{c}{[Al/H]} & Ref. &
\multicolumn{2}{c}{[Mg/H]$_{\rm lit}$} & \multicolumn{2}{c}{[Mg/H]} & Ref. &
\multicolumn{2}{c}{[Si/H]$_{\rm lit}$} & \multicolumn{2}{c}{[Si/H]} & Ref. &
\multicolumn{2}{c}{[Ca/H]$_{\rm lit}$} & \multicolumn{2}{c}{[Ca/H]} & Ref. \\
\noalign{\smallskip}\hline\noalign{\smallskip}
   UZ\,Sct &	0&.79	 &    0&.79    &   TS	&    0&.60    &    0&.60    &   TS   &    0&.32    &	0&.32    &   TS   &    0&.45 &    0&.45 &   TS &    0&.22 &    0&.22 &   TS \\
 V367\,Sct &	0&.30	 &    0&.30    &   TS	&    0&.33    &    0&.33    &   TS   & $-$0&.01    & $-$0&.01    &   TS   &    0&.16 &    0&.16 &   TS &    0&.01 &    0&.01 &   TS \\
    X\,Sct &	0&.41	 &    0&.41    &   TS	&    0&.34    &    0&.34    &   TS   &    0&.12    &	0&.12    &   TS   &    0&.32 &    0&.32 &   TS &    0&.14 &    0&.14 &   TS \\
    Y\,Sct &	0&.38	 &    0&.48    & LIII	&    0&.28    &    0&.31    & LIII   &    0&.08    &	0&.18    & LIII   &    0&.16 &    0&.22 & LIII &    0&.03 &    0&.11 & LIII \\
    Z\,Sct &	0&.71	 &    0&.71    &   TS	&    0&.51    &    0&.51    &   TS   &    0&.14    &	0&.14    &   TS   &    0&.33 &    0&.33 &   TS &    0&.13 &    0&.13 &   TS \\
   AA\,Ser &	0&.98	 &    0&.98    &   TS	&    0&.50    &    0&.50    &   TS   &    0&.28    &	0&.28    &   TS   &    0&.47 &    0&.47 &   TS &    0&.16 &    0&.16 &   TS \\
   BQ\,Ser &	0&.12	 &    0&.25    &  LII	&    0&.14    &    0&.22    &  LII   & $-$0&.14    &	0&.09    &  LII   &    0&.07 &    0&.18 &  LII & $-$0&.05 &    0&.06 &  LII \\
   CR\,Ser &	0&.61	 &    0&.61    &   TS	&    0&.29    &    0&.29    &   TS   &    0&.32    &	0&.32    &   TS   &    0&.28 &    0&.28 &   TS &    0&.11 &    0&.11 &   TS \\
   DV\,Ser &	0&.72	 &    0&.82    & LIII	&    0&.59    &    0&.62    & LIII   &    0&.67    &	0&.77    & LIII   &    0&.46 &    0&.52 & LIII &    0&.37 &    0&.45 & LIII \\
   DG\,Sge &	0&.38	 &    0&.48    & LIII	&    0&.10    &    0&.13    & LIII   &    0&.14    &	0&.24    & LIII   &    0&.18 &    0&.24 & LIII &    0&.09 &    0&.17 & LIII \\
   GX\,Sge &	0&.27	 &    0&.37    & LIII	&    0&.31    &    0&.34    & LIII   &    0&.16    &	0&.26    & LIII   &    0&.21 &    0&.27 & LIII &    0&.01 &    0&.09 & LIII \\
   GY\,Sge &	0&.29	 &    0&.39    & LIII	&    0&.24    &    0&.27    & LIII   &    0&.15    &	0&.25    & LIII   &    0&.26 &    0&.32 & LIII &    0&.17 &    0&.25 & LIII \\
    S\,Sge &	0&.20	 &    0&.33    &  LII	&    0&.11    &    0&.19    &  LII   & $-$0&.17    &	0&.06    &  LII   &    0&.10 &    0&.21 &  LII & $-$0&.03 &    0&.08 &  LII \\
   AP\,Sgr &	0&.47	 &    0&.60    &  LII	&    0&.08    &    0&.16    &  LII   &    0&.25    &	0&.48    &  LII   &    0&.27 &    0&.38 &  LII &    0&.24 &    0&.35 &  LII \\
   AV\,Sgr &	0&.87	 &    0&.87    &   TS	&    0&.50    &    0&.50    &   TS   &    0&.73    &	0&.69    &  LEM   &    0&.47 &    0&.47 &   TS &    0&.18 &    0&.18 &   TS \\
   AY\,Sgr &	0&.32	 &    0&.32    &   TS	&    0&.19    &    0&.19    &   TS   &    0&.20    &	0&.20    &   TS   &    0&.23 &    0&.23 &   TS &    0&.11 &    0&.11 &   TS \\
   BB\,Sgr &	0&.36	 &    0&.49    &  LII	&    0&.15    &    0&.23    &  LII   &     &\ldots &     &\ldots & \ldots &    0&.14 &    0&.25 &  LII &    0&.12 &    0&.23 &  LII \\
    U\,Sgr &	0&.21	 &    0&.34    &  LII	&    0&.21    &    0&.29    &  LII   & $-$0&.17    &	0&.06    &  LII   &    0&.08 &    0&.19 &  LII & $-$0&.04 &    0&.07 &  LII \\
V1954\,Sgr &	0&.62	 &    0&.62    &   TS	&    0&.27    &    0&.27    &   TS   &    0&.17    &	0&.17    &   TS   &    0&.47 &    0&.47 &   TS &    0&.25 &    0&.25 &   TS \\
 V350\,Sgr &	0&.23	 &    0&.36    &  LII	&    0&.32    &    0&.40    &  LII   &    0&.19    &	0&.42    &  LII   &    0&.13 &    0&.24 &  LII &    0&.08 &    0&.19 &  LII \\
 V773\,Sgr &	0&.30	 &    0&.30    &   TS	&    0&.05    &    0&.05    &   TS   &    0&.20    &	0&.20    &   TS   &    0&.30 &    0&.30 &   TS &    0&.14 &    0&.14 &   TS \\
   VY\,Sgr &	0&.78	 &    0&.78    &   TS	&    0&.66    &    0&.66    &   TS   &    0&.36    &	0&.36    &   TS   &    0&.28 &    0&.28 &   TS &    0&.18 &    0&.18 &   TS \\
    W\,Sgr &	0&.18	 &    0&.31    &  LII	&    0&.07    &    0&.15    &  LII   & $-$0&.25    & $-$0&.02    &  LII   &    0&.05 &    0&.16 &  LII & $-$0&.07 &    0&.04 &  LII \\
   WZ\,Sgr &	0&.59	 &    0&.59    &   TS	&    0&.57    &    0&.57    &   TS   &    0&.02    &	0&.02    &   TS   &    0&.30 &    0&.30 &   TS &    0&.17 &    0&.17 &   TS \\
   XX\,Sgr &	0&.29	 &    0&.29    &   TS	&    0&.24    &    0&.24    &   TS   & $-$0&.01    & $-$0&.01    &   TS   &    0&.11 &    0&.11 &   TS &    0&.00 &    0&.00 &   TS \\
    Y\,Sgr &	0&.27	 &    0&.40    &  LII	&    0&.23    &    0&.31    &  LII   & $-$0&.08    &	0&.15    &  LII   &    0&.09 &    0&.20 &  LII &    0&.05 &    0&.16 &  LII \\
   YZ\,Sgr &	0&.25	 &    0&.38    &  LII	&    0&.17    &    0&.25    &  LII   & $-$0&.17    &	0&.06    &  LII   &    0&.11 &    0&.22 &  LII & $-$0&.02 &    0&.09 &  LII \\
   AE\,Tau &	0&.05	 &    0&.15    & LIII	& $-$0&.06    & $-$0&.03    & LIII   & $-$0&.15    & $-$0&.05    & LIII   & $-$0&.11 & $-$0&.05 & LIII & $-$0&.14 & $-$0&.06 & LIII \\
   AV\,Tau &	0&.17	 &    0&.19    &  LEM	&    0&.09    &    0&.13    &  LEM   &    0&.12    &	0&.08    &  LEM   &    0&.04 &    0&.09 &  LEM & $-$0&.03 &    0&.02 &  LEM \\
   EF\,Tau & $-$0&.48	 & $-$0&.35    &  LII	& $-$0&.46    & $-$0&.38    &  LII   & $-$0&.74    & $-$0&.51    &  LII   & $-$0&.65 & $-$0&.54 &  LII & $-$0&.62 & $-$0&.51 &  LII \\
   EU\,Tau &	0&.24	 &    0&.37    &  LII	& $-$0&.01    &    0&.07    &  LII   & $-$0&.28    & $-$0&.05    &  LII   &    0&.04 &    0&.15 &  LII & $-$0&.05 &    0&.06 &  LII \\
   ST\,Tau &	0&.29	 &    0&.39    & LIII	&    0&.11    &    0&.15    &  LEM   &    0&.08    &	0&.04    &  LEM   &    0&.08 &    0&.13 &  LEM &    0&.06 &    0&.11 &  LEM \\
   SZ\,Tau &	0&.27	 &    0&.40    &  LII	&    0&.10    &    0&.18    &  LII   & $-$0&.18    &	0&.05    &  LII   &    0&.07 &    0&.18 &  LII & $-$0&.02 &    0&.09 &  LII \\
   LR\,TrA &	0&.15	 &    0&.25    & LIII	&    0&.36    &    0&.39    & LIII   &    0&.47    &	0&.57    & LIII   &    0&.28 &    0&.34 & LIII &    0&.22 &    0&.30 & LIII \\
    R\,TrA &	0&.49	 &    0&.59    & LIII	&    0&.12    &    0&.15    & LIII   &    0&.21    &	0&.31    & LIII   &    0&.23 &    0&.29 & LIII &    0&.10 &    0&.18 & LIII \\
    S\,TrA &	0&.41	 &    0&.51    & LIII	&    0&.16    &    0&.19    & LIII   &    0&.28    &	0&.38    & LIII   &    0&.14 &    0&.20 & LIII & $-$0&.03 &    0&.05 & LIII \\
   AE\,Vel &	0&.14	 &    0&.24    & LIII	&    0&.20    &    0&.23    & LIII   &    0&.06    &	0&.16    & LIII   &    0&.03 &    0&.09 & LIII & $-$0&.20 & $-$0&.12 & LIII \\
   AH\,Vel &	0&.53	 &    0&.55    &  LEM	&    0&.16    &    0&.20    &  LEM   &    0&.14    &	0&.10    &  LEM   &    0&.12 &    0&.17 &  LEM &    0&.04 &    0&.09 &  LEM \\
   AX\,Vel &	 &\ldots &     &\ldots & \ldots &    0&.12    &    0&.16    &  LEM   &    0&.13    &	0&.09    &  LEM   &    0&.20 &    0&.25 &  LEM &    0&.03 &    0&.08 &  LEM \\
   BG\,Vel &	0&.21	 &    0&.31    & LIII	& $-$0&.11    & $-$0&.07    &  LEM   &    0&.10    &	0&.20    & LIII   &    0&.09 &    0&.14 &  LEM &    0&.08 &    0&.13 &  LEM \\
   CS\,Vel &	0&.27	 &    0&.37    & LIII	&    0&.26    &    0&.29    & LIII   &    0&.08    &	0&.18    & LIII   &    0&.12 &    0&.18 & LIII & $-$0&.08 & $-$0&.00 & LIII \\
   CX\,Vel &	0&.33	 &    0&.43    & LIII	&    0&.21    &    0&.24    & LIII   &    0&.52    &	0&.62    & LIII   &    0&.17 &    0&.23 & LIII &    0&.10 &    0&.18 & LIII \\
   DK\,Vel &	0&.21	 &    0&.31    & LIII	&    0&.11    &    0&.14    & LIII   &    0&.10    &	0&.20    & LIII   &    0&.15 &    0&.21 & LIII &    0&.02 &    0&.10 & LIII \\
   DR\,Vel &	0&.29	 &    0&.31    &  LEM	&    0&.26    &    0&.30    &  LEM   &    0&.21    &	0&.17    &  LEM   &    0&.20 &    0&.25 &  LEM &    0&.26 &    0&.31 &  LEM \\
   EX\,Vel &	0&.15	 &    0&.25    & LIII	&    0&.15    &    0&.18    & LIII   &    0&.07    &	0&.17    & LIII   &    0&.08 &    0&.14 & LIII &    0&.01 &    0&.09 & LIII \\
   EZ\,Vel &	0&.17	 &    0&.17    &   TS	&    0&.01    &    0&.01    &   TS   & $-$0&.03    & $-$0&.03    &   TS   &    0&.02 &    0&.07 &  LEM & $-$0&.08 & $-$0&.08 &   TS \\
   FG\,Vel &	0&.17	 &    0&.27    & LIII	&    0&.15    &    0&.18    & LIII   &    0&.01    &	0&.11    & LIII   &    0&.02 &    0&.08 & LIII & $-$0&.18 & $-$0&.10 & LIII \\
   FN\,Vel &	0&.14	 &    0&.24    & LIII	&    0&.15    &    0&.18    & LIII   &    0&.23    &	0&.33    & LIII   &    0&.14 &    0&.20 & LIII &    0&.09 &    0&.17 & LIII \\
   RY\,Vel &	0&.44	 &    0&.46    &  LEM	&    0&.40    &    0&.44    &  LEM   &    0&.01    &	0&.11    & LIII   &    0&.18 &    0&.23 &  LEM &    0&.13 &    0&.18 &  LEM \\
   RZ\,Vel &	0&.35	 &    0&.45    & LIII	&    0&.45    &    0&.48    & LIII   &    0&.43    &	0&.39    &  LEM   &    0&.28 &    0&.33 &  LEM & $-$0&.23 & $-$0&.18 &  LEM \\
   ST\,Vel &	0&.28	 &    0&.38    & LIII	&    0&.09    &    0&.13    &  LEM   &    0&.26    &	0&.22    &  LEM   &    0&.12 &    0&.17 &  LEM &    0&.09 &    0&.14 &  LEM \\
   SV\,Vel &	0&.39	 &    0&.49    & LIII	&    0&.10    &    0&.13    & LIII   &    0&.11    &	0&.21    & LIII   &    0&.17 &    0&.23 & LIII &    0&.10 &    0&.18 & LIII \\
   SW\,Vel &	0&.21	 &    0&.23    &  LEM	&    0&.28    &    0&.32    &  LEM   &    0&.28    &	0&.24    &  LEM   & $-$0&.03 &    0&.02 &  LEM &    0&.05 &    0&.10 &  LEM \\
   SX\,Vel &	0&.32	 &    0&.34    &  LEM	& $-$0&.10    & $-$0&.06    &  LEM   & $-$0&.03    &	0&.07    & LIII   &    0&.13 &    0&.18 &  LEM & $-$0&.01 &    0&.04 &  LEM \\
    T\,Vel &	0&.33	 &    0&.35    &  LEM	&    0&.29    &    0&.33    &  LEM   &    0&.32    &	0&.28    &  LEM   &    0&.28 &    0&.33 &  LEM &    0&.29 &    0&.34 &  LEM \\
    V\,Vel & $-$0&.01	 &    0&.01    &  LEM	& $-$0&.06    & $-$0&.02    &  LEM   &    0&.05    &	0&.01    &  LEM   & $-$0&.12 & $-$0&.07 &  LEM & $-$0&.11 & $-$0&.06 &  LEM \\
   XX\,Vel &	0&.21	 &    0&.31    & LIII	&    0&.17    &    0&.20    & LIII   &    0&.11    &	0&.21    & LIII   &    0&.14 &    0&.20 & LIII &    0&.00 &    0&.08 & LIII \\
   AS\,Vul &	0&.62	 &    0&.72    & LIII	&    0&.42    &    0&.45    & LIII   &    0&.29    &	0&.39    & LIII   &    0&.30 &    0&.36 & LIII &    0&.27 &    0&.35 & LIII \\
   DG\,Vul &	0&.42	 &    0&.52    & LIII	&    0&.21    &    0&.24    & LIII   &    0&.26    &	0&.36    & LIII   &    0&.21 &    0&.27 & LIII &    0&.05 &    0&.13 & LIII \\
    S\,Vul &	0&.22	 &    0&.32    & LIII	&    0&.27    &    0&.30    & LIII   &    0&.06    &	0&.16    & LIII   &    0&.11 &    0&.17 & LIII &    0&.03 &    0&.11 & LIII \\
   SV\,Vul &	0&.05	 &    0&.18    &  LII	&    0&.12    &    0&.20    &  LII   & $-$0&.17    &	0&.06    &  LII   &    0&.04 &    0&.15 &  LII & $-$0&.04 &    0&.07 &  LII \\
    T\,Vul &	0&.15	 &    0&.28    &  LII	&    0&.10    &    0&.18    &  LII   & $-$0&.13    &	0&.10    &  LII   &    0&.04 &    0&.15 &  LII & $-$0&.01 &    0&.10 &  LII \\
    U\,Vul &	0&.28	 &    0&.38    & LIII	&    0&.17    &    0&.20    & LIII   &    0&.09    &	0&.19    & LIII   &    0&.16 &    0&.22 & LIII &    0&.01 &    0&.09 & LIII \\
    X\,Vul &	0&.17	 &    0&.30    &  LII	&    0&.14    &    0&.22    &  LII   & $-$0&.20    &	0&.03    &  LII   &    0&.09 &    0&.20 &  LII & $-$0&.02 &    0&.09 &  LII \\
\hline
\end{tabular}}
\end{table*}

\end{document}